\documentclass[acmsmall]{acmart}

\AtBeginDocument{%
  }

\setcopyright{acmcopyright}
\acmJournal{TOIS}
\acmYear{2025} \acmVolume{44} \acmNumber{1} \acmArticle{23} \acmMonth{12} \acmPrice{}\acmDOI{10.1145/3772275}

\usepackage{booktabs} 
\usepackage{xspace}
\usepackage{enumitem}
\usepackage{subcaption}
\usepackage{multirow}

\newcommand{\idest}{i.e.,\xspace}
\newcommand{\eg}{e.g.,\xspace}

\newcommand{\palpha}{\ensuremath{\text{P}^3\!\alpha}\xspace}
\newcommand{\pbeta}{\ensuremath{\text{RP}^3\!\beta}\xspace}
\newcommand{\iALS}{iALS\xspace}
\newcommand{\EASER}{EASE$^R$\xspace}

\newcommand{\tableArticleResult}[5]{
\begin{table}[h]
    \caption{Results for #2 on the #4 dataset. The baselines are highlighted in \textbf{bold} if they outperform our results for #2. Results for #2 are highlighted in \textbf{bold} only if they outperform all baselines. Values are \underline{underlined} if they are better than the results for #2 that were reported in the original paper.}
    \label{tab:#2-#3-result}
    \footnotesize
    \centering
    
    \ifnum#5=0
        \input{tables/#1_#3_article_metrics_latex_results.txt}
    \else
        \resizebox{\linewidth}{!}{%
        \input{tables/#1_#3_article_metrics_latex_results.txt}
    }    
    \fi
\end{table}
}

\usepackage[title]{appendix}

\begin{document}

\title{Reproducibility and Artifact Consistency of the SIGIR 2022 Recommender Systems Papers Based on Message Passing}

\author{Maurizio {Ferrari Dacrema}}
\orcid{0000-0001-7103-2788}
\affiliation{%
    \department{ReMAP Lab}
    \institution{Politecnico di Milano}
    \city{Milano}
    \country{Italy}
}
\email{maurizio.ferrari@polimi.it}

\author{Michael Benigni}
\orcid{0009-0002-2810-5800}
\affiliation{%
    \department{ReMAP Lab}
    \institution{Politecnico di Milano}
    \country{Italy}
    \city{Milano}
}
\email{michael.benigni@polimi.it}

\author{Nicola Ferro}
\orcid{0000-0001-9219-6239}
\affiliation{%
  \institution{Università degli Studi di Padova}
  \country{Italy}
  \city{Padova}
}
\email{ferro@dei.unipd.it}


\begin{abstract}
Graph-based techniques relying on neural networks and embeddings have gained attention as a way to develop Recommender Systems (RS) with several papers on the topic presented at SIGIR 2022 and 2023. Given the importance of ensuring that published research is methodologically sound and reproducible, in this paper we analyze 10 graph-based RS papers, most of which were published at SIGIR 2022, and assess their impact on subsequent work published in SIGIR 2023.
Our analysis reveals several critical points that require attention: (i) the prevalence of bad practices, such as erroneous data splits or information leakage between training and testing data, which call into question the validity of the results; (ii) frequent inconsistencies between the provided artifacts (source code and data) and their descriptions in the paper, causing uncertainty about what is actually being evaluated; and (iii) the preference for new or complex baselines that are weaker compared to simpler ones, creating the impression of continuous improvement even when, particularly for the Amazon-Book dataset, the state-of-the-art has significantly worsened. Due to these issues, we are unable to confirm the claims made in most of the papers that we examined and attempted to reproduce.
\end{abstract}




\makeatletter
\let\origaddcontentsline\addcontentsline
\let\addcontentsline\@gobblethree
\maketitle
\let\addcontentsline\origaddcontentsline
\makeatother

\makeatletter
\begingroup
  \let\addcontentsline\@gobblethree
  \section{Introduction}
\label{sec:introduction}

Reproducibility is a central issue in several areas of science that face the so-called \emph{reproducibility crisis}~\cite{OSC2015}. As reported by~\citet{Baker2016}, approximately $70\%$ of researchers in physics and engineering are unable to reproduce someone else's experiments, and roughly $50\%$ fail to reproduce even their own experiments. Computational and data-intensive sciences~\citep{zz-DagstuhlSeminar16041,NAP2019} are no exception, and this is also true for \emph{Information Retrieval} (IR) and \emph{Recommender Systems} (RS)~\cite{BreuerEtAl2020,DBLP:conf/recsys/DacremaCJ19,Ferro2016d,Lin2022}, especially considering how both fields are heavily reliant on machine learning approaches today~\cite{LucicEtAl2022} and how reproducibility is a concern for machine learning itself~\cite{Gibney2020,PineauEtAl2021}. 

Defining what \emph{reproducibility} means for IR and RS is still an open issue. However, despite the terminological debate, it is generally understood as the ability to obtain results very similar to those reported in the paper, whether in terms of the absolute values of the effectiveness scores or the actual labels (or recommendation lists) produced by the model \cite{BreuerEtAl2020,MaistroEtAl2023,DBLP:conf/sigir/Breuer0FMSSS20}. One of the first issues that arises is ensuring that the original source code and data artifacts provided by the authors are, in fact, \emph{consistent} with what was described in the paper. There can be several sources of inconsistencies, and indeed we report several examples of this, from erroneous data splits (an issue already identified by \citet{DBLP:journals/tois/DacremaBCJ21}) to changes in the model implementation after the publication of the paper. These types of inconsistencies are problematic because they hide \emph{what} is actually being \emph{reproduced}, whether it is what is described in the paper or something else, and may easily propagate to subsequent publications if other researchers use those artifacts themselves.
Once the consistency of the artifacts has been established, the experimental evaluation can proceed to obtain the results and determine whether they reproduce what is reported in the original paper according to the desired criteria. This approach focuses on the reproduced method but does not consider the context in which the result was presented. Indeed, in machine learning research the results are generally interpreted in a \emph{relative} manner; that is, the absolute value of an effectiveness metric does not provide much information on its own but becomes meaningful when compared with the same effectiveness score measured for other methods. Typically, this is used to support claims that the proposed method is able to outperform the state-of-the-art in a given scenario.
This opens a new challenge that goes beyond reproducing the absolute results, that is, confirming the \emph{conclusions} that the proposed method is indeed competitive with the state-of-the-art. This requires reevaluating several aspects of the experimental protocol, from the training procedure used for the proposed method to the way in which the state-of-the-art baselines were selected and optimized. 
Several previous studies over the years have shown that a large number of papers present methods that are, in fact, not competitive against simple baselines due to various bad practices, such as poor baseline optimization, information leakage, and anomalous data splits
~\cite{ArmstrongEtAl2009b,DBLP:conf/recsys/DacremaCJ19,DBLP:journals/tois/DacremaBCJ21,DBLP:conf/recsys/RendleKZA20,KharazmiEtAl2016,LvEtAl2021,YangEtAl2019,DBLP:conf/recsys/AnelliBNP21,DBLP:conf/cikm/DacremaPCJ20,DBLP:conf/sigir/ShehzadDJ25,DBLP:journals/corr/abs-2505-09364}. 
Indeed, back in 2009 ~\citet{ArmstrongEtAl2009b} raised the issue by observing how new methods in IR were not competitive against some older and simpler baselines. More than 10 years later, this was found to still be the case by \citet{KharazmiEtAl2016} and \citet{LvEtAl2021}.  Similar issues have been reported in the RS domain by \citet{DBLP:conf/recsys/DacremaCJ19,DBLP:journals/tois/DacremaBCJ21}, who observed that it was not possible to reproduce several deep learning techniques in the RS domain and that most of them were not competitive with a set of simple fine-tuned baselines when applied to the traditional top-n collaborative filtering task.

In the last few years, graph-based techniques for neural networks and embeddings~\cite{AbadalEtAl2022,BarrosEtAl2023,RossiEtAl2021} have become a very active and impactful area of research with many ramifications including RS~\cite{WuEtAl2023}, which is the focus of this paper. To the best of our knowledge, \citet{DBLP:conf/sigir/0001DWLZ020} published the first paper on these topics at a SIGIR conference, which subsequently attracted attention and inspired follow-up work in later years, particularly at SIGIR 2022~\cite{DBLP:conf/sigir/PengSM22,DBLP:conf/sigir/YuY00CN22,DBLP:conf/sigir/TianXLYZ22,DBLP:conf/sigir/WuCSTC22,DBLP:conf/sigir/XiaHXZYH22,DBLP:conf/sigir/0002ZCZG22,DBLP:conf/sigir/FanL0ZT022,DBLP:conf/sigir/YangHXL22,DBLP:conf/sigir/0001DWLZ020,DBLP:conf/sigir/LiuWZS22} and SIGIR 2023 \cite{ZhuEtAl2023b, WangEtAl2023, ZhuEtAl2023, YangEtAl2023, RenEtAl2023, HeEtAl2023, LiEtAl2023b,LiuEtAl2023, ChoiEtAl2023, WeiEtAl2023, RenEtAl2023b}.

Given the general interest in graph-based techniques for RS, the surge of papers on this topic at SIGIR, and the previous experiences and concerns of the IR and RS communities regarding reproducibility and weak baselines, the question of whether the reproducibility of the results and reliability of the experimental methodology has improved naturally arises and papers addressing these questions are starting to appear, \eg 
\citet{DBLP:conf/recsys/AnelliMPBSN23}.
Therefore, in this paper, we investigate the following research questions about a set of graph-based RS papers published at SIGIR 2022:
 \begin{itemize}[nosep,left= 0pt]
    \item \textbf{Can we reproduce the results reported in the original papers using the same data splits and source code?} Our results indicate that slightly less than half of the reported results can be reproduced, with large variations across papers ranging from 0\% to 66\%, consistent with observations from prior studies in related fields.
    \item \textbf{Are the experimental methodologies used in the original papers, as well as the publicly available artifacts, correct and consistent?}  While 90\% of the papers provided publicly available artifacts, we identified several major inconsistencies related to anomalous data splitting. The consistency of model implementations is generally good, with rare exceptions. However, the early-stopping process is often inconsistent with the descriptions in the papers, and in some cases even relies on test data to select the number of training epochs.
    \item \textbf{Can we confirm that the proposed methods are competitive against simple yet robust baselines, beyond those reported in the original papers?} We observe that most methods fail to outperform simple baselines, particularly on the Amazon-Book dataset, where message-passing models fall substantially behind. In some cases, the analyzed method remains highly competitive, being outperformed by only a small margin by a single baseline. However, we also find several instances where a simple ItemKNN largely outperforms them.
\end{itemize}

Moreover, we asked ourselves about the impact that (ir)reproducible papers might have on follow-up research. To address this, we surveyed papers published the following year at SIGIR 2023~\cite{ZhuEtAl2023b, WangEtAl2023, ZhuEtAl2023, LiuEtAl2023, ChoiEtAl2023, WeiEtAl2023, YangEtAl2023, RenEtAl2023, RenEtAl2023b, HeEtAl2023, LiEtAl2023b} to qualitatively understand whether and how the analyzed SIGIR 2022 papers influenced their evaluations and findings. This analysis reveals that a comparison is almost impossible, due to how different the results are even for papers that adopt similar evaluation protocols, which is surprising and worrying.

The paper is organized as follows. In Section \ref{sec:research-method}, we describe our research method and how we attempted to reproduce the selected papers. The results of our analysis on methodology, consistency and the re-execution of the experiments including additional baselines are presented in Section \ref{sec:analysis}. 
We finally summarize our findings along several dimensions and discuss the implications of our research in Section \ref{sec:discussion}.

\section{Research Method}
\label{sec:research-method}

\subsection{Terminology}
\label{subsec:terminology}

While over the years there has been a lot of discussion, which is still ongoing, on how reproducibility and replicability should be defined~\cite{DeRoure2014,Plesser2018}, we follow the definitions by the updated \emph{ACM Policy on Artifact and Review Badging},\footnote{\url{https://www.acm.org/publications/policies/artifact-review-and-badging-current}} which are aligned to those of the International Vocabulary for Metrology (VIM)~\cite{VIM2008} and the NISO definitions for reproducibility badging~\cite{NISO-RP-31-2021}. These are also the definitions adopted by the \emph{ACM SIGIR Artifact Badging} committee:\footnote{\url{https://sigir.org/general-information/acm-sigir-artifact-badging/}} 

\begin{itemize}[left= 0pt]

    \item \textbf{Reproducibility (Different team, same experimental setup)}:
            The main results of the paper have been obtained in a subsequent study by a person or team other than the authors, using, in part, artifacts provided by the author.

    \item \textbf{Replicability (Different team, different experimental setup)}:
            The main results of the paper have been independently obtained in a subsequent study by a person or team other than the authors, without the use of author-supplied artifacts.
\end{itemize}

This study has two main components. First, a \emph{consistency} analysis of the original source code and data artifacts provided by the original paper, as well as the methodology adopted for the evaluation. Note that previous studies on reproducibility typically do not conduct an in-depth consistency analysis of the artifacts. Second, an assessment of the \emph{reproducibility} (same artifacts/experimental setup) of the results, using the original source code, the same datasets, the experimental setup, and the same optimal hyperparameters for the proposed methods. However, there are instances where we also performed some \emph{replicability} (different artifacts/experimental setup) of the results. This occurred when it was necessary to fix the original source code, replace an erroneous data split after identifying potential issues such as anomalous distributions or information leakage among training/validation/test sets, apply our own early-stopping methodology when the original one was not properly specified or implemented, or include our own baselines as a common reference for comparison across all the analyzed papers.
However, in the remainder of the paper, we will use only the term reproducibility for the sake of readability, as it is the main focus of this study. Note that in defining the criteria for determining whether a paper was successfully reproduced or not, as further detailed in Section~\ref{subsec:reproducibility}, we also considered the guidelines of the \emph{ACM SIGIR Artifact Badging} committee for awarding this kind of badge.

We did not perform only a reproducibility analysis; since we also verified the source code, its availability, documentation, as well as its consistency with the paper, we conducted, in the terms of the \emph{ACM SIGIR Artifact Badging} committee: 
\begin{itemize}[left= 0pt]
    \item \textbf{Artifacts Evaluated – Functional}:
            The artifacts associated with the research are found to be documented, consistent, complete, exercisable, and include appropriate evidence of verification and validation.

    \item \textbf{Artifacts Evaluated – Reusable and Available}:
            The artifacts associated with the paper are of a quality that significantly exceeds minimal functionality. That is, they have all the qualities of the Artifacts Evaluated – Functional level, but, in addition, they are very carefully documented and well-structured to the extent that reuse and repurposing are facilitated.
\end{itemize}

Note that, as explained earlier, different papers used different datasets or the same datasets but with different preprocessing or splitting. As a consequence, most of the methods examined can not be directly compared to determine which is more effective and under which conditions, even when the experiment uses the same dataset. In order to address this issue, we also conduct an experiment similar to what done in \cite{DBLP:conf/recsys/AnelliMPBSN23}, where we independently optimize all the hyperparameters of the methods we attempt to reproduce on a limited set of datasets. Although such an analysis on all the originally reported datasets or on a more ample set of datasets and conditions would be appropriate and interesting, it focuses on the \emph{generalizability} of the results and therefore falls outside the scope of this study and is left for future work. Moreover, such an analysis would be extremely computationally expensive. However, since we fine-tuned our baselines for each of the datasets used, they still provide a reference to qualitatively assess the relative merits of the methods examined.

\subsection{Selection of Candidate Papers}
\label{subsec:selection}

In this study, we focus on RS based on message passing. Since the number of papers published on the topic is very large, we chose to select those published at one of the most prominent conferences in the field, SIGIR.

\subsubsection*{Reproducibility Study on the SIGIR 2022 Papers}

The selection of papers was carried out through the following process. First, a list of candidate papers was retrieved by scanning the \emph{Collaborative Filtering} and \emph{Recommender Systems} sessions in the proceedings of SIGIR 2022~\cite{zz-SIGIR2022}. 
A paper was considered a candidate for reproduction if it (i) proposed a graph-based recommender technique with message passing and (ii) focused on the top-n recommendation problem. 
We did not select papers that develop \eg session-based or reinforcement learning methods, nor papers that propose more general methods not strongly connected to graph-based approaches, \eg \cite{DBLP:conf/sigir/Gao0HCZFZ22} or that use message passing but as a secondary component, \eg \cite{DBLP:conf/sigir/Zou0MWQ0C22}. 
Since all the selected papers are strongly based on LightGCN \cite{DBLP:conf/sigir/0001DWLZ020}, a method that has had a substantial impact on the community, we included it in our analysis, even though it was published earlier at SIGIR 2020.
After this screening process, we ended up with a collection of ten candidate papers \cite{DBLP:conf/sigir/PengSM22,DBLP:conf/sigir/YuY00CN22,DBLP:conf/sigir/TianXLYZ22,DBLP:conf/sigir/WuCSTC22,DBLP:conf/sigir/XiaHXZYH22,DBLP:conf/sigir/0002ZCZG22,DBLP:conf/sigir/FanL0ZT022,DBLP:conf/sigir/YangHXL22,DBLP:conf/sigir/0001DWLZ020,DBLP:conf/sigir/LiuWZS22}.

We then checked for the availability of artifacts, \idest source code and data, provided by the original authors. 
If the source code was not publicly available, we contacted the authors via email. We were able to retrieve the artifacts for all the ten candidate papers.\footnote{Note that the GitHub repository of \citet{DBLP:conf/sigir/LiuWZS22} remained empty until January 2023, six months after SIGIR 2022.}

\subsubsection*{Qualitative Impact on SIGIR 2023 Papers} The selection of papers was carried out through the following process. First, a list of candidate papers was retrieved by scanning the \emph{Collaborative Filtering}, \emph{Collaborative Filtering and Graph Neural Approaches}, and \emph{Knowledge Graphs for Recommendation} sessions in the proceedings of SIGIR 2023~\cite{zz-SIGIR2023}. A paper was considered a candidate for the qualitative analysis if it focused on the same task selected for the SIGIR 2022 papers and used one of the SIGIR 2022 papers as a baseline in the experiments, along with LightGCN. After this screening process, we ended up with a collection of eleven papers~\cite{ZhuEtAl2023b, WangEtAl2023, ZhuEtAl2023, LiuEtAl2023, ChoiEtAl2023, WeiEtAl2023, YangEtAl2023, RenEtAl2023, RenEtAl2023b, HeEtAl2023, LiEtAl2023b} for the qualitative analysis.

\subsection{Consistency and Reproducibility Analysis}
\label{subsec:reproducibility}

The first stage of the analysis is to assess whether the original source code and data artifacts are \emph{consistent} with what is described in the paper. To do so, we performed a manual inspection of the source code as well as some simple analyses on the data splits (\eg we compared the number of occurrences the items have in the training and test data). When assessing consistency, we must remain mindful that the provided source code is often a simplified version of the original experimental pipeline, which may only include the implementation or correct configuration for one of the experiments reported in the paper.\footnote{Providing all the source code should, in principle, enhance reproducibility. However, full experimental pipelines can be quite long, particularly when they rely on complex structured libraries. In such cases, if they are not well-organized and supported by adequate documentation, it may be challenging for a researcher to discern the complete experimental protocol. This could make it very difficult to identify inconsistencies between the paper and the artifacts, as well as to recognize important details that were omitted. Consequently, while the results may be reproducible, the transparency of the process is not necessarily improved.} Therefore, we consider the original artifacts to be \emph{consistent} when:
\begin{itemize}[left= 0pt]
    \item There are no apparent anomalies in the statistical properties of the data splits.
    \item The core algorithm and the details of the training process are consistent with the description in the paper.
\end{itemize}

For the second stage of the analysis, a candidate paper must meet several conditions to be considered successfully reproduced:
\begin{itemize}[left= 0pt]
    \item The provided source code can be executed successfully and is correct. 
    Note that we fix minor errors when necessary and explicitly report when we do.\footnote{For example, we correct the source code whenever early-stopping is performed on test data.}
    \item At least one of the datasets used in the original evaluation is available, either because the original training-test split is provided or because the dataset is publicly available and the paper contains sufficient detail to perform the data preprocessing and splitting.
    \item It is possible to \emph{closely} reproduce the numerical results reported in the original paper by using the provided artifacts.
\end{itemize}

Regarding the first requirement, an important aspect to consider is whether reproducing a paper also requires us to reproduce possible methodological mistakes that were made in the original implementation, \eg information leakage. In those instances, we argue that running experiments which contain errors has little scientific value, therefore we correct errors when present and describe them in our analysis. 
This issue typically arises when selecting the optimal number of epochs to train a machine learning model. Often published papers report the optimal number of epochs but do not explain how it was obtained or, in some cases, the provided implementation performs early-stopping on the test data.

In order to ensure that our analysis is done with a consistent setup, we integrate all the original implementations into our own evaluation framework \cite{DBLP:journals/tois/DacremaBCJ21}. This typically involves leaving the original model unchanged while using our implementation for early-stopping and evaluation. For efficiency and consistency across implementations, we replace the original data sampling implementation with one developed in Cython,\footnote{Cython is an extension of Python that allows to include static types and compile the code for substantially improved performance \url{https://cython.org/}} ensuring that the original sampling strategy is not altered.

Finally, the notion of what \emph{closely reproduced} means is still an open issue. \citet{BreuerEtAl2020} proposed a set of measures, among which relative distance among the effectiveness scores, which we adopt also here. However, neither \citet{BreuerEtAl2020} nor subsequent work by~\citet{MaistroEtAl2023} provided guidance on what ranges of effectiveness delta should be considered indicative of successful reproduction. 
Therefore, we follow a simple rule-of-thumb, where a relative difference of 2\% or less in at least one metric on a dataset is used to determine that the results have been successfully reproduced on that dataset. This definition aims to categorize as non-reproducible only those methods that produce results very different from those reported in the original paper, while acknowledging that results can sometimes be affected by factors that are difficult to control (\eg stochastic behavior of the method, different hardware configurations, etc.).
If the results have been successfully reproduced on some but not all of the datasets used in the paper, we consider that paper to be only partially reproduced.

Overall, we were able to conduct this analysis for nine out of the ten candidate papers, as the artifacts provided by \citet{DBLP:conf/sigir/LiuWZS22} did not meet our requirements.
The artifacts include various scripts, but they lack instructions on how to execute them, the sequence to follow, and the appropriate environment settings. Additionally, much of the relevant source code appears to be commented out. We also observed inconsistencies in the hard-coded paths, with some referencing the Last-FM dataset while others the Amazon-Book dataset, which is not mentioned in the paper. The state of the source code, the absence of execution instructions, and the missing preprocessed data files prevented us from assessing the consistency of the processing procedure and running the experiments. We contacted the authors for assistance but did not receive a response. 

\subsection{Baselines}
\label{subsec:baselines} 
In order to provide a comprehensive view of the possible strong baselines, we selected a representative set of methods from different families, ranging from the nearly 30-year-old user-based \emph{k-Nearest Neighbors} (KNN) to more recent neural and graph-based methods. These methods were identified as highly effective in previous comparative studies \cite{DBLP:journals/tois/DacremaBCJ21,DBLP:conf/recsys/AnelliMPBSN23} as well as based our own experience. In the main paper we focus on the following ones: 
\begin{itemize}[left = 0pt]
    \item \textbf{TopPop}: non-personalized method recommending to all users the most popular items the user has not yet interacted with.
    \item \textbf{UserKNN}: user-based nearest-neighbor algorithm~\cite{DBLP:conf/cscw/ResnickISBR94}, with cosine similarity and shrinkage~\cite{bell2007improved}.
    \item \textbf{ItemKNN}: item-based nearest-neighbor algorithm~\cite{DBLP:conf/www/SarwarKKR01}, with cosine similarity and shrinkage~\cite{bell2007improved}.
    \item \textbf{GF-CF}: a graph-based method that is based on a low-pass filter and has a closed form solution~\cite{DBLP:conf/cikm/ShenWZSZLL21}.\footnote{Note that occasionally the results for \textbf{GF-CF} may be missing due to its memory requirements exceeding the 64GB available on our server.}
    \item \textbf{SLIM}: item-based model that uses linear regression to compute the item similarity~\cite{DBLP:conf/icdm/NingK11}.\footnote{In particular we optimize the ElasticNet loss.}
    \item \textbf{MF-BPR}: matrix factorization method based on the \emph{Bayesian Personalized Ranking} (BPR) loss~\cite{DBLP:conf/uai/RendleFGS09}.
    \item \textbf{\iALS}: matrix factorization method for ranking tasks based on alternating least-squares \cite{DBLP:conf/icdm/HuKV08}.
    \item \textbf{\pbeta}: graph-based method that uses a two-steps random walk from users to items and vice-versa, where transition probabilities are computed from the normalized ratings~\cite{DBLP:journals/tiis/PaudelCNB17}.
    \item \textbf{NegHOSLIM (EN)}: linear full-rank model similar to SLIM, which includes higher-order interactions as input-features~\cite{DBLP:conf/recsys/SteckL21}.\footnote{Due to the large memory requirement we trained it by using an ElasticNet loss (EN) instead of the originally proposed one, in a similar way as SLIM.}
    \item \textbf{MultVAE}: variational autoencoder that assumes a multinomial likelihood for user-item interactions \cite{DBLP:conf/www/LiangKHJ18}.
\end{itemize}

The full results, available in the additional material, include additional baselines for a total of 21 collaborative models (Random, Global Effects, \palpha, \EASER, SLIM-BPR, NegHOSLIM, SVDpp, PureSVD, NMF, LightFM) and 6 content-based or hybrid models (ItemKNN, UserKNN and LightFM both content-based and hybrid).

\paragraph{Hyperparameter Optimization}
In order to ensure that the baseline algorithms are properly optimized we select their hyperparameters with a Bayesian search \cite{DBLP:conf/nips/Hernandez-LobatoHG14}, implemented by Scikit-Optimize.\footnote{\url{https://scikit-optimize.github.io/}} The search explores a total of 50 cases, with the first 16 used as initial random points. Once the optimal hyperparameters are determined, including the number of epochs, the final model is fitted on the union of training and validation data using these optimal hyperparameters. The considered hyperparameter ranges and distributions are the same as those in \citet{DBLP:journals/tois/DacremaBCJ21} and are listed in the additional material.

\paragraph{Early-stopping}
In order to select the optimal number of epochs for both the candidate algorithm as well as for baselines based on iterative optimization, we rely on the widely used early-stopping. The model is trained on the training data, and its effectiveness is evaluated on the validation data every 5 epochs. If the model's effectiveness does not improve for 5 consecutive evaluations, the training is stopped, and the epoch number associated with the best-performing model is selected. Note that we apply this early-stopping method to our baselines and also in some additional experiments with the candidate algorithms to validate the reliability of how the optimal number of epochs was determined.

\section{Detailed Analysis}
\label{sec:analysis}
For each paper, we first summarize its contributions, then analyze it along four dimensions: (i) the datasets originally used for evaluation; (ii) methodological issues and the consistency between the artifacts and the paper; (iii) the reproducibility of the results based on the provided artifacts; and (iv) the competitiveness of the method against baselines.

The reported results are the product of extensive experiments. 
We report the results of approximately 800 trained models, which required the fitting of approximately 25.000 models during the hyperparameter optimization phase. Overall the experiments required a total computation time of 4 years. 
Due to the extensive number of experiments and datasets, we report only one table for each paper selecting the dataset in which we obtained the worse outcome in terms of the reproducibility of the results. Notice that this does not necessarily mean \emph{worse results}, but rather the results that are furthest from those published.\footnote{There are two exceptions to this: (i) MovieLens 100k for GDE, because the dataset is very small, and since there are no other datasets in which we could successfully run the experiment but not reproduce the results, we select Gowalla, which is the dataset with the highest number of interactions and successfully reproduced results; (ii) Last-FM for HAKG because we had to apply a significant alteration to the dataset, and therefore we select the dataset with the second worst results in terms of reproducibility.}
To complement the main paper, we also provide extensive additional information in the appendix: the full results for each dataset with up to 26 baseline algorithms, the description and statistics of the datasets as well as a table listing the hyperparameter values of the analyzed methods that we used in our experiments, specifying for each where they were described (\idest paper or source code).\footnote{\url{https://github.com/remaplab/TOIS25_Reproducibility-SIGIR22-GCN}}
Each table groups baselines into two categories: those with closed-form solutions and those that require iterative training.

\subsection{LightGCN: Simplifying and Powering Graph Convolution Network for Recommendation}
\label{sec:method_LightGCN}

\citet{DBLP:conf/sigir/0001DWLZ020} propose LightGCN, a graph-based collaborative filtering method where user and item embeddings are propagated according to the graph adjacency matrix, via message passing.
Six of the other methods we attempted to reproduce in this study (all except SimGCL and HAKG) used LightGCN as a baseline.

\paragraph{Datasets}
LightGCN is evaluated on three datasets: Amazon-Book, Gowalla 
and Yelp2018. All datasets are preprocessed with a 10-core selection. The evaluation procedure is similar to that adopted by \citet{DBLP:conf/sigir/Wang0WFC19}, the data splitting is performed with user-wise random holdout as 72\% training, 8\% validation and 20\% test for all the datasets.

\paragraph{Consistency and Methodology} 
The GitHub repository\footnote{\url{https://github.com/gusye1234/LightGCN-PyTorch}} contains both the implementation and the training-test data split. 
The provided material is \emph{not fully consistent} with what is described in the paper. 

The first issue we observed is that the training and test sets do not exhibit the distribution expected from a user-wise random holdout split. Figure \ref{fig:LightGCN_yelp2018_popularity_plot} shows the popularity distribution of the items in the training and test sets for both the original data split and a new data split generated by us following the  user-wise random holdout procedure described in the paper. The popularity is normalized by dividing it by that of the most popular item in the corresponding set. Items are then sorted by decreasing popularity in the training data. In a user-wise random holdout split, the popularity distributions of the training and test sets should, on average, be similar however, this is not the case for the original split. Note that, as pointed out by \citet{DBLP:conf/recsys/AnelliMPBSN23} the arXiv version of \citet{DBLP:conf/sigir/Wang0WFC19} which published the original data split, includes an updated version of the results indicating that the issue has been fixed.\footnote{The new version with corrected data split and results is dated July 2020, see \url{https://arxiv.org/abs/1905.08108}} Nonetheless, the original erroneous split has been used by LightGCN and several other papers. Similar issues with anomalous data splits have also been observed in other papers by previous reproducibility studies \cite{DBLP:journals/tois/DacremaBCJ21}. To assess the extent of the anomaly, we consider three statistical measures. First, we compute the Gini Index of the popularity distributions to quantify the strength of their popularity bias.
We observe that the Gini Index of the original split is quite similar to that of our split, indicating that the two splits do not exhibit different popularity biases. For example, considering Yelp2018 the Gini Index of the original split is 0.53 for the training set and 0.56 for the test set; while the split generated by us has 0.51 for the training set and 0.54 for the test set.
As a second step we compute the Kendall's $\tau$ and Pearson correlation coefficients between the number of interactions of each item in the training and test sets. The Kendall's $\tau$ correlation, given two lists, measures the percentage of couples of elements that are in the same order in both lists. In this context, we aim to confirm that if an item $a$ is more popular than item $b$ in the training set, it also has more interactions than $b$ in the test set. However, a drawback of this metric is that if the full dataset contains many items with a similar number of interactions, the stochastic nature of the holdout split could introduce significant noise, resulting in low Kendall's $\tau$ values. The Pearson correlation, on the other hand, measures the linear correlation in the number of interactions, making it robust to the type of noise that Kendall's $\tau$ is sensitive to, but it is less intuitive to interpret. In a random holdout split we expect the training and test sets to exhibit a Pearson correlation close to 1 and a high Kendall's $\tau$. 
The results indicate strong discrepancies. In the original Amazon-Book training-test sets the relative ordering of items differs substantially (Kendall's $\tau$ 0.17, Pearson Correlation 0.50) whereas  in our split the relative ordering is more consistent, and the item popularities are highly correlated (Kendall's $\tau$ 0.52, Pearson Correlation 0.95). On Yelp2018 the findings are similar but less pronounced, with the original split (Kendall's $\tau$ 0.37, Pearson Correlation 0.78) showing worse correlations when compared to ours (Kendall's $\tau$ 0.59, Pearson Correlation 0.96). On Gowalla the gap is smallest, with the original split (Kendall's $\tau$ 0.25, Pearson Correlation 0.85) displaying a distribution very similar to ours (Kendall's $\tau$ 0.34, Pearson Correlation 0.96).  
These unusual data splits are not justified. For the reproducibility part of our study, we retain the original training-test splits. However, when evaluating competitiveness against baselines, we also conduct experiments on our newly generated training-test splits. We chose to retain the original erroneous splits in our experiments for two reasons: (i) to enable a direct comparison with the results in the LightGCN paper, and (ii) because two other papers in our study (SimGCL and GTN) used the same datasets and these inconsistent splits.

\begin{figure}[h!]
\begin{subfigure}{.48\textwidth}
  \centering
    \includegraphics[width=\linewidth]{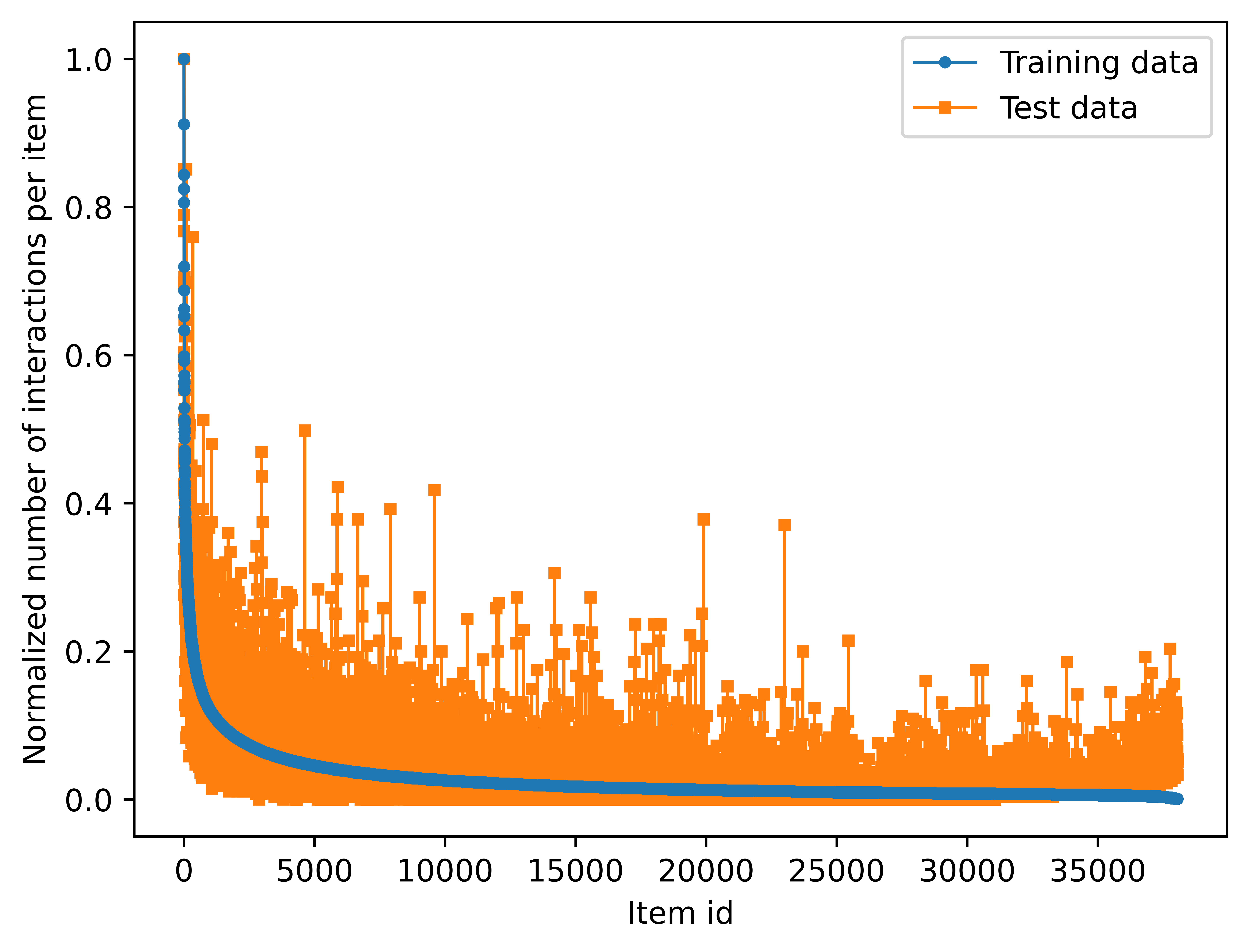}
    \caption{Normalized popularity distributions of the original training and test data splits.}
    \label{fig:LightGCN_yelp2018_original_popularity_plot}
\end{subfigure}%
\hfill
\begin{subfigure}{.48\textwidth}
  \centering
\includegraphics[width=\linewidth]{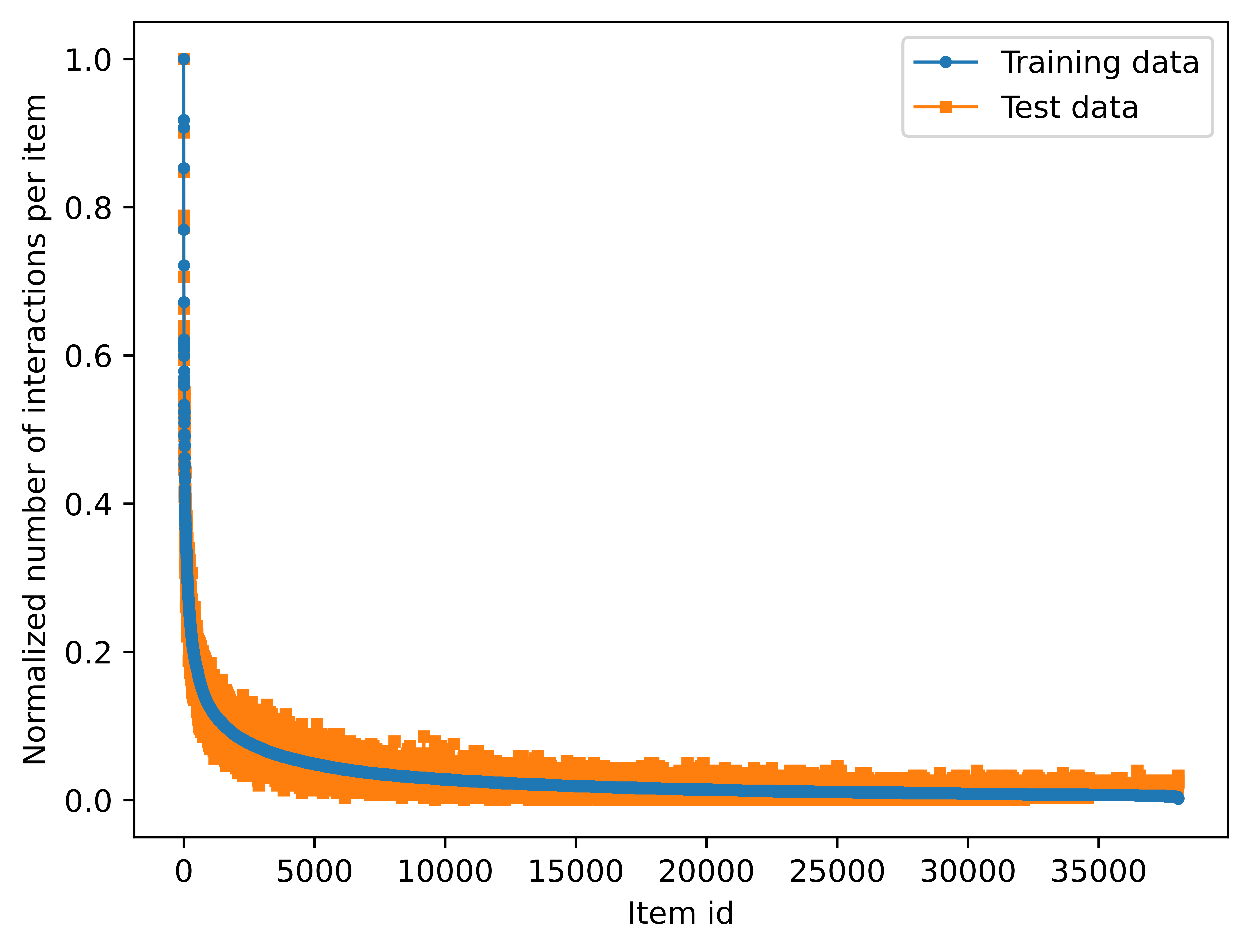}
    \caption{Normalized popularity distributions of the training and test data splits randomly generated by us.}
    \label{fig:LightGCN_yelp2018_ours_popularity_plot}
\end{subfigure}
\caption{Normalized popularity distributions of the training and test data splits for Yelp2018 used in the LightGCN paper, the value 1 corresponds to the most popular item in that split. Figure \ref{fig:LightGCN_yelp2018_ours_popularity_plot} shows the expected popularity distribution for a random holdout data split, with the normalized values on both training and validation being on average similar. Figure \ref{fig:LightGCN_yelp2018_original_popularity_plot} shows instead the distribution in the original data splits, as can be seen the training and test distributions are different.}
\label{fig:LightGCN_yelp2018_popularity_plot}
\end{figure}

A second issue is that the LightGCN paper refers to a previous paper for the experimental methodology \cite{DBLP:conf/sigir/Wang0WFC19}, which specifies performing early-stopping if the Recall@20 does not improve for 50 successive epochs. However, the provided implementation does not perform early-stopping, rather it only evaluates and prints the results on the test data every 5 epochs. Neither the paper nor the repository material provide details or insights on how many training epochs were used for the evaluation.  To address this, we applied both the original methodology and our own early-stopping approach when training the model, to validate the results.

\paragraph{Reproducibility}
In our experiments we could \emph{partially reproduce} the results reported in the original paper. In particular, we could closely reproduce the results for both Gowalla and Amazon-Book, while on Yelp2018  (see Table \ref{tab:LightGCN-yelp2018_original-result}) our results were approximately 5\% lower than those reported. Regarding early-stopping, both the original and our approaches produced very similar results.

\paragraph{Baselines}
LightGCN demonstrates inferior effectiveness compared to our set of baselines across all datasets, both in terms of the results reported in the original paper and those obtained by our experiments.
For example, in our experiments, MultVAE, \pbeta and GF-CF (the latter two being simple graph-based baselines) outperform all versions of LightGCN on all datasets (see the results for Yelp2018 in Table \ref{tab:LightGCN-yelp2018_original-result}). Particularly striking are the results for Amazon-Book, where LightGCN lags significantly behind the baselines, achieving an NDCG of 0.0315 while the simple ItemKNN reaches almost twice that value, 0.0624.
The results are consistent between the original training-test split and the one generated by us. However, for the Amazon-Book dataset, the split generated by us yields much higher absolute metric values. For example, the ItemKNN baseline achieves an NDCG@20 of 0.0624 in the original split and 0.1680 in the split generated by us. This aligns with our earlier observation that, in the original split, the data distribution of the test set differs from that of the training set. Consequently, it is not surprising that all models perform worse in absolute terms on the original split.

\tableArticleResult{LightGCN}{LightGCN}{yelp2018_original}{Yelp2018}{0}

\subsection{Less is More: Reweighting Important Spectral Graph Features for Recommendation}
\label{sec:method_GDE}
\citet{DBLP:conf/sigir/PengSM22} propose \emph{Graph Denoising Encoder} (GDE). The paper analyses graph convolution in the spectral domain and observe that only a limited number of spectral graph features significantly contribute to model effectiveness, specifically the highest and lowest frequencies (\idest eigenvalues), while intermediate frequencies are less important. This effect is attributed to the different semantics of these frequencies, with higher frequencies representing differences between users and lower frequencies representing commonalities. Based on this observation they introduce GDE, which acts as a band-pass filter by selecting high and low frequencies while removing intermediate ones.

\paragraph{Datasets}
GDE is evaluated on five datasets: MovieLens 100k, MovieLens 1M, CiteULike-a, Pinterest, and Gowalla. The data splitting is a random holdout of interactions sampled globally, 20\% for training and 80\% for testing. Note that, as opposed to what commonly done, the training data is much smaller than the testing data. The validation set is created by splitting 5\% of the training set. 
Moreover, the paper only states that all interactions are made implicit with a value of 1, but does not report the preprocessing applied to the Gowalla and Pinterest datasets, which are much smaller than their original versions.

\paragraph{Methodological Issues}
The GitHub repository\footnote{\url{https://github.com/tanatosuu/GDE}} contains both the implementation and the training-test data split. 
The provided material is \emph{fully consistent} with what is described in the paper. 

During our experiments we observed that GDE is numerically unstable, frequently failing to train properly, particularly on certain datasets. We believe this instability is related to an issue in how the lowest eigenvalues are computed. 
Consider a matrix containing the user-item interactions defined as $R\in\mathbb{R}^{u \times i}$, where $u$ is the number of users and $i$ is the number of items. GDE requires computing both the highest and the lowest eigenvalues, along with their corresponding eigenvectors, for the item-item similarity matrix $S_{I-I}=R^TR$ and the user-user similarity matrix $S_{U-U}=RR^T$. While computing the highest eigenvalues can be done rather efficiently, computing the \emph{lowest} ones for such large matrices is a much more computationally intensive task and is subject to numerical rounding errors. The original implementation uses the LOBPCG method \cite{DBLP:journals/siamsc/Knyazev01}\footnote{The implementation is based on the \texttt{lobpcg} function from \texttt{PyTorch}, see the reference documentation here \url{https://pytorch.org/docs/stable/generated/torch.lobpcg.html}} which is very fast but assumes the input matrix to be \emph{positive definite}, \idest all its eigenvalues are strictly positive. Crucially, this requirement is never verified and, as we will show, it is not met.
Any matrix $A$ that can be represented as $A=B^TB$ is at least \emph{positive semi-definite}. However, to be strictly  \emph{positive definite} an additional condition must be met, matrix $B$ must have linearly independent columns \cite[Ch.\ 7, p.\ 396]{strang2014differential}. The number of linearly independent columns and rows of a matrix is its \emph{rank}, which can never be larger than the smallest of its dimensions, \idest $rank(R) \leq min(u,i)$. This immediately implies that at least one of the two similarity matrices will \emph{not} be positive definite, simply because $R$ is rectangular. In a typical scenario $u>i$ hence $S_{U-U}$ will not be positive definite. To validate our statements, we compute the rank of the matrix $R$ containing the training set by applying the SVD method, \idest the rank of the matrix is equal to the number of non-zero singular values.\footnote{We applied the function \texttt{matrix\_rank} from \texttt{PyTorch}, see the reference documentation here \url{https://pytorch.org/docs/stable/generated/torch.linalg.matrix_rank.html}} 
The results are that on MovieLens 100k and CiteULike-a there are fewer than 10 linearly dependent rows or columns, while for MovieLens 1M and Pinterest there are approximately 500 and for Gowalla almost 900.\footnote{
The detailed results are the following: 
MovieLens 100k (users=943, items=1682, rank=940), 
MovieLens 1M (users=6040, items=3952, rank=3401), 
CiteULike-a (users=5551, items=16980, rank=5544), 
Pinterest (users=37501, items=9836, rank=9303) and
 Gowalla (users=29858, items=40981, rank=28965).
}

This issue has two important consequences: (i) since the data violates the assumptions of the method used to compute the eigenvectors its results are prone to be erroneous, potentially invalidating the findings reported in the original paper; (ii) a large number of the smallest eigenvalues that GDE tries to use, possibly \emph{all of them}, will be zero.\footnote{Their value will not be exactly zero due to the limits of machine precision. Manual inspection reveals that, on MovieLens 100k, several of the smallest eigenvalues have an absolute value of approximately $10^{-9}$.} Therefore, they will not represent the high frequency signals that GDE aims to leverage. Indeed, by examining the hyperparameter values provided in the original GitHub repository, we observe that using the lowest eigenvalues is beneficial only for the MovieLens 100k and 1M datasets, while it is not advantageous for the other datasets. We believe the combination of these two effects explains the unstable behavior observed for GDE.
Computing the lowest eigenvalues using the full SVD decomposition is impractical due to its potentially large memory requirements and likely large numerical errors. While it may be possible to address this issue by rethinking how this stage of the computation is carried out, possibly leveraging other algebraic properties, this would require a redesign of the method that goes beyond the scope of this paper.

We also identified other methodological issues. In the original data splits, a small number of interactions (11 for CiteULike-a, which is 0.02\% of the training set, and 7884 for Pinterest, which is 3.94\% of the training set) appear in both the training and test sets.
In our experiments we removed these interactions from the training set to avoid any overlap and prevent the possibility of information leakage.
Furthermore, although the source code reports the optimal number of training epochs for each dataset, the paper does not explain how these numbers were determined. The original implementation does not apply early-stopping, but rather trains for a large number of epochs and periodically evaluates the model on the test data. To address this, we conducted two experiments: one using the reported number of epochs and the other applying our own early-stopping methodology.
We also observe that, due to the data splitting methodology, only 1\% of the data is used for validation, which could result in noisy hyperparameter tuning and early-stopping.
Lastly, the paper does not mention that the source code employs a definition of Recall where the denominator is not the number of relevant items for the user but rather the minimum between this value and the length of the recommendation list. While this definition has been used in previous studies \cite{DBLP:conf/www/LiangKHJ18}, and is intended to normalize the metric in cases where the number of relevant items exceeds the length of the recommendation list, the existence of two competing definitions of a metric poses obstacles to reproducibility when the paper does not explicitly state which definition is being used.
For our experiments, we use both the normalized definition of Recall used in the GDE paper to enable a direct comparison, as well as the more common non-normalized definition of Recall.

\paragraph{Reproducibility}
In our experiments we could \emph{partially reproduce} the results reported in the original paper. In particular, we could closely reproduce the results for both MovieLens 1M and Gowalla (see Table \ref{tab:GDE-gowalla-result}). On MovieLens 100k the results were lower and slightly beyond the 2\% threshold, with a normalized Recall of 0.5196 compared to the reported 0.5400. On the remaining CiteULike-a and Pinterest datasets GDE exhibited numerical instabilities that caused the training to fail after a few epochs.\footnote{This issue was also raised by other practitioners on the GDE GitHub repository.}
In our experiments with our early-stopping, we could only reproduce the results for MovieLens 1M.

\paragraph{Baselines}
Our early-stopping runs of GDE show competitive results only on the two MovieLens datasets compared to our baselines. 
On the other hand, our runs of GDE with the number of epochs from the paper are competitive on all datasets except CiteULike-a and Pinterest, where convergence is not achieved and the normalized Recall is around 0.002. Given the unstable nature of GDE, and its fast training time, we decided to perform a new hyperparameter optimization. With our new set of hyperparameters, GDE was able to reach convergence on Pinterest and outperform our baselines as well, with a normalized Recall of 0.1082 compared to the 0.1067 reached by the best performing baseline \iALS.
Finally, the results reported in the paper for GDE are competitive with our baselines across all datasets. The results obtained with a non-normalized implementation of Recall are consistent with those obtained with the normalized definition used in the original source code.

\tableArticleResult{GDE}{GDE}{gowalla}{Gowalla}{0}

\subsection{Are Graph Augmentations Necessary? Simple Graph Contrastive Learning for Recommendation}
\label{sec:method_SimGCL}
\citet{DBLP:conf/sigir/YuY00CN22} propose \emph{Simple Graph Contrastive Learning} (SimGCL). The paper argues that an important step in a typical contrastive learning pipeline is to perform graph augmentation by applying perturbations to the adjacency matrix used by message passing, \idest the user-item interaction matrix. The contrastive loss then pushes the original and augmented graphs toward similar latent representations.
The paper claims that the main contribution to the model effectiveness in constrastive learning based models does not come from the graph augmentation (\eg random edge dropout) but rather comes from the constrastive learning loss function InfoNCE. The InfoNCE loss effect is to increase the separation between positive and negative samples for each user. SimGCL generates contrastive views by applying random perturbations of the embeddings instead of graph augmentations.

\paragraph{Datasets}
SimGCL is evaluated on three datasets: Yelp2018 and Amazon-Book, with the same training-test split of LightGCN \cite{DBLP:conf/sigir/0001DWLZ020}, and Douban Book. Both Yelp2018 and Amazon-Book are preprocessed with a 10-core selection, while for Douban Book only interactions with rating of at least 4 are retained.
For all datasets the data splitting is a user-wise random holdout, 72\% training, 8\% validation and 20\% test.

\paragraph{Methodological Issues}
The GitHub repository\footnote{We use the PyTorch implementation provided by the authors \url{https://github.com/Coder-Yu/SELFRec}} includes both the implementation and the training-test data split. 
However, the provided materials are only \emph{partially consistent} with what is described in the paper. 

The first issue is that the paper relies on the same data splits used in the LightGCN paper for the Yelp2018 and Amazon-Book datasets which, as shown in Section \ref{sec:method_LightGCN}, are not the result of a correct user-wise random holdout split.
We decided to keep these splits in our experiments for the same reasons as for LightGCN, but we also conducted experiments on the new user-wise random holdout data splits we generated following the procedure described in the paper.
Secondly, there is a discrepancy between the optimal values of the hyperparameters used by the authors in their experiments and the optimal values obtained from the sensitivity analysis of the same hyperparameters, as reported in the paper. For example, the hyperparameter sensitivity analysis refers to a model with 2 graph convolution layers ($K=2$), whereas it was previously shown that the best model requires 3 convolution layers. Additionally, the paper states that the optimal noise level $\epsilon$ is 0.1 yet the sensitivity analysis plots indicate values 0.05 for Yelp2018, 0.1 for Amazon-Book, and 0.2  for Douban Book. These inconsistencies, albeit small, create uncertainty about the correct configuration that should be used to reproduce the results reported in the paper.
Finally, the number of training epochs is determined based on a convergence plot that shows the Recall and BPR loss, but the paper does not specify whether this plot is computed on the training, validation, or test data. Due to this, we conducted two experiments, one training the model using the reported number of epochs and the other applying our own early-stopping methodology. 

\paragraph{Reproducibility}
In our experiments we could \emph{partially reproduce} the results reported in the original paper. In particular, we could reproduce the results on Amazon-Book and Yelp2018. However, on Douban Books our effectiveness was about 10\% lower than what was reported in the paper (see Table \ref{tab:SimGCL-doubanbook_original-result}).
SimGCL achieved almost identical results regardless of whether the model was trained using the reported number of epochs or with our early-stopping methodology, across all datasets and data splits.

\paragraph{Baselines}
On Yelp2018, SimGCL performs competitively against almost all our simple baselines with an NDCG@20 of 0.0594, with only MultVAE achieving the better result of 0.0602.  
On Douban Book and Amazon-Book, however, SimGCL performs significantly worse than our baselines, both in our runs and using the higher results reported in the original paper. 
For instance, on Douban Book, the NDCG@20 for SLIM is 0.2226 while SimGCL reaches 0.1583 according to the original paper and 0.1445 in our run. Since SimGCL uses the same training-test splits as LightGCN, we can directly compare the two methods by looking in particular at NDCG@20. On Yelp2018, we observe a 15\% improvement of SimGCL  (0.0594) over LightGCN (0.0506). For Amazon-Book, the improvement is 20\% on our data split (0.1047 for SimGCL and 0.0862 for LightGCN) and a surprising 80\% on the original LightGCN data split (0.0402 for SimGCL and 0.0315 for LightGCN. See Section \ref{sec:method_LightGCN} for an analysis of its anomalous distribution).

\tableArticleResult{SimGCL}{SimGCL}{doubanbook_original}{Douban Book}{0}

\subsection{Learning to Denoise Unreliable Interactions for Graph Collaborative Filtering}
\label{sec:method_RGCF}
\citet{DBLP:conf/sigir/TianXLYZ22} presents \emph{Robust Graph Collaborative Filtering} (RGCF). RGCF comprises two main steps. First, a graph denoising module removes interactions, from the adjacency matrix used in the graph convolution, that are estimated by the model itself as noisy, while assigning a reliability weight to the remaining interactions. 
Second, a diversity-preserving module builds new interaction graphs (\idest adjacency matrix) based on the denoised one by adding new edges derived from the trained model's predictions.  
The model is trained using BPR with an additional contrastive loss, \idest InfoNCE, that pulls the representations of nodes learned from the augmented graphs closer to each other.

\paragraph{Datasets}
RGCF is evaluated on three datasets: Yelp2018, Amazon-Book and MovieLens 1M. Both Yelp2018 and Amazon-Book are preprocessed with a 15-core selection, while for  MovieLens 1M only interactions whose rating is at least 4 are retained and associated to an implicit rating of 1. The data splitting is a random holdout of interactions sampled globally, 80\% training, 10\% validation and 10\% test. 

\paragraph{Consistency and Methodology} 
The GitHub repository\footnote{\url{https://github.com/ChangxinTian/RGCF}} contains only the implementation but does not contain the training-test data split. 
The provided implementation is \emph{fully consistent} with what is described in the paper. 

Since the provided materials do not include the training-test data split, we applied the described preprocessing on all three datasets but could not reproduce exactly the data statistics reported in the paper. 
Due to this, we ran RGCF with the optimal hyperparameters and also conducted a new hyperparameter search.
Unfortunately, we were able to use the model only on MovieLens 1M. On the other datasets, the gradient update step in PyTorch caused a memory spike that exceeded the 24GB available on our RTX 3090 GPU, preventing us from running the model, while the computational time on a CPU was prohibitive.\footnote{The issue appears to be related to how PyTorch handles gradient updates rather than the original RGCF implementation. The RGCF paper does not provide details about the hardware configuration.
We found that on our i9-9900K CPU with 16 cores and 64GB of RAM, it is only feasible to run the experiment for Yelp2018, which, however, requires 1.5 hours per epoch, resulting in an estimated total runtime of 31 days. 
} The implementation correctly uses early-stopping on the validation set to determine the optimal number of training epochs.

\paragraph{Reproducibility}
In our experiments we could \emph{partially reproduce} the results reported in the original paper.
Considering that our experiments could not use the original training-test split, the results we obtain include an additional level of variance due to the stochastic nature of the data splitting process. On MovieLens 1M, the only dataset we could use, our experiments yielded better results than those reported in the paper (see Table \ref{tab:RGCF-movielens1m-result}) and can be considered fully reproduced, as the HR metric is within 2\% of the values reported. 
Furthermore, we observed that applying our early-stopping approach increased the model's effectiveness, and performing a new hyperparameter optimization resulted in additional improvements.

\paragraph{Baselines}
According to the results on MovieLens 1M (see Table \ref{tab:RGCF-movielens1m-result}), the only dataset usable for this reproducibility study, RGCF exhibits worse effectiveness than SLIM and GF-CF, with a gap of up to 10\%. Our version, using the newly optimized hyperparameters, shows improved effectiveness but remains below several simple baselines. 

\tableArticleResult{RGCF}{RGCF}{movielens1m}{MovieLens 1M}{0}

\subsection{INMO: A Model-Agnostic and Scalable Module for Inductive Collaborative Filtering}
\label{sec:method_INMO}
\citet{DBLP:conf/sigir/WuCSTC22} presents \emph{Inductive Embedding Module for collaborative filtering} (INMO), which aims to improve the effectiveness of matrix factorization models for new users. The paper focuses on matrix factorization models that are \emph{transductive} (\idest not model based, such as SVD++, MF-BPR etc..) and proposes an \emph{inductive} (model-based) representation of users and items as a function of the embeddings of a selected subset of template users and items. The paper also explores strategies to select these templates.

\paragraph{Datasets}
INMO is evaluated on three datasets: Yelp2018, Amazon-Book and Gowalla. Both Yelp2018 and Amazon-Book are preprocessed by selecting only the interactions with a rating of at least 4, associating them to an implicit rating of 1, followed by a 10-core selection. No information is provided on the preprocessing of Gowalla. The data splitting is a user-wise random holdout, 70\% training, 10\% validation and 20\% test.

\paragraph{Methodological Issues}
The GitHub repository\footnote{\url{https://github.com/WuYunfan/igcn_cf}} contains the implementation, the training-test data split, and, uniquely among the algorithms analyzed in this study, it also includes the implementation of the baseline models along with their own hyperparameter optimization code.
The provided material is \emph{fully consistent} with what is described in the paper.\footnote{We note that the paper uses the BPR loss and defines it by using the log sigmoid function:$-ln (\sigma(x))$, while the implementation instead uses the $\mbox{softplus}(-x) = ln(1+e^{-x})$ function. This occurs in a few of the papers we analyze and we wish to point out that the two formulations are mathematically identical: $-ln (\sigma(x)) = -ln\left(1/(1+e^{-x}) \right) = ln(1+e^{-x}) = \mbox{softplus}(-x)$.} 
As a minor note, the paper reports a number of interactions for all datasets that is approximately 15\% higher than that of the provided data.
The number of epochs is correctly selected using early-stopping, where training stops if the NDCG@20 on the validation data does not improve for 50 successive epochs.

\paragraph{Reproducibility}
In our experiments we could \emph{partially reproduce} the results reported in the original paper.
In particular, we could reproduce the results on Yelp2018 within less than 1\% as well as for one of the metrics on Amazon-Book, while for Gowalla (see Table \ref{tab:INMO-gowalla-result}) we obtained results that are approximately 2.5-5\% lower.

\paragraph{Baselines}
On the Yelp2018 dataset, INMO has an NDCG of 0.0647 that consistently outperforms most simple baselines with the sole exception of MultVAE, which reaches 0.0670. 
However, on the Gowalla dataset (see Table \ref{tab:INMO-gowalla-result}), INMO is competitive against some of the baselines but not with \pbeta, GF-CF, SLIM, NegHOSLIM and MultVAE, which demonstrate comparable effectiveness.
On the Amazon-Book dataset, INMO is considerably less effective, with an NDCG of 0.0934, 32\% lower than that of the best-performing baselines \pbeta (0.1402) or SLIM (0.1451).

\tableArticleResult{INMO}{INMO}{gowalla}{Gowalla}{0}

\subsection{Hypergraph Contrastive Collaborative Filtering}
\label{sec:method_HCCF}
\citet{DBLP:conf/sigir/XiaHXZYH22} 
proposed \emph{Hypergraph Contrastive Collaborative Filtering} (HCCF), which extends LightGCN's message-passing approach on the user-item adjacency matrix. In addition, HCCF incorporates a layer of message-passing on a learnable hypergraph adjacency matrix, which is decomposed into the product of two lower-dimensional matrices. The model also includes a step called Hierarchical Hypergraph Mapping, which applies message-passing on the learned hypergraph adjacency matrix. The model is trained using contrastive learning.

\paragraph{Datasets}
HCCF is evaluated on three datasets: Yelp2018, Amazon-Book and MovieLens 10M. Both Yelp2018 and MovieLens 10M are preprocessed with a 10-core selection, while Amazon-Book with a 20-core one. The data splitting is a random holdout, 70\% training, 10\% validation and 20\% test, but the paper does not specify if the sampling is done globally or user-wise.

\paragraph{Consistency and Methodology} 
The GitHub repository\footnote{\url{https://github.com/akaxlh/HCCF}} provides two versions of the implementation (one in TensorFlow 1, which is now obsolete, and one in PyTorch) and the data split for training, validation, and testing. 
The provided material is \emph{partially consistent} with what is described in the paper. 

Both the PyTorch and TensorFlow versions include a setting,\footnote{This setting is referred to as \texttt{trnNum} in the original source code.} which is not described in the paper, that limits the number of training samples per epoch to $10^4$. Given that the datasets contain between $1.5\cdot10^6$ and $10^7$ interactions, this setting has a significant impact on the training process, as the model requires a much higher number of epochs to converge, thereby affecting the early-stopping process.
There is also an inconsistency between the optimal hyperparameters used in the source code and the search space reported in the paper.
For example, the contrastive loss weight $\lambda_1$ has optimal values of $10^{-6}$ for MovieLens 10M and $10^{-7}$ for Amazon-Book, despite the smallest value in the reported search space being $10^{-5}$.
Finally, the criteria used to select the optimal number of epochs is not described.

Lastly, the PyTorch implementation of HCCF differs from the TensorFlow version, as well as from the version described in the paper, in several ways: (i) when computing the contrastive loss, the PyTorch implementation uses the embeddings obtained through the graph convolution but does not update them; (ii) only a single layer is used instead of the multi-layer hypergraph convolution; and (iii) the learned hypergraph adjacency matrix is excluded from the hypergraph convolution.
According to the description provided in the GitHub repository, these modifications were introduced to improve the model's effectiveness on sparse data, which partially contradicts the original claims of the paper. 
In our experiments, we updated the PyTorch implementation to ensure its consistency with the HCCF model described in the paper. Note that we do not report the results for the simplified model because it is effectively a different algorithm.

\paragraph{Reproducibility}
In our experiments we could \emph{partially reproduce} the results reported in the original paper.
In particular, we could reproduce the results on Yelp2018 and achieved a \emph{substantial improvement} of about 70\% on MovieLens 10M (see Table \ref{tab:HCCF-movielens10m-result}). However, HCCF failed to converge on Amazon-Book, exhibiting effectiveness comparable to a random recommender. This issue also occurred with the TensorFlow version of HCCF.\footnote{We observed that adjusting the regularization hyperparameters allowed the model to exhibit limited effectiveness, however performing a new hyperparameter optimization for non reproducible models goes beyond the scope of this paper.} The substantial improvement observed on MovieLens 10M may be explained by the fact that the original training limited the number of samples drawn per epoch, whereas in our experiments, we used all available samples. Since MovieLens 10M is the dataset with the highest number of interactions, this difference is likely more pronounced.

\paragraph{Baselines}
In all datasets, HCCF performs largely below the baselines, both when considering the results reported in the paper and those from our experiments.  
On MovieLens 10M (see Table \ref{tab:HCCF-movielens10m-result}), the best-performing baseline achieves results that are 16\% better than the best results we obtained for HCCF. On Yelp2018, the best-performing baseline MultVAE reaches an NDCG@40 of 0.1264, while HCCF of 0.0592, outperforming it by more than twice.
On Amazon-Book, where our run of the model failed to converge, the best-performing baseline achieves results that are 546\% better than the original result reported in the paper, with an NDCG@40 of 0.1803 compared to 0.0330 for HCCF.

\tableArticleResult{HCCF}{HCCF}{movielens10m}{MovieLens 10M}{0}

\subsection{HAKG: Hierarchy-Aware Knowledge Gated Network for Recommendation}
\label{sec:method_HAKG}
\citet{DBLP:conf/sigir/0002ZCZG22} proposes \emph{Hierarchy-Aware Knowledge Gated Network} (HAKG), 
which integrates both collaborative interactions and knowledge-based graphs. 
The paper aims to leverage the hierarchical structure of knowledge graphs and the “higher order” relations in collaborative data through the use of hyperbolic embeddings.

\paragraph{Datasets}
HAKG is evaluated on three datasets: Alibaba-iFashion, Yelp2018 and Last-FM. All datasets are preprocessed with a 10-core selection.
For Last-FM the split is the same used by KGAT~\cite{DBLP:conf/kdd/Wang00LC19} (user-wise random holdout, 
 72\% training, 8\% validation and 20\% test). 
For the other two datasets the split is 80\% training, 10\% validation and 10\% test, but the paper does not specify if the sampling is done globally or user-wise.

\paragraph{Consistency and Methodology} 
The GitHub repository\footnote{\url{https://github.com/zealscott/HAKG}} contains both the implementation and the training-test data split. 
The provided material is \emph{not consistent} with what is described in the paper.

First, in the Last-FM data splits, a substantial number of interactions (169782 which is 13.17\% of the training data) appear in both the training and test data. 
This erroneous split originates from a previous paper \cite{DBLP:conf/kdd/Wang00LC19}. 
In our experiments, we removed these interactions from the training set to avoid any overlap and prevent the possibility of information leakage. 
Furthermore, the training-test split of the Yelp2018 dataset exhibits an item popularity distribution that is not consistent with a user-wise random holdout split, see Figure \ref{fig:HAKG_yelp2018_original_popularity_plot}. Using the same procedure described in Section \ref{sec:method_LightGCN}, we computed three statistics. The Gini Index of the item popularity in the entire dataset is 0.58, for the original training data is 0.59, while for the test data is 0.63, indicating that the distribution is more unbalanced. Comparing the item popularity between the training and test data, we observe that the Kendall's $\tau$ is 0.38 and the Pearson Correlation is 0.83. In a user-wise random holdout data split both values should be much higher.\footnote{In a new user-wise random holdout split generated by us, we obtain a Kendall's $\tau$ of 0.59 and a Pearson Correlation of 0.96.}
Unfortunately, we are not able to run HAKG on Yelp2018 due to its memory requirements exceeding the 24GB available on our RTX 3090 GPU and its prohibitive computational time on CPU.\footnote{The HAKG paper does not provide details on the hardware configuration. We attempted to run the experiment on a i9-9900K CPU with 16 cores and 64GB of RAM. However, on Yelp2018 the training time for each epoch is approximately 7 hours resulting in an estimated runtime of between 30 days (for 100 epochs) and 300 days (for 1000 epochs).  
For this reason it is not possible to run this experiment.}
As a minor note, while the paper states the splitting of the data is 80\% training, 10\% validation and 10\% test, the provided data split is actually 72\% training, 8\% validation and 20\% test.
Finally, the provided source code performs early-stopping using the test data, which introduces information leakage. In our experiments, we corrected this issue and performed early-stopping exclusively using the validation data.

\begin{figure}[h!]
    \centering
    \includegraphics[width=0.5\linewidth]{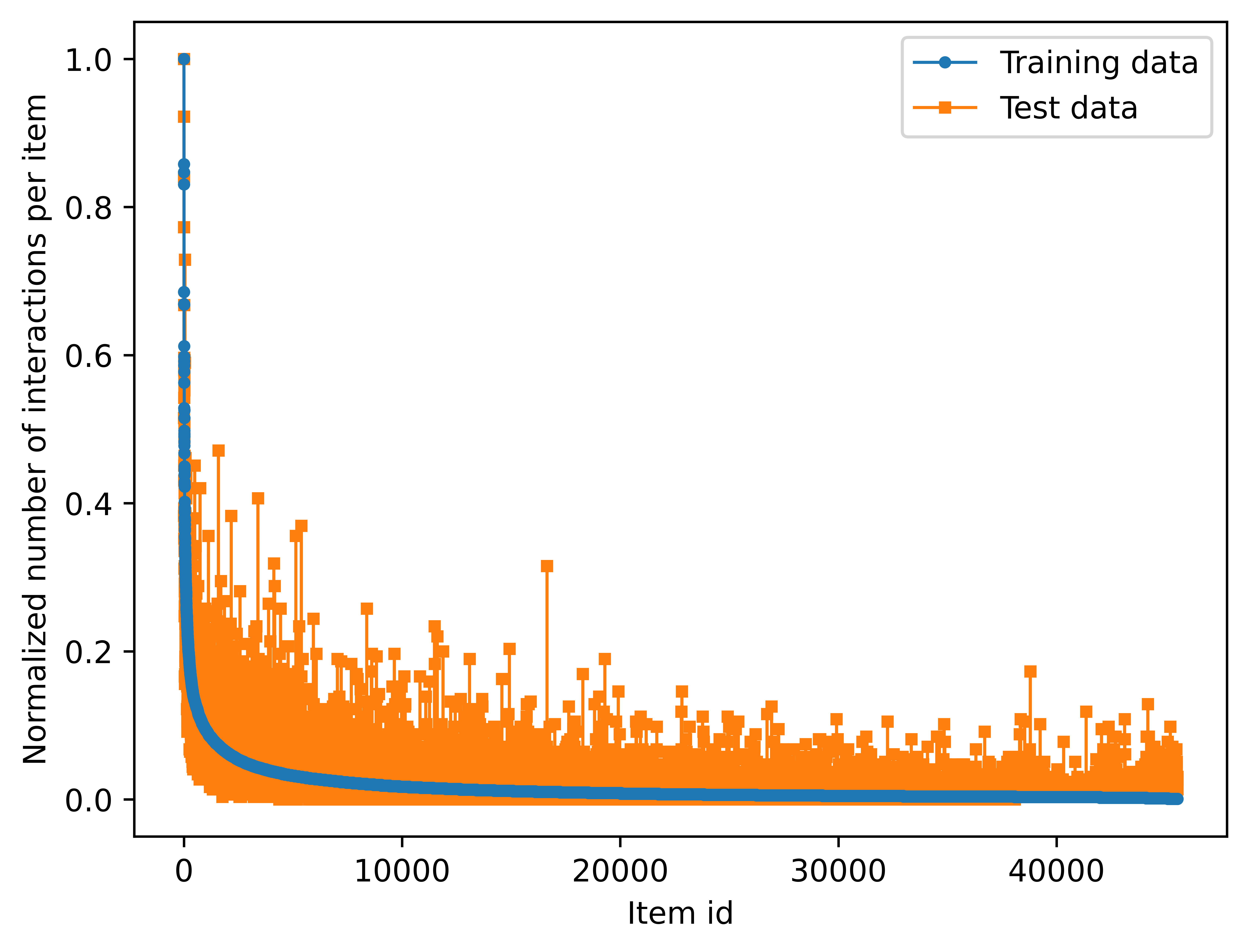}
    \captionof{figure}{Normalized popularity distributions of the original training and test data splits for Yelp2018 used by the HAKG paper.}\label{fig:HAKG_yelp2018_original_popularity_plot}
\end{figure}

\paragraph{Reproducibility}
In our experiments we could \emph{not reproduce} the results reported in the original paper on any of the datasets.
In particular, we obtained results that were 7\% lower than those reported in the paper on the Alibaba-iFashion dataset (see Table \ref{tab:HAKG-alibaba-ifashion_original-result}), possibly due to the early-stopping on the test data used in the original source code. On the other hand, we achieved nearly twice as high results on the Last-FM dataset, 0.1644 compared to the reported 0.0931. This unusual result on Last-FM may be attributed to the inconsistent data split used in the paper, which we corrected for our experiments. Although it may seem counterintuitive that removing information leakage would lead to an increase in effectiveness for all models, this effect arises because, for this task, items already interacted with by a user are not recommended to them again. This prevented all models from recommending several items present in the user's test data that had already been observed by the user. 
We were not able to run HAKG on Yelp2018 due to memory limitations and therefore we were unable to evaluate their reproducibility on this dataset.

\paragraph{Baselines}
On both the Alibaba-iFashion and Last-FM datasets, HAKG performs worse than many simple baselines.  This is despite the significant improvements we observe on Last-FM, where we achieved an effectiveness that is nearly double the results reported in the paper, our run of HAKG has an NDCG@20 of 0.1644 compared to 0.2014 for \pbeta and 0.2078 for SLIM.
On the Alibaba-iFashion dataset (see Table \ref{tab:HAKG-alibaba-ifashion_original-result}), while our run of HAKG falls below several baselines, the original results reported in the paper would have been competitive against all baselines except MultVAE. However, the use of test data as part of the early-stopping process in the provided source code raises concerns about the reliability of this result.

\tableArticleResult{HAKG}{HAKG}{alibaba-ifashion_original}{Alibaba-iFashion}{0}

\subsection{Graph Trend Filtering Networks for Recommendation}
\label{sec:method_GTN}
\citet{DBLP:conf/sigir/FanL0ZT022} introduce \emph{Graph Trend Filtering Networks for Recommendation} (GTN), which proposes an adaptive method for assessing the reliability of interactions. 
To achieve this, a smoothness constraint is applied to the embeddings, penalizing interactions between users and items with very different embeddings. 
The paper further proposes to use the Proximal Alternating Predictor-Corrector method and formulates an iterative solver that operates through three steps.

\paragraph{Datasets}
GTN is evaluated on four datasets: Yelp2018, Amazon-Book and Gowalla (using the same training-test split of LightGCN \cite{DBLP:conf/sigir/0001DWLZ020}) and Last-FM (using the same training-test split of KGAT \cite{DBLP:conf/kdd/Wang00LC19}). All datasets are preprocessed with a 10-core selection and splitted with a user-wise random holdout, 
 72\% training, 8\% validation and 20\% test.

\paragraph{Consistency and Methodology} 
The GitHub repository\footnote{\url{https://github.com/xiangwang1223/knowledge_graph_attention_network}} contains both the implementation and the training-test data split. 
The provided material is \emph{not consistent} with what is described in the paper. 

First, none of the datasets have splits that follow the procedure described in the paper. The splits of the Amazon-Book, Yelp2018 and Gowalla datasets are not correct random holdout splits, as we also reported for LightGCN in Section \ref{sec:method_LightGCN}. For consistency, we report results for both the original splits and the correct splits we generated for LightGCN using the procedure described in the paper.
Furthermore, the Last-FM dataset exhibits the same issue previously described in Section \ref{sec:method_HAKG} for HAKG, with significant overlap between test and training data.
In our experiments we have removed these overlapping interactions from the training set.
Finally, the paper does not provide details on how the optimal number of epochs is determined. Additionally, the model's implementation does not employ early-stopping, rather it evaluates the model on the test data every 5 epochs and prints the results.

\paragraph{Reproducibility}
In our experiments we could \emph{partially reproduce} the results reported in the original paper.
In particular, we could reproduce the results on Yelp2018 and Gowalla, while on Amazon-Book we obtained results that are 10\% better than those reported in the paper (see Table \ref{tab:GTN-amazon-book_original-result}).
For the Last-FM dataset  we obtained an NDCG@20 of 0.1776 achieving a 90\% improvement over the 0.0857 reported in the paper. This large discrepancy is again due to the removal of the overlap between the training and test data, as discussed in Section \ref{sec:method_LightGCN}.

\paragraph{Baselines}
On Yelp2018, GTN reaches an NDCG@20 of 0.0559, outperforming all baselines with the exception of MultVAE (0.0590) and GF-CF (0.0568). On Last-FM, GTN is not competitive with the baselines with an NDCG@20 of 0.1776 compared to 0.1838 for ItemKNN, and on Amazon-Book it falls substantially behind (see Table \ref{tab:GTN-amazon-book_original-result}), achieving nearly half the NDCG of a simple ItemKNN.
On Gowalla, GTN outperforms the baselines only on NDCG but is outperformed by MultVAE on Recall. 
These results are consistent for the data splits we generate ourselves.

\tableArticleResult{GTN}{GTN}{amazon-book_original}{Amazon-Book}{0}

\subsection{Knowledge Graph Contrastive Learning for Recommendation}
\label{sec:method_KGCL}
\citet{DBLP:conf/sigir/YangHXL22} present the \emph{Knowledge Graph Contrastive Learning framework} (KGCL) framework, 
which aims to mitigate the impact of noisy knowledge bases. This is achieved by incorporating a knowledge graph augmentation schema to guide the contrastive learning process. KGCL employs a parameterized attention matrix on the concatenation of user and item embeddings to estimate the relevance. Additionally, it utilizes a translation-aware loss function to handle relations within the knowledge base.

\paragraph{Datasets}
KGCL is evaluated on three datasets: Yelp2018 (using the same training-test split as in LightGCN and HAKG \cite{DBLP:conf/sigir/0002ZCZG22} but a different knowledge base), Amazon-Book and MIND. Both Yelp2018 and Amazon-Book are preprocessed with a 10-core selection, while MIND only contains users with at least 5 interactions within a specific six weeks time frame.
The data splitting is not described in the paper but based on the cited papers we assume is user-wise a random holdout, 72\% training, 8\% validation and 20\% test, with the exception of MIND which uses global sampling.

\paragraph{Consistency and Methodology} 
The GitHub repository\footnote{\url{https://github.com/yuh-yang/KGCL-SIGIR22}} contains both the implementation and the training-test data split. 
The provided material is \emph{partially consistent} with what is described in the paper. 

First, the splits of the Amazon-Book and Yelp2018 datasets are not the result of a correct random holdout splits, as we also reported for LightGCN in Section \ref{sec:method_LightGCN}.
More importantly, the provided implementation performs early-stopping using the test data, which introduces information leakage. In our experiments, we performed early-stopping using the validation data.

\paragraph{Reproducibility}
In our experiments we could \emph{partially reproduce} the results reported in the original paper. In particular, we could reproduce the results on Amazon-Book, while for MIND (see Table \ref{tab:KGCL-MIND_original-result}) and Yelp2018 we obtained lower results within 6\%.

\paragraph{Baselines}
Many of our simple baselines exhibit better effectiveness than KGCL. 
The original results reported in the paper would have been mostly competitive with the baselines on Yelp2018, with a reported NDCG@20 of 0.0493, outperformed only by the 0.0521 of MultVAE. However, the use of test data as part of the early-stopping process in the provided source code, combined with the anomalous data splits, raise concerns about the reliability of this result.

\tableArticleResult{KGCL}{KGCL}{MIND_original}{MIND}{0}

\begin{table}[h]
    \caption{Results for all the analyzed methods and baselines on the Amazon-Book dataset (Table \ref{tab:ALL-amazon_book-result}) and the Yelp2018 dataset (Table \ref{tab:ALL-yelp-result}). The baselines are highlighted in \textbf{bold} if they outperform our results for all the SIGIR 2022 methods we analyze. Results for the SIGIR 2022 methods are highlighted in \textbf{bold} only if they outperform all baselines.}
    \label{tab:ALL-both-result}
    \begin{minipage}{.45\textwidth}
        \caption{Experimental results for the Amazon-Book dataset.}
        \label{tab:ALL-amazon_book-result}
        \footnotesize
        \centering
        \begin{tabular}{l|cc|c|}
\toprule
{} & \multicolumn{2}{c}{Cutoff  20} \vline & \multirow{2}{*}{\shortstack{Cases\\Explored}} \\
{} &              Recall &             NDCG &            \\
\midrule
TopPop               &           0.0370 &           0.0168  &  -   \\
UserKNN CF    &           0.2301 &           0.1371 &  50    \\
ItemKNN CF    &           0.2436 &  \textbf{0.1496}  &  50   \\
\pbeta              &           0.2474 &  \textbf{0.1488}  &  50   \\
GF-CF       &                - &                -  &  -   \\
\midrule
SLIM                &           \textbf{0.2511} &  \textbf{0.1563}  &  50   \\
NegHOSLIM (EN) &           0.2472 &  \textbf{0.1536} &  50    \\
MF-BPR               &           0.1633 &           0.0911  &  50   \\
\iALS                 &           0.2378 &           0.1356  &  50   \\
MultVAE             &           0.2485 &  \textbf{0.1473} & 50\\
\midrule
GDE                  &           0.0004 &           0.0002 &  50   \\
GTN                  &           0.1852 &           0.0996 &  15   \\
HAKG                 &                - &                - &   -  \\
RGCF                 &                - &                - &   -  \\
HCCF                 &           0.1328 &           0.0651 &  50   \\
INMO                 &  \textbf{0.2511} &           0.1456 &  45  \\
KGCL                 &           0.2425 &           0.1403 &  13   \\
SimGCL               &           0.2441 &           0.1412 &  38   \\
LightGCN             &           0.2442 &           0.1407 &  34   \\
\bottomrule
\end{tabular}
   
    \end{minipage}%
    \quad
    \begin{minipage}{.45\textwidth}
        \caption{Experimental results for the Yelp2018 dataset.}
        \label{tab:ALL-yelp-result}
        \footnotesize
        \centering
        \begin{tabular}{l|cc|c|}
\toprule
{} & \multicolumn{2}{c}{Cutoff  20} \vline & \multirow{2}{*}{\shortstack{Cases\\Explored}} \\
{} &              Recall &             NDCG &            \\
\midrule
TopPop               &           0.0213 &           0.0132 &  -   \\
UserKNN CF     &           0.0988 &           0.0668 &  50   \\
ItemKNN CF     &           0.1057 &           0.0718 &  50   \\
\pbeta              &           0.1043 &           0.0693 &  50   \\
GF-CF       &                - &                - &  -    \\
\midrule
SLIM      &           0.1007 &           0.0700 &  50   \\
NegHOSLIM (EN) &           0.0994 &           0.0667 &  50   \\
MF-BPR               &           0.0556 &           0.0358 &  50   \\
IALS                 &           0.1120 &           0.0750 &  50   \\
MultVAE      &           0.1188 &           0.0796 &  50   \\
\midrule
GDE                  &           0.0834 &           0.0535 &  50   \\
GTN                  &           0.1000 &           0.0643 &  17   \\
HAKG                 &                - &                - &  -   \\
RGCF                 &                - &                - &  -   \\
HCCF                 &           0.0610 &           0.0399 &  50   \\
INMO                 &  \textbf{0.1219} &  \textbf{0.0809} &  28   \\
KGCL                 &           0.1125 &           0.0745 & 24\\
SimGCL               &  \textbf{0.1222} &  \textbf{0.0822} &  30   \\
LightGCN             &           0.1172 &           0.0781 &  30   \\
\bottomrule
\end{tabular}

    \end{minipage}
\end{table}

\subsection{Comparison Between the SIGIR 2022 Analyzed Methods}
\label{sec:comparison_all}
To provide a more complete picture of the effectiveness of the GNN models we analyze and to separate this assessment from potential confounding factors related to hyperparameter optimization, which, as we have pointed out, is generally extremely limited, we select two datasets and conduct an independent hyperparameter optimization. It is important to note that this type of comparison comes with a significant computational cost, as all the examined methods require iterative training. The experiments we report in this section required approximately six months of GPU time, making this type of evaluation infeasible for many academic or independent researchers.

\paragraph{Datasets}
We selected the Amazon-Book and Yelp2018 datasets provided in \cite{DBLP:conf/kdd/Wang00LC19} because they are the only datasets among those we analyzed that also include the required knowledge base for HAKG and KGCL.

\paragraph{Optimization} The hyperparameter optimization and early-stopping follow the same approach used for the baseline methods, as described in Section \ref{subsec:baselines}. The metric optimized is NDCG@20. The additional material contains the list of hyperparameters, their ranges and distributions.

\paragraph{Results}
The results of the experiments, along with the number of hyperparameter sets explored within the allotted 14 days, are reported in Table \ref{tab:ALL-both-result}. Some results are missing when the method exceeded either the 64GB of RAM available on our server or the 24GB available on our RTX 3090 GPUs. While we attempted to modify the hyperparameter search space to reduce the memory requirements we could not successfully conduct the optimization for those methods. The additional material reports the optimized values of the hyperparameters for all baselines and GNN algorithms.

The results for the Amazon-Book dataset, reported in Table \ref{tab:ALL-amazon_book-result}, once again show that none of the GNN methods outperform our selection of simple baselines. However, we observe that INMO matches SLIM in terms of Recall and that, for some methods (INMO, KGCL, SimGCL, and LightGCN), the effectiveness gap is much smaller compared to the previous results we obtained in our reproducibility analysis, see Section \ref{sec:method_LightGCN}, \ref{sec:method_SimGCL}, \ref{sec:method_INMO}, and \ref{sec:method_KGCL}. On the other hand, both HCCF and GTN prove ineffective. While HCCF completed the allotted 50 hyperparameter trials, GTN was limited to only 15 due to its long training time which may partly explain its poor effectiveness. Again, as we previously observed in Section \ref{sec:method_GDE}, GDE exhibits numerical instabilities that prevented it from training properly. Note that slight differences in preprocessing and data splitting make it impossible to directly compare the absolute values of effectiveness metrics across many of the experiments we reported in the previous sections. For example, in the analysis of SimGCL (see Section \ref{sec:method_SimGCL}), its results with the original hyperparameters showed an NDCG@20 of 0.1047, whereas SLIM achieved 0.1816, a substantial difference. However, in this comparison experiment on Amzon-Book, the two methods are within 2\%. A similar trend is observed for INMO, which previously had a result of 0.0934 compared to SLIM's 0.1451, a 55\% difference, while in this experiment the results differ by 7\%; KGCL, which reached 0.0794 compared to SLIM's 0.1031, a 30\% difference, while in this experiment results differ by 11\%; and LightGCN, which originally reached 0.0862 compared to SLIM's 0.1838, a whopping 113\% difference, while here the results differ by 11\%. Given that the data split used here does not appear to exhibit anomalies, we argue that a major reason for this large difference is the ineffectiveness of the original hyperparameter tuning. Since the weak baselines used for comparison did not present a sufficient challenge, there was little incentive to properly fine-tune the methods. This further reinforces the observation that the experimental practices adopted in the community are often ineffective and unreliable, and allow researchers to support many contradicting conclusions depending on minor changes in the experimental protocol. As discussed by \citet{DBLP:conf/recsys/ShehzadJ23}, \emph{"Everyone's a Winner"}, any method can appear competitive if the analysis is conducted carelessly, and current peer review practices do not seem capable of detecting such cases.

The results for the Yelp2018 dataset, reported in Table \ref{tab:ALL-yelp-result}, present a different picture, with two GNN-based methods, INMO and SimGCL, outperforming all baselines. This is partially aligned with our previous analysis of Section \ref{sec:method_SimGCL}, where SimGCL was the best method on the original (erroneous) Yelp2018 data split and the second-best method after MultVAE on the split we generated. Similarly, INMO, which was originally the second-best method after MultVAE, see Section \ref{sec:method_INMO}, shows improved effectiveness after our optimization. Once again, GDE and HCCF fall significantly behind, with HCCF in particular achieving only half the Recall of the best-performing method, consistent with our original experiments in Sections \ref{sec:method_GDE} and \ref{sec:method_HCCF}. Notably, GTN exhibits very weak effectiveness in this setting, despite ranking as the third-best method after MultVAE and GF-CF in our previous reproducibility experiments in Section \ref{sec:method_GTN}. This drop in effectiveness can be attributed to our hyperparameter optimization being truncated after only 17 trials, as the method exceeded the allotted 14 days.

Overall, the only conclusive statement we can make is that there is little consistency between the results of the experiments conducting according to the original papers and those we conducted in this section. For the Amazon-Book dataset, our optimized GNN models often performed much better than originally reported, though not enough to outperform MultVAE or SLIM, highlighting the insufficiency of the original hyperparameter optimization. A similar, though less pronounced, trend was observed on Yelp2018, where some progress over the baselines could be seen. 
It seems, therefore, that unless transparent hyperparameter optimization is given greater attention, the results reported in research papers may have little meaning. Overall, if the reported baselines are weak and poorly optimized then the evaluation is conducted in a scenario where the recommendation problem itself is simply not challenging enough. When that is the case, there is no need to push our methods and models forward and, despite the large number of papers published each year, the field risks a long phase of stagnation.

\section{Discussion}
\label{sec:discussion}

In this section, we summarize our findings across the following dimensions: artifacts, methodological issues, reproducibility, and baselines.

\subsection{Artifacts} 

\paragraph{Availability}
Our study reveals a substantial improvement in the accessibility of the original artifacts, source code, and data splits compared to prior studies~\cite{DBLP:journals/tois/DacremaBCJ21, 10.1145/3360311, stodden2018empirical}. 
Nearly all of the ten papers analyzed in this study provide the essential artifacts online, with only one paper providing incomplete and non-executable ones. For recommender systems research, the availability of original artifacts has risen from less than 50\%, as reported by~\citet{DBLP:journals/tois/DacremaBCJ21} for papers published between 2015 and 2018, to 90\% in our study.

\paragraph{Consistency of the data artifacts} 
However, when evaluating the \emph{consistency} of the artifacts with the descriptions in the papers, several issues were identified. Table \ref{tab:anaysis_dataset_artifacts} summarizes the datasets used in each paper we analyzed, the availability of the original training-test data splits, and their consistency with the descriptions provided in the respective papers. 
Out of the nine papers that provided complete artifacts, five (LightGCN, SimGCL, HAKG, GTN and KGCL) shared training-test splits that did not appear to be consistent with a correct random holdout splitting process as described in the papers. 
These unusual splits were not motivated in the papers and could have introduced biases in the evaluation. 
Non-uniform splits have the potential to alter the distribution or popularity bias of the test set, which may affect the relative effectiveness of the algorithms evaluated.
Furthermore, we found it concerning that three papers (GDE, HAKG, and GTN) used at least one dataset where the training and test sets partially overlapped. This was particularly evident for the Last-FM dataset, whose split was sourced from a prior study \cite{DBLP:conf/kdd/Wang00LC19}. 
These issues lead to evaluations conducted under anomalous conditions as well as potential information leakage between training and testing, both of which could call into question the validity of the results presented in the papers.
While reusing datasets and splits from previous publications can facilitate the comparison of results across different studies, it is essential for authors to carefully examine the quality of the splits before utilizing them. In some cases, the paper did not explicitly state that the training-test splits were taken from previous studies, but we were able to discover this through an automated data inspection. We emphasize the importance of explicitly stating when a training-test split is reused from prior work, as this transparency aids in comparing results and identifying potential issues. A further step to streamline the use of datasets and ensure reliance on commonly agreed and verified splits could involve adopting the concept of, or even directly utilizing, \texttt{ir\_datasets}\footnote{\url{https://ir-datasets.com/}}~\cite{MacAvaneyEtAl2021}, which provides a standardized interface and API for numerous datasets in Information Retrieval.
While relying on shared and commonly available data splits has advantages, it also has disadvantages. A potential risk is that models may overfit to a specific dataset, since research contributions are generally evaluated based on their ability to outperform prior work on a given test set. This can lead to implicit selection bias, where models are optimized to exploit patterns specific to a dataset rather than on their ability to generalize. Furthermore, the more research studies iterate on the same benchmark data split, the higher the risk of overfitting not only on the dataset but also on the specific train-test split \cite{DBLP:conf/sigir/CraswellMYCVS21}. This phenomenon is known as \emph{leaderboard chasing} and has been widely discussed in the Information Retrieval community \cite{DBLP:journals/dagstuhl-reports/BauerCFFF23}, which leverages several benchmarking datasets some of which, \idest MS MARCO, have been in use for a decade. Indeed, previous research by \citet{DBLP:conf/sigir/LinCCMY21} observed how several of the best results for the MS MARCO leaderboard are not statistically different. Addressing this issue requires a combination of practices, including the use of multiple test sets, careful statistical validation, and, when feasible, blind evaluation settings where test labels remain hidden from researchers until final submission \cite{DBLP:journals/dagstuhl-reports/BauerCFFF23}.

As a further issue, papers often do not clearly specify which approach was used to perform a random holdout. In the papers we analyzed, we found two commonly used approaches: a \emph{global} one, where interactions are randomly sampled from the entire dataset, and a \emph{user-wise} one, where the sampling process is performed independently for each user. Splitting the dataset using these two approaches, even when applying the same split percentages, can lead to different distributions, especially if the data is very sparse. For example, consider a user with only two interactions. Depending on the strategy adopted, these interactions may end up in different splits, possibly leaving the user with no iterations in either the training or the test data. If many such cases occur, the data distribution between a global and a user-wise split may differ substantially. Therefore, omitting this information in the paper is an obstacle to correctly reproducing the experimental conditions.

\paragraph{Consistency of the source code artifacts} 
When considering the consistency of the source code artifacts, our findings are much more positive, with all papers providing implementations of the proposed model that are consistent with what is described in the papers. A summary of our assessment on the consistency of the artifacts is reported in Table \ref{tab:anaysis_baseline_competitiveness}. However, when the training process is examined, it is not uncommon to find discrepancies. For instance, some papers state that the number of epochs is selected using early-stopping, yet their implementation does not include it. Furthermore, only one paper (INMO) provides an implementation of the hyperparameter optimization used for the proposed model. The remaining nine papers partially list the optimal hyperparameters in the paper and partially in the source code, creating fragmentation.
Lastly, HCCF employs hyperparameters outside the search space reported in the paper and limits the number of samples drawn per epoch in a manner that is not documented in the paper.
While these inconsistencies might be regarded as minor, they create additional obstacles for the research community in conducting reliable reproducibility studies, even when artifacts are made available.

\paragraph{Documentation} Among the potential obstacles to conducting a reproducibility study, a lack of documentation for the provided artifacts can be a significant challenge. When source code is shared without adequate documentation, it can be extremely difficult to determine how to use it and resolve any issues that arise especially if it contains a complex experimental pipeline involving multiple stages and scripts. In this regard, our findings are largely positive, for nine out of ten candidate papers the artifacts were sufficiently well-organized and documented to allow us to conduct the experiments. The only exception was the artifacts provided by \citet{DBLP:conf/sigir/LiuWZS22}, which did not meet this threshold. The source code lacked instructions on the required steps and contained several hard-coded paths for preprocessed data, which were also undocumented, making it impossible for us to proceed.

\paragraph{ACM SIGIR Badges}
Overall, we can be conclude that only two out of the ten reviewed papers (GDE and RGCF) meet the requirements for the less demanding of the ACM SIGIR Badges, specifically the \emph{Artifact Evaluated - Functional} badge. In all other cases, the provided artifacts were \emph{inconsistent} in various ways with the contents of the respective papers.

\begin{table}
    \caption{Summary of the analysis on the data artifacts summarizing the preprocessing applied, as well as whether the original training-test split was available and consistent with the description in the paper.}
    \label{tab:anaysis_dataset_artifacts}
    \begin{minipage}{\textwidth}
    \centering
    \footnotesize
    \resizebox{\textwidth}{!}{  
    \begin{tabular}{llcclccc}
    \toprule
        \multirow{2}{*}{Paper} 
        & \multirow{2}{*}{Datasets} 
        & \multirow{2}{*}{K-Cores} 
        & \multicolumn{2}{c}{Data split protocol}
        & \multicolumn{2}{c}{Training-test split} \\

        &  &  & Sampling & train-validation-test & Is available? & Is consistent? \\
    \midrule
        LightGCN     
        & Amazon-Book, Gowalla, Yelp2018
        & 10
        & user-wise
        & 72\% - 8\% - 20\%
        & Yes
        & No
        \\
        GDE 
        & CiteULike-a, Gowalla, MovieLens 100k and 1M, Pinterest
        & none
        & global
        & 19\% - 1\% - 80\%
        & Yes
        & Yes\footnote{Negligible number of interaction overlap between training and test data for CiteULike-a and Pinterest.}
        \\        
        SimGCL   
        & Amazon-Book, Douban Book, Yelp2018
        & 10\footnote{With the exception of Douban Book, where all interactions with a rating of at lest 4 are retained.}
        & user-wise
        & 70\% - 10\% - 20\%
        & Yes
        & Partially\footnote{\label{foot:inconsistency_lightgcn}The data split for Amazon-Book, Gowalla and Yelp2018 is not consistent and is the same used in LightGCN.}
        \\
        RGCF    
        & Amazon-Book, MovieLens 1M, Yelp2018
        & 15\footnote{With the exception of MovieLens 1M, where no k-core is applied.}
        & global
        & 80\% - 10\% - 10\%
        & No
        & -
        \\
        INMO      
        & Amazon-Book, Gowalla, Yelp2018
        & 10\footnote{This is preceded by selecting only the interactions with a rating of at least 4. No information is provided on the preprocessing of Gowalla.}
        & user-wise
        & 70\% - 10\% - 20\%
        & Yes
        & Yes
        \\
        HCCF    
        & Amazon-Book, MovieLens 10M, Yelp2018
        & 10 or 20\footnote{Both Yelp2018 and
MovieLens 10M are preprocessed with a 10-core selection, while Amazon-Book with a 20-core one.}
        & not described
        & 70\% - 10\% - 20\%
        & Yes
        & Yes
        \\
        HAKG    
        & Alibaba-iFashion, Last-FM, Yelp2018 
        & 10
        & not described
        & 80\% - 10\% - 10\% \footnote{With the exception of Last-FM which uses a user-wise sampling, 72\% - 8\% - 20\%.}
        & Yes
        & Partially\footnote{\label{foot:inconsistency_lastfm}The data split for Last-FM comes from a previous paper \cite{DBLP:conf/kdd/Wang00LC19} outside the scope of this study.}
        \\
        GTN     
        & Amazon-Book, Gowalla, Last-FM, Yelp2018
        & 10
        & user-wise
        & 72\% - 8\% - 20\%
        & Yes
        & No\footref{foot:inconsistency_lightgcn} \footref{foot:inconsistency_lastfm}
        \\
        KGCL     
        & Amazon-Book, MIND, Yelp2018
        & 10\footnote{With the exception of MIND.}
        & user-wise\footnote{With the exception of MIND that uses global sampling.}
        & 72\% - 8\% - 20\%
        & Yes
        & Partially\footref{foot:inconsistency_lightgcn}
        \\
    \bottomrule
    \end{tabular}
    }
    \end{minipage}
\end{table}

\subsection{Methodological Issues} 
When examining the experimental protocols adopted in the papers, several methodological issues emerge. 

\paragraph{Evaluation Procedure}
The design of the evaluation procedure involves several decisions, such as selecting datasets and preprocessing methods, determining which effectiveness measures to report, and the cutoffs. Previous studies observed that it was very rare to find two papers employing identical evaluation procedures, even when addressing the same task \cite{DBLP:journals/tois/DacremaBCJ21}. In our analysis, we note a slight improvement in this regard, as 8 out of 9 papers report results on commonly used datasets like Amazon-Book or Yelp2018, as shown in Table \ref{tab:anaysis_dataset_artifacts}. Conducting part of the experimental analysis on one or two commonly used datasets can greatly improve the transparency of results and facilitates comparisons across papers. 
Unfortunately, nearly all the papers apply different preprocessing strategies, making direct comparisons of their results almost impossible.
For example, while eight out of nine papers utilize the Yelp2018 dataset, it is used with six slightly different preprocessing and splitting techniques, most of which are unique, with the exception of a user-wise holdout with quotas for training, validation and test sets of 72\% - 8\% - 20\% and 70\% - 10\% - 20\% that occur twice.
Similarly, although seven papers use Amazon-Book, the dataset is preprocessed and split in five slightly different ways with the most common being a user-wise holdout split with quotas 72\% - 8\% - 20\% that occurs for three papers. 
The use of inconsistent and unmotivated data preprocessing raises questions about whether a specific preprocessing was chosen based on the goals of the paper and assumptions related to the scenario and task of interest, or whether it was treated as a hyperparameter to be explored in search of a good result, potentially influenced by confirmation bias.
Indeed, even a seemingly inconspicuous preprocessing step can significantly affect how the results should be interpreted. Consider for example the second most frequently used dataset, Amazon-Book. In its original form it contains $2.3 M$ items, $8 M$ users and $22.5 M$ interactions corresponding to a density of $10^{-6}$, however these numbers change drastically if we apply different commonly used preprocessing strategies:  
5-core ($367 k$ items, $603 k$ users, $8.8 M$ interactions, density of $4\cdot10^{-5}$), 
10-core ($128 k$ items, $158 k$ users, $4.7 M$ interactions, density of $2.3\cdot10^{-4}$), or 
20-core ($38 k$ items, $35 k$ users, $1.9 M$ interactions, density of $1.4\cdot10^{-3}$). When applying 10-core preprocessing, the most commonly adopted one (see Table \ref{tab:anaysis_dataset_artifacts}), the density of the dataset increases by a factor of $50$ and the statistics change so drastically that, for all practical purposes, these become entirely different datasets. In such cases, the name of the dataset creates a misleading perception of ``familiarity'' in the evaluation, which may not be accurate at all. Indeed, hidden within this seemingly minor preprocessing detail could be significant limitations of the proposed model, such as performing well only on very dense data, requiring a substantial amount of memory, or exhibiting very limited scalability. 
As a final observation, which will be further expanded in the following Sections, despite Amazon-Book being the second most used dataset, it is the one where the original and reproduced results of the analyzed message-passing methods perform worse compared to the simple baselines often by a very large margin, while the difference is less pronounced in our independent optimization, see Section \ref{sec:comparison_all}. It is surprising that this could happen for such a frequently used dataset without anyone seeming to take notice.

\paragraph{Model Optimization}
A second very common methodological issue is a lack of transparency regarding how the hyperparameters of the proposed method, as well as its number of training epochs, are selected. 
As shown in Table \ref{tab:anaysis_code_artifacts}, even though the methods have numerous hyperparameters, up to 18 for KGCL, only a small subset is reported as being tuned. The remaining hyperparameters are typically set to fixed values, often without explanation or with only a generic reference to ensuring a \emph{fair} comparison, without clarifying why such settings should be considered fair. This lack of transparency makes it generally impossible to determine whether the hyperparameter tuning process was conducted reliably. This becomes problematic when the proposed model's results are compared with those of baselines and when it obscures potential limitations of the model, such as requiring particularly careful fine-tuning to achieve competitive results. Indeed, it is known that improper optimization can drastically alter the outcomes of an experiment, often being the difference between supporting a conclusion or its opposite \cite{DBLP:conf/recsys/ShehzadJ23}. Our independent optimization confirms this observation, yielding results that are substantially different from those obtained with the original hyperparameters and showing a much smaller relative difference compared to the best baselines (see Section \ref{sec:comparison_all}). While the inclusion of ablation studies, which are often reported for a limited number of hyperparameters (2–3), helps and is positive, that alone is not sufficient to ensure transparency.

A particular hyperparameter is the number of epochs for which the model should be trained. This is the only hyperparameter for which it is sometimes possible to identify the selection criteria within the source code artifacts and therefore assess the consistency with what is described in the paper. 
Several issues emerge in this regard as well. Only two of the analyzed papers (RGCF and INMO) provided implementations with correct early-stopping based on validation data. In two other cases (HAKG  and KGCL), the implementations performed early-stopping based on test data, which introduces information leakage and results in an overestimation of the model's effectiveness. For five of the papers we analyzed, there is little to no information on how the number of training epochs was determined, either in the paper or in the source code. 
Some of these papers include a plot showing how the model converges during training, but with limited details, such as whether the convergence is measured on validation or test data. Only one paper (GDE) specifies the number of training epochs for each dataset, but it provides no information on how these numbers were determined.
However, all papers compute and plot the loss function on the test set during training, despite there being no reason to do so, and at the risk of researchers using this information to fine-tune the training process.

\paragraph{Sampled Metrics}
On a positive note, we observe that not a single paper has applied \emph{sampled metrics} in their evaluation. Typically the evaluation is performed by ranking all available items in the dataset, while sampled metrics only rank a small number of them (often 1000 or 100) aiming to reduce the computational cost of the evaluation. The use of sampled metrics had become commonplace a few years ago but was later found by \citet{DBLP:conf/kdd/KricheneR20} to produce results that are inconsistent with the traditional evaluation. They concluded:  "\emph{It has shown that most metrics are inconsistent under sampling and can lead to false discoveries. ... For this reason, sampling should be avoided as much as possible during evaluation}". Since then, some efforts have been made to address this issue, but the broader community has largely reacted by abandoning the use of sampled metrics. This constitutes a good example of the community self-healing, by reacting to an issue in the way experiments were conducted and moving away from methods that were shown to produce unreliable results.

\begin{table*}
    \caption{Summary of the analysis on the optimization of the model hyperparameters, in particular on how many of the method hyperparameters were stated to be tuned, whether the number of training epochs was selected with early-stopping and whether the provided implementation uses test data during training.}
    \label{tab:anaysis_code_artifacts}
    \begin{minipage}{\textwidth}
    \centering
    \footnotesize
    \begin{tabular}{lccl}
    \toprule
    Paper 
    & \shortstack{N. of Hyperparameters\\tuned}    
    & \shortstack{Paper mentions\\early-stopping?} 
    & Implementation uses test data during training?\\
    \midrule
        LightGCN   
        & 1 of 9
        & No
        & No, but computes results during early-stopping
        \\
        GDE 
        & 3 of 11
        & Yes
        & No, but computes results during early-stopping
        \\        
        SimGCL 
        & 2 of 10
        & No
        & No
        \\
        RGCF    
        & Not stated
        & Yes
        & No
        \\
        INMO
        & 2 of 12
        & Yes
        & No
        \\
        HCCF 
        & 4 of 14
        & No
        & No, but computes results during early-stopping
        \\
        HAKG    
        & 3 of 8
        & Yes
        & \textbf{Yes, to select optimal epochs during early-stopping}
        \\
        GTN     
        & 1 of 12
        & No
        & No, but computes results during early-stopping
        \\
        KGCL    
        & 3 of 18
        & Yes
        & \textbf{Yes, to select optimal epochs during early-stopping}
        \\
    \bottomrule
    \end{tabular}
    \end{minipage}
\end{table*}

\subsection{Reproducibility} 
We could partially reproduce the results for only five of the papers we analyze, namely GDE, SimGCL, RGCF, INMO, and KGCL. For two algorithms (RGCF and HAKG), we were not able to run experiments on all datasets due to high computational resource requirements. Table \ref{tab:anaysis_baseline_competitiveness} summarizes the percentage of individual effectiveness metrics that we could reproduce. As shown, there is wide variation, ranging from 0\% to 66\% depending on the method. While a reproducibility rate of 50\% is relatively low, it aligns with the outcomes of previous studies in physics and engineering~\cite{Baker2016}.

One particular issue we would like to raise is that the computational requirements of new methods have been steadily growing for several years. Consider how, in 2009, the Netflix Prize provided a dataset containing 100 million interactions. Yet now, despite continuous technological advancements that have substantially increased the computational capabilities of modern hardware, a large portion of research is conducted on datasets that contain less than 5\% of that number. For example, in the data splits used in this study, most datasets (Amazon-Book, Pinterest, Gowalla, MIND, etc.) contain between 1 and 3 million interactions, while some (CiteULike and Douban Book) have fewer than a million. Furthermore, almost all of these datasets include far fewer than 100,000 items and users.
While this is not a methodological issue per-se, despite the fact it could be argued that as the data becomes smaller the recommendation problem also becomes \emph{simpler} (\eg many niche users will disappear), it points to the fact that new methods are so computationally expensive that they are impractical to run on the hardware resources that are available to academics. When this is the case, they will also be too computationally demanding for real-world use on datasets of the scale of the Netflix Prize.
This issue limits the ability to conduct reproducibility analyses when hardware constraints make running the experiments impossible or impractical, as we found for RGCF and HAKG. One possible approach to mitigate this issue, when authors are aware that their proposed method has significant computational requirements, is to provide smaller datasets that enable researchers with limited hardware to replicate key findings, preferably from publicly available benchmarks. However, it is important to recognize that \emph{even smaller} datasets will be even less representative of real-world scenarios, furthermore they may not be optimal for training the model effectively, as they could contain dynamics that are \emph{too simple}. This recommendation also applies to industrial research, where publicly releasing datasets may not be feasible. In such cases, it is advisable to provide results for at least one publicly available dataset. However, it is important to note that this dataset may not contain all the information the model should leverage and, as such, can only serve as an imperfect proxy for evaluating the model’s effectiveness. A complementary strategy is to share pre-trained models, allowing researchers to validate results without requiring full-scale retraining. While this approach can reduce computational burdens, it comes at the cost of limiting the assessment of the reproducibility of the training and optimization process. Additionally, it introduces the risk of data leakage, as pre-trained models may encode information from datasets that should not be publicly accessible. Lastly, providing full experimental pipelines is advisable and should be encouraged, but it is essential to ensure they are accompanied by adequate documentation and instructions so that other researchers can attempt reproducibility analysis even years after the paper has been published. Overall, while ensuring full reproducibility for independent researchers may not always be feasible due to the high computational requirements of particular models structures or domains, ensuring the ability of other researchers to conduct at least a partial reproducibility analysis under reasonable constraints remains a valuable goal.

The issue of large computational cost also relates to the expectation that such a reproduction should be conducted during the review process. Given the number of papers a reviewer must evaluate, it is unreasonable to assume they will attempt to run the provided source code to assess their reproducibility or even check whether it is executable at all. At best, reviewers can assess the clarity and completeness of the provided artifacts and experimental setup, but verifying the results independently remains a broader challenge for the research community. In this respect, it is worth citing the work of \citet{DBLP:conf/ecir/LinZ20} titled \emph{"Reproducibility is a Process, Not an Achievement: The Replicability of IR Reproducibility Experiments"}, as reproducibility should indeed become a frequent and integral part of a larger discussion any healthy scientific community should have, critically reassessing its achievements and continuously reevaluating its practices. To this end, new approaches are needed. For example, the adoption of Registered Reports,\footnote{\url{https://www.cos.io/initiatives/registered-reports}} where the methodology is reviewed prior to conducting the experiments allowing the researcher to invest their efforts on the experiments only after the methodology has been approved. This approach could ensure greater methodological transparency and improved rigor by reducing confirmation bias pressure on researchers.  Furthermore, shifting the primary review focus from the results to the motivation and methodology would encourage research that is more grounded in hypotheses, which are often overlooked today.

\begin{table*}
    \caption{Summary of the analysis on the consistency, reproducibility of the results, and the comparison against our set of baselines. In particular, we report a judgment on artifact consistency that combines both the data and the source code, how many of the individual results could be reproduced, how many outperform all baselines and how many baselines are on a given dataset always better than the analyzed method.}
    \label{tab:anaysis_baseline_competitiveness}
    \begin{minipage}{\textwidth}
    \centering
    \footnotesize
    \begin{tabular}{l|cr|rc}
    \toprule
    Paper    
    & \shortstack{Artifact\\consistency}    
    & \shortstack{N. of results\\reproduced}
    & \shortstack{N. of results better\\than all baselines}
    & \shortstack{N. of better baselines\\on each dataset (min - max)} \\
    \midrule
        LightGCN 
        & Partial
        & 4/6 (66\%)
        & 0/10 (0\%)
        & 1 to 10
        \\    
        GDE 
        & Full
        & 4/10 (40\%)
        & 8/10 (80\%)
        & 0 to 1
        \\        
        SimGCL 
        & Partial
        & 3/6 (50\%)
        & 0/10 (0\%)
        & 1 to 11
        \\     
        RGCF   
        & Full
        & 1/4 (25\%)
        & 0/4 (0\%)
        & 6
        \\     
        INMO
        & Full
        & 4/9 (44\%)
        & 0/9 (0\%)
        & 1 to 8
        \\     
        HCCF 
        & Full
        & 2/12 (17\%)
        & 0/12 (0\%)
        & 9 to 13
        \\     
        HAKG 
        & Partial
        & 0/4 (0\%)
        & 0/4 (0\%)
        & 5 to 13
        \\     
        GTN 
        & Partial
        & 4/8 (50\%)
        & 1/12 (8\%)
        & 0 to 10 
        \\     
        KGCL  
        & Partial
        & 2/6 (33\%)
        & 0/6 (0\%)
        & 6 to 12
        \\      
    \bottomrule
    \end{tabular}
    \end{minipage}
\end{table*}

\subsection{Baselines} 
The competitiveness of the message passing methods we analyzed compared to simple baselines appears to be relatively weak. Table \ref{tab:anaysis_baseline_competitiveness} presents, for each paper, the percentage of individual metrics in which the best result we obtained for the proposed model outperformed \emph{all} our baselines. Additionally, it reports the number of baselines that were \emph{always} better than the model across all metrics on a given dataset. 
We can observe that GDE is the only model able to outperform our set of baselines in several measurements (80\%), corresponding to all results reported in four out of five datasets. It should be noted however that GDE uses a particular data split where the quota of the data used for training (20\%) is much smaller than the testing one (80\%), which may put it in a somewhat different scenario compared to the other methods.
We also report the minimum and maximum number of baselines that consistently outperform the proposed model across all measurements on a dataset. This number provides a more complete perspective on the relative effectiveness of the model. For example, when GDE is not competitive against the baselines, there is only one baseline that outperforms it, indicating that its effectiveness is good. Similarly, for GTN the number of consistently better baselines has a minimum of 0 because, in one dataset, GTN outperforms all baselines on one measurement but not on others, meaning no single baseline is consistently better than GTN across all metrics. In some cases (SimGCL, INMO, LightGCN) we observe datasets where only one baseline outperforms the proposed model. This baseline is typically MultVAE, which emerges as the single strongest baseline in our study.  In these cases, while the message passing methods are not the best overall, they still exhibit relatively strong results. However, for the same methods, there are other datasets where between 8 and 11 baselines outperform them, indicating that their effectiveness is inconsistent and can fall severely behind. 

Particularly striking are the results for the Amazon-Book dataset, where most of the analyzed algorithms perform considerably worse than simple baselines. The most notable example is LightGCN, which achieves only \emph{half} the effectiveness of the baselines. This is especially surprising considering that Amazon-Book is the second most commonly used dataset, after Yelp2018.
It is to be expected, and indeed quite normal, that certain methods or even multiple methods based on the same underlying architecture may not perform well on specific datasets.  However, the current common practice of overlooking stronger baselines creates a misleading impression that the effectiveness of state-of-the-art methods is always improving. In many cases, as shown in our analysis, this is not true. For Amazon-Book, the effectiveness of what are considered state-of-the-art methods has declined substantially. The results of our independent hyperparameter optimization in Section \ref{sec:comparison_all} confirm this, showing how better optimization, or in some cases any optimization at all, can significantly enhance the effectiveness of the methods. However, the state-of-the-art baselines used in the analyzed papers for Amazon-Book were so weak that there was no incentive to perform even this basic form of optimization.
We also suggest that including at least one example of a negative result, along with an explanation of how that dataset or task differs from others, should be encouraged as a valuable scientific practice. This approach would offer valuable insights to the research community, highlighting where the proposed methods are most effective, identifying areas where further development is necessary, and scenarios where the method may not be suitable at all.

Overall, the competitiveness of message passing algorithms against simple baselines appears limited, with only GDE in its unique data split outperforming the baselines on most measurements, and only INMO and SimGCL outperforming the baselines on Yelp2018 in our independent optimization. While some methods, such as GTN, achieve results close to the best-performing baseline (MultVAE), their effectiveness varies greatly across datasets. This suggests that message passing algorithms may be effective only in specific scenarios or that further work is needed to improve their generalizability.
It is worth noting, however, that all the analyzed methods apply message passing on relatively similar and simple graph structures, typically the traditional bipartite graph derived from user-item interaction data, with only two methods (HAKG and KGCL) extending this approach to more complex knowledge graphs. In this context it may be relevant to draw an analogy from the experience reported by \citet{DBLP:journals/aim/SteckBELRB21} at Netflix. In the paper, the authors explained that, during Netflix's early experiments with deep learning, it proved challenging to achieve results that outperformed traditional baselines, a situation similar to that reported by \citet{DBLP:journals/tois/DacremaBCJ21}. The breakthrough occurred when the model was enriched with several new features, enabling it to leverage the strengths of deep learning. It is possible that message passing methods are in a similar position when applied for traditional collaborative filtering problems, where the strengths of graph-based approaches may not be utilized effectively, resulting in these methods being less competitive than much simpler baselines. Identifying scenarios where these methods can consistently and robustly demonstrate their advantages, and determining whether further adaptations are necessary to align them with the specific graph topologies and semantics of recommendation problems, remains an open question that only further research and industrial experience can answer. However, as we have discussed, many experimental practices that are currently commonplace and accepted at high-level venues hinder the community's ability to achieve this goal.

\subsection{Impact on Follow-up Research}

In this section, we conduct a qualitative analysis of how the SIGIR 2022 papers we analyzed influenced subsequent SIGIR 2023 papers that used them as baselines, and whether some of the issues we identified in 2022 persisted in 2023.
As opposed to the previous analysis, which focused on evaluating the consistency of artifacts and attempting to reproduce results, this section takes a more qualitative approach, relying on the descriptions provided in each paper.

As described in Section \ref{sec:research-method}, we identified eleven papers from SIGIR 2023, listed in Table \ref{tab:baselines_SIGIR23} alongside the SIGIR 2022 papers they used as baselines. In this section, we will first analyze them along the same dimensions we used for the SIGIR 2022 papers, namely the artifacts availability, evaluation procedure and methodological issues. Finally, we will discuss how the methods from SIGIR 2022 have been used as baseline and attempt to compare their results.

\begin{table}
    \caption{Overview of the papers published in SIGIR 2023 that meet our selection criteria described in Section \ref{sec:research-method}, reporting which of the SIGIR 2022 we analyzed they use as baseline, as well as a summary of the preprocessing applied.}
    \label{tab:baselines_SIGIR23}
    \begin{minipage}{\textwidth}
    \centering
    \footnotesize
    \resizebox{\textwidth}{!}{  
    \begin{tabular}{lllclc}
    \toprule
        \multirow{2}{*}{Paper} 
        & \multirow{2}{*}{Baselines} 
        & \multirow{2}{*}{Datasets} 
        & \multirow{2}{*}{K-Cores} 
        & \multicolumn{2}{c}{Data split protocol}\\
        &  & &  & Sampling & train-validation-test \\
    \midrule
        KRDN \cite{ZhuEtAl2023b} 
        & SimGCL, KGCL 
        & Alibaba-iFashion, Last-FM, Yelp2018
        & not described
        & not described
        & not described
        \\
        MixGCN \cite{WangEtAl2023} 
        & KGCL
        & MovieLens 1M, Last-FM, Book Crossing
        & 1
        & global
        & 60\% - 20\% - 20\%\footnote{The data includes additional \emph{fake} interactions.}
        \\
        AdaMCL \cite{ZhuEtAl2023}     
        & LightGCN, SimGCL, HCCF
        & Yelp, Amazon-Book, Gowalla, Alibaba-iFashion
        & 15\footnote{The preprocessing is described in another cited paper.}
        & not described
        & 80\% - 10\% - 10\%
        \\
        TriSIM4Rec \cite{LiuEtAl2023}   
        & LightGCN
        & Amazon-Video, Amazon-Game, MovieLens 1M and 100k
        & 1
        & global, time-based
        & 80\% - 10\% - 10\%
        \\
        BSPM \cite{ChoiEtAl2023} 
        & LightGCN, GTN
        & Gowalla, Yelp2018, Amazon-Book
        & not described
        & not described
        & not described\footnote{The paper states that the data and train/test split is the same as previous studies, but does not explicitly state which ones.} 
        \\
        CoRML \cite{WeiEtAl2023} 
        & SimGCL
        & Pinterest, Gowalla, Yelp2018, MovieLens 20M
        & 1\footnote{For MovieLens 20M only users with at least 5 interactions are retained.}
        & not described
        & 60\% - 20\% - 20\%
        \\
        VGCL \cite{YangEtAl2023} 
        & LightGCN, SimGCL
        & Douban-Book, Dianping, MovieLens 25M
        & 1\footnote{Only users with at least 10 interactions are retained.}
        & not described
        & 80\% - 0\% - 20\%
        \\
        DCCF \cite{RenEtAl2023} 
        & LightGCN, HCCF
        & Gowalla, Amazon-Book, Tmall
        & not described
        & not described
        & not described
        \\
        DGMAE \cite{RenEtAl2023b} 
        & LightGCN
        & Youshu, NetEase, Alibaba-iFashion
        & 1
        & not described
        & 70\% - 10\% - 20\%
        \\
        CGCL \cite{HeEtAl2023} 
        & LightGCN
        & Gowalla, Yelp2018, Amazon-Book
        & 15\footnote{The preprocessing is described in another cited paper.}
        & not described
        & 80\% - 10\% - 10\%
        \\
        GFormer \cite{LiEtAl2023b} 
        & LightGCN, HCCF
        & Yelp, Alibaba-iFashion, Last-FM
        & 1
        & not described
        & 70\% - 5\% - 25\%
        \\
    \bottomrule
    \end{tabular}
    }
    \end{minipage}
\end{table}

\paragraph{Artifacts Availability}
While the proportion of papers that provide publicly available artifacts is higher than observed in other studies and well above 50\%, it fluctuates over the years. For SIGIR 2022 the proportion was 100\%, whereas for SIGIR 2023 it dropped to 63\%, with seven out of eleven papers providing artifacts \cite{ZhuEtAl2023b, ZhuEtAl2023, ChoiEtAl2023, WeiEtAl2023, RenEtAl2023, HeEtAl2023, LiEtAl2023b}. 
We should also note that if the consistency of the artifacts with the descriptions in the papers were evaluated, the number of consistent artifacts would likely decrease, highlighting the need for continued efforts in this area. On a positive note, all seven papers providing artifacts included the data splits or used implementations written with RecBole, an open-source Python library. While relying on open-access libraries introduces potential issues related to the correctness and consistency of the implementations, as is the case with any third-party method implementation, see for example \citet{DBLP:conf/recsys/HidasiC23}, relying on a limited number of well-maintained libraries can help mitigate this problem. Over time, as issues are identified and corrected, such libraries can contribute significantly to improving the overall quality and reproducibility of research.

\paragraph{Evaluation Procedure}
Similarly to what was observed in 2022, the evaluation protocols in the SIGIR 2023 papers vary significantly, see Table \ref{tab:baselines_SIGIR23}. Overall, 18 different datasets are used, with Yelp and Gowalla being the most common ones, appearing in 6 and 5 out of eleven papers, respectively. 
Evaluating methods on common datasets and relying on consistent preprocessing and splitting steps would facilitate easier comparisons of the results reported across different papers. However, we again observe that even when papers use the same dataset, the preprocessing and splitting steps often differ. For example, the most frequently used dataset is Yelp2018, which is however preprocessed in three different ways. 
A further persistent issue is that detailed information on the data processing may be missing. For instance, three out of eleven papers \cite{ZhuEtAl2023b, RenEtAl2023, ChoiEtAl2023} do not report the percentage of interactions used for the training-test split, and three out of eleven do not provide details on how the data is preprocessed. There is also a particular instance, in paper \cite{ChoiEtAl2023}, where it is stated that to maintain fairness with ``previous studies'' the same datasets and training test splits are used. Unfortunately, the paper does not reference which previous studies are being referred to, so the reader is left to speculate.

\paragraph{Model Optimization}
When analyzing how the model optimization is conducted, we again see strong similarities with what was observed in SIGIR 2022.
In most of the SIGIR 2023 papers we analyzed, some form of hyperparameter tuning on a validation set is present. However, many hyperparameters are often fixed a-priori without any particular justification, and most optimizations rely on notoriously inefficient grid searches that explore only a limited number of cases. Typically, key hyperparameters like the learning rate, embedding size, and batch size, critical components of iterative methods, are  fixed seemingly arbitrarily or even omitted from the descriptions altogether \cite{ZhuEtAl2023b, WangEtAl2023, ZhuEtAl2023, WeiEtAl2023, YangEtAl2023, RenEtAl2023, RenEtAl2023b, HeEtAl2023, LiEtAl2023b}. These hyperparameters should be carefully tuned, as different configurations can yield to substantially different results for different methods. 

Another persistent issue is the lack of transparency in how methods with iterative training determine the number of epochs. Among the papers we analyzed,  aside from one article \cite{ChoiEtAl2023} that proposes a method not requiring any training and two articles \cite{ZhuEtAl2023, HeEtAl2023} that properly describe the use of early-stopping, one paper \cite{LiuEtAl2023} provides incomplete description, while the remaining papers offer no information at all  \cite{ZhuEtAl2023b, WangEtAl2023, WeiEtAl2023, YangEtAl2023, RenEtAl2023, RenEtAl2023b, LiEtAl2023b}. 

Regarding the optimization of baselines reported in the papers, it is apparent that insufficient attention is given to clarifying how hyperparameters are selected. Two papers \cite{WangEtAl2023, ChoiEtAl2023} adopt the common yet bad practice of using the baseline hyperparameters values reported in the original papers proposing them, even when, as we very frequently observed, the experimental procedure differs significantly, including the use of different datasets. While tuning hyperparameters based on the methodology in the original paper proposing the method can generally be considered good practice, as done in \cite{ZhuEtAl2023b}, the frequent occurrences where this tuning was poorly done, ignoring even very important hyperparameters, highlight the need for caution given the importance of this step \cite{DBLP:conf/recsys/ShehzadJ23}. Indeed, this tuning should be approached critically to avoid repeating any methodological errors. All other papers simply state that each hyperparameter is tuned properly without providing meaningful details, apart from occasional mentions of the search space for the number of GCN layers in \cite{RenEtAl2023, LiEtAl2023b}. In one case \cite{ChoiEtAl2023}, no information is provided regarding the optimization of baseline hyperparameters.

\paragraph{Papers from SIGIR 2022 used as baselines}
Our goal for this section is to assess to what extent irreproducible or weak methods from SIGIR 2022 have been used as baselines in SIGIR 2023. Table \ref{tab:baselines_SIGIR23} lists, for each of the eleven SIGIR 2023 papers we identified, the SIGIR 2022 papers they used as baselines.
It is clear that LightGCN is the most influential model, included as baseline in eight out of eleven papers. The second most frequently used baseline is SimGCL, included in four papers, followed by HCCF, which is used in three. In all of these cases, the methods are presented as representing the state-of-the-art, which, as we have observed in this study is not true with very limited exceptions. In our independent optimization, see Section \ref{sec:comparison_all}, HCCF and LightGCN were consistently below the baseline models and SimGCL exhibited competitive results only on Yelp2018.
Unfortunately, it is challenging to compare and cross-check the results for the baselines with those reported in the papers that proposed them. For example, one might be interested to assess the consistency in the effectiveness of LightGCN, SimGCL, and HCCF, the three most-used baselines. In their original papers, LightGCN and SimGCL employ splits of 72\%-8\%-20\% on Gowalla, Amazon-Book, and Yelp2018. HCCF, on the other hand, is evaluated on Yelp, Amazon-Book, and MovieLens 10M with a 70\%-10\%-20\% split. When comparing this with the evaluation procedure used in the SIGIR 2023 papers we find that none adopts a data split consistent with those used in the original baseline papers. Furthermore, one paper \cite{WangEtAl2023} uses a 5-fold cross-validation instead of a single hold-out training-test set, while two papers \citet{YangEtAl2023, HeEtAl2023} repeat the experiments 5 times on the same test set. Although these are good practices which allow to measure the variance of the results, they further complicate direct comparisons between papers especially when the variance is not reported.
On one hand it is somewhat beneficial that the erroneous preprocessing and splitting steps used by LightGCN and SimGCL do not appear to be adopted in the SIGIR 2023 papers, except possibly \cite{ChoiEtAl2023}.\footnote{It should be noted that this cannot be confirmed without directly verifying the content of the data splits.} However, this comes at the significant cost of losing any hope of comparing results across methods and introduces the continuous risk of new mistakes emerging and propagating before they are identified. Indeed, the outcome of this part of the analysis is that it is not possible to compare the results of the SIGIR 2022 papers we analyzed with those reported as baselines in the SIGIR 2023 papers, which we find worrying.

A further step we can take is to examine the results more qualitatively to assess whether the method produces \emph{similar} results under \emph{similar} data splitting and preprocessing. However, we find this to be highly challenging. Based on the descriptions provided in the papers, we can identify only a handful of cases where such a comparison is possible. While additional comparable protocols may exist, the lack of sufficiently detailed descriptions in the papers means that discovering them would require manually analyzing the provided data. This type of investigative work is an unreasonable expectation for a researcher simply interested in understanding or building upon a paper. The first case is \citet{WeiEtAl2023}, which applies a similar data split to SimGCL \cite{DBLP:conf/sigir/YuY00CN22}, holding out 20\% of the data for testing but using different partitions for training and validation. The only dataset the two papers have in common is Yelp2018, and the only shared baseline is SimGCL itself. We observe a notable discrepancy in the results for SimGCL, with its NDCG@20 equal to 0.0601 in the original paper but 0.0795 in \cite{WeiEtAl2023}. Due to the limited details provided on hyperparameter optimization and the use of fixed values for some hyperparameters, we are unable to precisely determine whether this difference stems from variations in the data splits or differences in the optimization process.
A more successful comparison can be made with \citet{YangEtAl2023}, which reports the results of SimGCL and LightGCN on the Douban-Book dataset. Their NDCG@20 values are nearly identical up to the second decimal place. For example, SimGCL achieves 0.1583 in the original paper and 0.1540 in \cite{YangEtAl2023}, while LightGCN reaches 0.1272 in \cite{DBLP:conf/sigir/YuY00CN22} and 0.1278 in \cite{YangEtAl2023}. The results for another baseline, MultVAE, are similarly close, with 0.1103 in SimGCL and 0.1155 in \cite{YangEtAl2023}. However, all of these values remain significantly lower than the 0.1694 we obtained for MultVAE in our experiments, which would also slightly outperform the 0.1638 reported for the newly proposed algorithm.
The last paper with a similar data split is \citet{LiEtAl2023b}, but once again, we observe effectiveness measurements with substantially different absolute values. For example, the original NDCG@20 for HCCF on the Yelp2018 dataset was 0.0510, while \cite{LiEtAl2023b} reports a lower value of 0.0391. A similar discrepancy is found for LightGCN, where the original result on the (erroneous) Yelp2018 split was 0.0530, whereas in \cite{LiEtAl2023b}, it is 0.0373.

Overall, the outcome of this analysis is that while the SIGIR 2022 papers we examined have been used as baselines to a limited extent, with the exception of the popular LightGCN which however was published at SIGIR 2020, comparing their results with subsequent work from SIGIR 2023 is not feasible. This is due to several factors: (1) the rarity of papers that use the same experimental protocol or data splits in a transparent manner; (2) even when the experimental protocol appears similar, relying on comparable data splits and preprocessing, the reported results may differ significantly, with insufficient details in the papers to pinpoint the cause; and (3) relative comparisons between baselines are also difficult, as each paper tends to report a unique selection of methods with little overlap.

Despite being a research field that heavily relies on experimental findings, the growing body of research on reproducibility shows that these published findings often do not reflect the true effectiveness of a method and therefore provide little guidance for navigating the vast body of available literature. One of the great successes of the scientific method has been moving beyond an \emph{authoritative} conception of science, where arguments were deemed valid based on the prestige of the scholar presenting them. However, in a landscape where reliable points of reference are scarce, results are inconsistent and weak models are frequently published, even at top-tier venues, there is a growing risk that readers will be driven to mainly rely on the reputation of researchers or labs as a proxy to assess the correctness of the paper. This is the opposite of what the scientific process is meant to achieve.

\section{Conclusion}
This study examines the reproducibility of results reported in nine graph-based Recommender Systems papers published at SIGIR 2022. The focus is in particular on the consistency of the provided source code and data artifacts with the descriptions in the papers, the correctness of the experimental methodologies, the reproducibility of the published results, and the competitiveness of the proposed methods against robust baselines. Furthermore, this study explores how these papers may have impacted subsequent work published at SIGIR 2023. Overall, the analysis required considerable experimental efforts, involving the fitting of approximately 25.000 models with a total computation time of 4 years.

The findings of this study highlight significant issues related to artifact consistency and the propagation of poor practices in subsequent publications. Specifically, many of the available training-test data artifacts were clearly the result of an erroneous splitting procedure and were, at the very least, inconsistent with the descriptions provided in the corresponding papers. These data splits, sometimes reused from previous papers, lead to anomalous and questionable evaluation results. When the data split does not appear anomalous, it is typically a unique split created by the authors by applying a specific combination of preprocessing and training-validation-test split percentages, often not clearly described, making comparisons between papers impossible. Indeed, the phenomenon is so prevalent that it is not possible to compare results even between the papers published at SIGIR 2022 and 2023, which we find worrying considering how much of the research literature is based on empirical results. 
Our analysis also uncovered several bad practices, ranging from arbitrarily setting the values of critical hyperparameters to selecting the number of epochs based on the test data causing information leakage. 

When assessing the reproducibility of results, we found it to be poor. Only three out of nine papers were reproducible for at least 50\% of their results, with one paper being entirely irreproducible. Lastly, in terms of competitiveness against baselines, most of the analyzed methods demonstrated instances where they were reasonably competitive, alongside cases where they were not. Surprisingly, on the frequently used Amazon-Book dataset most message-passing methods are substantially below simple baselines despite claiming state-of-the-art results in the original papers. This phenomenon has been observed multiple times in previous studies focusing on different methods and remains a persistent issue. It is likely the result of several contributing factors that require further investigation.

Overall, to address these issues, it is imperative for future research to adopt more rigorous standards for artifact documentation and experimental methodology. Ensuring that the provided source code, datasets, and experimental procedures are thoroughly documented and consistent with their description in the paper is of utmost importance. However, it is important to acknowledge that conducting such detailed analyses and reproductions is time-consuming and requires meticulous effort. As such, expecting this level of scrutiny during the peer-review phase is unrealistic, as it would place an unsustainable burden on reviewers. To mitigate this, new approaches are needed. For example, the adoption of Registered Reports and open review tools could improve transparency. Additionally, encouraging the use of robust and simple baselines over more complex but often weaker ones in many scenarios would lead to more accurate evaluations. Lastly, encouraging the publication and discussion of negative results would help provide a more comprehensive understanding of the strengths and limitations of proposed approaches.

In conclusion, this study underscores the need for improved practices in the publication and evaluation of research in Recommender Systems. It is by addressing these issues that the community can improve the reliability of the published research findings and ensure more robust advancements in the field.

  \bibliographystyle{ACM-Reference-Format}
  \bibliography{main-references,zz-proceedings}


\begin{thebibliography}{85}


\ifx \showCODEN    \undefined \def \showCODEN     #1{\unskip}     \fi
\ifx \showISBNx    \undefined \def \showISBNx     #1{\unskip}     \fi
\ifx \showISBNxiii \undefined \def \showISBNxiii  #1{\unskip}     \fi
\ifx \showISSN     \undefined \def \showISSN      #1{\unskip}     \fi
\ifx \showLCCN     \undefined \def \showLCCN      #1{\unskip}     \fi
\ifx \shownote     \undefined \def \shownote      #1{#1}          \fi
\ifx \showarticletitle \undefined \def \showarticletitle #1{#1}   \fi
\ifx \showURL      \undefined \def \showURL       {\relax}        \fi
\providecommand\bibfield[2]{#2}
\providecommand\bibinfo[2]{#2}
\providecommand\natexlab[1]{#1}
\providecommand\showeprint[2][]{arXiv:#2}

\bibitem[Abadal et~al\mbox{.}(2022)]%
        {AbadalEtAl2022}
\bibfield{author}{\bibinfo{person}{S. Abadal}, \bibinfo{person}{A. Jain}, \bibinfo{person}{R. Guirado}, \bibinfo{person}{J. L{\'o}pez-Alonso}, {and} \bibinfo{person}{E. Alarc{\'o}n}.} \bibinfo{year}{2022}\natexlab{}.
\newblock \showarticletitle{{Computing Graph Neural Networks: A Survey from Algorithms to Accelerators}}.
\newblock \bibinfo{journal}{\emph{{ACM Computing Surveys (CSUR)}}} \bibinfo{volume}{54}, \bibinfo{number}{9} (\bibinfo{date}{December} \bibinfo{year}{2022}), \bibinfo{pages}{191:1--191:38}.
\newblock


\bibitem[Amig{\'o} et~al\mbox{.}(2022)]%
        {zz-SIGIR2022}
\bibfield{editor}{\bibinfo{person}{E. Amig{\'o}}, \bibinfo{person}{P. Castells}, \bibinfo{person}{J. Gonzalo}, \bibinfo{person}{B.~A. Carterette}, \bibinfo{person}{J.~S. Culpepper}, {and} \bibinfo{person}{G. Kazai}} (Eds.). \bibinfo{year}{2022}\natexlab{}.
\newblock \bibinfo{booktitle}{\emph{{Proc. 45th Annual International ACM SIGIR Conference on Research and Development in Information Retrieval (SIGIR 2022)}}}. \bibinfo{publisher}{{ACM Press, New York, USA}}.
\newblock


\bibitem[Anelli et~al\mbox{.}(2021)]%
        {DBLP:conf/recsys/AnelliBNP21}
\bibfield{author}{\bibinfo{person}{Vito~Walter Anelli}, \bibinfo{person}{Alejandro Bellog{\'{\i}}n}, \bibinfo{person}{Tommaso~Di Noia}, {and} \bibinfo{person}{Claudio Pomo}.} \bibinfo{year}{2021}\natexlab{}.
\newblock \showarticletitle{Reenvisioning the comparison between Neural Collaborative Filtering and Matrix Factorization}. In \bibinfo{booktitle}{\emph{RecSys '21: Fifteenth {ACM} Conference on Recommender Systems, Amsterdam, The Netherlands, 27 September 2021 - 1 October 2021}}, \bibfield{editor}{\bibinfo{person}{Humberto Jes{\'{u}}s~Corona Pamp{\'{\i}}n}, \bibinfo{person}{Martha~A. Larson}, \bibinfo{person}{Martijn~C. Willemsen}, \bibinfo{person}{Joseph~A. Konstan}, \bibinfo{person}{Julian~J. McAuley}, \bibinfo{person}{Jean Garcia{-}Gathright}, \bibinfo{person}{Bouke Huurnink}, {and} \bibinfo{person}{Even Oldridge}} (Eds.). \bibinfo{publisher}{{ACM}}, \bibinfo{pages}{521--529}.
\newblock
\href{https://doi.org/10.1145/3460231.3475944}{doi:\nolinkurl{10.1145/3460231.3475944}}


\bibitem[Anelli et~al\mbox{.}(2023)]%
        {DBLP:conf/recsys/AnelliMPBSN23}
\bibfield{author}{\bibinfo{person}{Vito~Walter Anelli}, \bibinfo{person}{Daniele Malitesta}, \bibinfo{person}{Claudio Pomo}, \bibinfo{person}{Alejandro Bellog{\'{\i}}n}, \bibinfo{person}{Eugenio~Di Sciascio}, {and} \bibinfo{person}{Tommaso~Di Noia}.} \bibinfo{year}{2023}\natexlab{}.
\newblock \showarticletitle{Challenging the Myth of Graph Collaborative Filtering: a Reasoned and Reproducibility-driven Analysis}. In \bibinfo{booktitle}{\emph{Proceedings of the 17th {ACM} Conference on Recommender Systems, RecSys 2023, Singapore, Singapore, September 18-22, 2023}}, \bibfield{editor}{\bibinfo{person}{Jie Zhang}, \bibinfo{person}{Li~Chen}, \bibinfo{person}{Shlomo Berkovsky}, \bibinfo{person}{Min Zhang}, \bibinfo{person}{Tommaso~Di Noia}, \bibinfo{person}{Justin Basilico}, \bibinfo{person}{Luiz Pizzato}, {and} \bibinfo{person}{Yang Song}} (Eds.). \bibinfo{publisher}{{ACM}}, \bibinfo{pages}{350--361}.
\newblock
\href{https://doi.org/10.1145/3604915.3609489}{doi:\nolinkurl{10.1145/3604915.3609489}}


\bibitem[Armstrong et~al\mbox{.}(2009)]%
        {ArmstrongEtAl2009b}
\bibfield{author}{\bibinfo{person}{T.~G. Armstrong}, \bibinfo{person}{A. Moffat}, \bibinfo{person}{W. Webber}, {and} \bibinfo{person}{J. Zobel}.} \bibinfo{year}{2009}\natexlab{}.
\newblock \showarticletitle{{Improvements That Don't Add Up: Ad-Hoc Retrieval Results Since 1998}}. In \bibinfo{booktitle}{\emph{{Proc. 18th International Conference on Information and Knowledge Management (CIKM 2009)}}}, \bibfield{editor}{\bibinfo{person}{D.~W.-L. Cheung}, \bibinfo{person}{I.-Y. Song}, \bibinfo{person}{W.~W. Chu}, \bibinfo{person}{X.~Hu}, {and} \bibinfo{person}{J.~J. Lin}} (Eds.). \bibinfo{publisher}{{ACM Press, New York, USA}}, \bibinfo{pages}{601--610}.
\newblock


\bibitem[Baker(2016)]%
        {Baker2016}
\bibfield{author}{\bibinfo{person}{M. Baker}.} \bibinfo{year}{2016}\natexlab{}.
\newblock \showarticletitle{{1,500 scientists lift the lid on reproducibility}}.
\newblock \bibinfo{journal}{\emph{{Nature}}}  \bibinfo{volume}{533} (\bibinfo{date}{May} \bibinfo{year}{2016}), \bibinfo{pages}{452--454}.
\newblock


\bibitem[Barros et~al\mbox{.}(2023)]%
        {BarrosEtAl2023}
\bibfield{author}{\bibinfo{person}{C.~D.~T. Barros}, \bibinfo{person}{M.~R.~F. Mendon{\c c}a}, \bibinfo{person}{A.~B. Vieira}, {and} \bibinfo{person}{A. Ziviani}.} \bibinfo{year}{2023}\natexlab{}.
\newblock \showarticletitle{{A Survey on Embedding Dynamic Graphs}}.
\newblock \bibinfo{journal}{\emph{{ACM Computing Surveys (CSUR)}}} \bibinfo{volume}{55}, \bibinfo{number}{1} (\bibinfo{date}{January} \bibinfo{year}{2023}), \bibinfo{pages}{10:1--10:37}.
\newblock


\bibitem[Bauer et~al\mbox{.}(2023)]%
        {DBLP:journals/dagstuhl-reports/BauerCFFF23}
\bibfield{author}{\bibinfo{person}{Christine Bauer}, \bibinfo{person}{Ben Carterette}, \bibinfo{person}{Nicola Ferro}, \bibinfo{person}{Norbert Fuhr}, {and} \bibinfo{person}{Guglielmo Faggioli}.} \bibinfo{year}{2023}\natexlab{}.
\newblock \showarticletitle{Frontiers of Information Access Experimentation for Research and Education (Dagstuhl Seminar 23031)}.
\newblock \bibinfo{journal}{\emph{Dagstuhl Reports}} \bibinfo{volume}{13}, \bibinfo{number}{1} (\bibinfo{year}{2023}), \bibinfo{pages}{68--154}.
\newblock
\href{https://doi.org/10.4230/DAGREP.13.1.68}{doi:\nolinkurl{10.4230/DAGREP.13.1.68}}


\bibitem[Bell and Koren(2007)]%
        {bell2007improved}
\bibfield{author}{\bibinfo{person}{Robert~M Bell} {and} \bibinfo{person}{Yehuda Koren}.} \bibinfo{year}{2007}\natexlab{}.
\newblock \showarticletitle{Improved neighborhood-based collaborative filtering}. In \bibinfo{booktitle}{\emph{KDD Cup and Workshop at the 13th ACM SIGKDD International Conference on Knowledge Discovery and Data Mining (KDD '07)}}. \bibinfo{pages}{7--14}.
\newblock


\bibitem[Benigni et~al\mbox{.}(2025)]%
        {DBLP:journals/corr/abs-2505-09364}
\bibfield{author}{\bibinfo{person}{Michael Benigni}, \bibinfo{person}{Maurizio {Ferrari Dacrema}}, {and} \bibinfo{person}{Dietmar Jannach}.} \bibinfo{year}{2025}\natexlab{}.
\newblock \showarticletitle{Diffusion Recommender Models and the Illusion of Progress: {A} Concerning Study of Reproducibility and a Conceptual Mismatch}.
\newblock \bibinfo{journal}{\emph{CoRR}}  \bibinfo{volume}{abs/2505.09364} (\bibinfo{year}{2025}).
\newblock
\showeprint[arXiv]{2505.09364}
\href{https://doi.org/10.48550/ARXIV.2505.09364}{doi:\nolinkurl{10.48550/ARXIV.2505.09364}}


\bibitem[Breuer et~al\mbox{.}(2020a)]%
        {BreuerEtAl2020}
\bibfield{author}{\bibinfo{person}{T. Breuer}, \bibinfo{person}{N. Ferro}, \bibinfo{person}{N. Fuhr}, \bibinfo{person}{M. Maistro}, \bibinfo{person}{T. Sakai}, \bibinfo{person}{P. Schaer}, {and} \bibinfo{person}{I. Soboroff}.} \bibinfo{year}{2020}\natexlab{a}.
\newblock \showarticletitle{{How to Measure the Reproducibility of System-oriented IR Experiments}}. In \bibinfo{booktitle}{\emph{{Proc. 43rd Annual International ACM SIGIR Conference on Research and Development in Information Retrieval (SIGIR 2020)}}}, \bibfield{editor}{\bibinfo{person}{Y.~Chang}, \bibinfo{person}{X.~Cheng}, \bibinfo{person}{J.~Huang}, \bibinfo{person}{Y.~Lu}, \bibinfo{person}{J.~Kamps}, \bibinfo{person}{V.~Murdock}, \bibinfo{person}{J.-R. Wen}, \bibinfo{person}{A.~Diriye}, \bibinfo{person}{J.~Guo}, {and} \bibinfo{person}{O.~Kurland}} (Eds.). \bibinfo{publisher}{{ACM Press, New York, USA}}, \bibinfo{pages}{349--358}.
\newblock


\bibitem[Breuer et~al\mbox{.}(2020b)]%
        {DBLP:conf/sigir/Breuer0FMSSS20}
\bibfield{author}{\bibinfo{person}{Timo Breuer}, \bibinfo{person}{Nicola Ferro}, \bibinfo{person}{Norbert Fuhr}, \bibinfo{person}{Maria Maistro}, \bibinfo{person}{Tetsuya Sakai}, \bibinfo{person}{Philipp Schaer}, {and} \bibinfo{person}{Ian Soboroff}.} \bibinfo{year}{2020}\natexlab{b}.
\newblock \showarticletitle{How to Measure the Reproducibility of System-oriented {IR} Experiments}. In \bibinfo{booktitle}{\emph{Proceedings of the 43rd International {ACM} {SIGIR} conference on research and development in Information Retrieval, {SIGIR} 2020, Virtual Event, China, July 25-30, 2020}}, \bibfield{editor}{\bibinfo{person}{Jimmy~X. Huang}, \bibinfo{person}{Yi~Chang}, \bibinfo{person}{Xueqi Cheng}, \bibinfo{person}{Jaap Kamps}, \bibinfo{person}{Vanessa Murdock}, \bibinfo{person}{Ji{-}Rong Wen}, {and} \bibinfo{person}{Yiqun Liu}} (Eds.). \bibinfo{publisher}{{ACM}}, \bibinfo{pages}{349--358}.
\newblock
\href{https://doi.org/10.1145/3397271.3401036}{doi:\nolinkurl{10.1145/3397271.3401036}}


\bibitem[Chen et~al\mbox{.}(2023)]%
        {zz-SIGIR2023}
\bibfield{editor}{\bibinfo{person}{H.-H. Chen}, \bibinfo{person}{W.-J. Duh}, \bibinfo{person}{H.-H. Huang}, \bibinfo{person}{M.~P. Kato}, \bibinfo{person}{J. Mothe}, {and} \bibinfo{person}{B. Poblete}} (Eds.). \bibinfo{year}{2023}\natexlab{}.
\newblock \bibinfo{booktitle}{\emph{{Proc. 46th Annual International ACM SIGIR Conference on Research and Development in Information Retrieval (SIGIR 2023)}}}. \bibinfo{publisher}{{ACM Press, New York, USA}}.
\newblock


\bibitem[Choi et~al\mbox{.}(2023)]%
        {ChoiEtAl2023}
\bibfield{author}{\bibinfo{person}{Jeongwhan Choi}, \bibinfo{person}{Seoyoung Hong}, \bibinfo{person}{Noseong Park}, {and} \bibinfo{person}{Sung{-}Bae Cho}.} \bibinfo{year}{2023}\natexlab{}.
\newblock \showarticletitle{Blurring-Sharpening Process Models for Collaborative Filtering}. In \bibinfo{booktitle}{\emph{Proceedings of the 46th International {ACM} {SIGIR} Conference on Research and Development in Information Retrieval, {SIGIR} 2023, Taipei, Taiwan, July 23-27, 2023}}, \bibfield{editor}{\bibinfo{person}{Hsin{-}Hsi Chen}, \bibinfo{person}{Wei{-}Jou~(Edward) Duh}, \bibinfo{person}{Hen{-}Hsen Huang}, \bibinfo{person}{Makoto~P. Kato}, \bibinfo{person}{Josiane Mothe}, {and} \bibinfo{person}{Barbara Poblete}} (Eds.). \bibinfo{publisher}{{ACM}}, \bibinfo{pages}{1096--1106}.
\newblock
\href{https://doi.org/10.1145/3539618.3591645}{doi:\nolinkurl{10.1145/3539618.3591645}}


\bibitem[Cockburn et~al\mbox{.}(2020)]%
        {10.1145/3360311}
\bibfield{author}{\bibinfo{person}{Andy Cockburn}, \bibinfo{person}{Pierre Dragicevic}, \bibinfo{person}{Lonni Besan\c{c}on}, {and} \bibinfo{person}{Carl Gutwin}.} \bibinfo{year}{2020}\natexlab{}.
\newblock \showarticletitle{Threats of a Replication Crisis in Empirical Computer Science}.
\newblock \bibinfo{journal}{\emph{Commun. ACM}} \bibinfo{volume}{63}, \bibinfo{number}{8} (\bibinfo{date}{jul} \bibinfo{year}{2020}), \bibinfo{pages}{70–79}.
\newblock
\showISSN{0001-0782}
\href{https://doi.org/10.1145/3360311}{doi:\nolinkurl{10.1145/3360311}}


\bibitem[Craswell et~al\mbox{.}(2021)]%
        {DBLP:conf/sigir/CraswellMYCVS21}
\bibfield{author}{\bibinfo{person}{Nick Craswell}, \bibinfo{person}{Bhaskar Mitra}, \bibinfo{person}{Emine Yilmaz}, \bibinfo{person}{Daniel Campos}, \bibinfo{person}{Ellen~M. Voorhees}, {and} \bibinfo{person}{Ian Soboroff}.} \bibinfo{year}{2021}\natexlab{}.
\newblock \showarticletitle{{TREC} Deep Learning Track: Reusable Test Collections in the Large Data Regime}. In \bibinfo{booktitle}{\emph{{SIGIR} '21: The 44th International {ACM} {SIGIR} Conference on Research and Development in Information Retrieval, Virtual Event, Canada, July 11-15, 2021}}, \bibfield{editor}{\bibinfo{person}{Fernando Diaz}, \bibinfo{person}{Chirag Shah}, \bibinfo{person}{Torsten Suel}, \bibinfo{person}{Pablo Castells}, \bibinfo{person}{Rosie Jones}, {and} \bibinfo{person}{Tetsuya Sakai}} (Eds.). \bibinfo{publisher}{{ACM}}, \bibinfo{pages}{2369--2375}.
\newblock
\href{https://doi.org/10.1145/3404835.3463249}{doi:\nolinkurl{10.1145/3404835.3463249}}


\bibitem[Cremonesi et~al\mbox{.}(2010)]%
        {DBLP:conf/recsys/CremonesiKT10}
\bibfield{author}{\bibinfo{person}{Paolo Cremonesi}, \bibinfo{person}{Yehuda Koren}, {and} \bibinfo{person}{Roberto Turrin}.} \bibinfo{year}{2010}\natexlab{}.
\newblock \showarticletitle{Performance of recommender algorithms on top-n recommendation tasks}. In \bibinfo{booktitle}{\emph{Proceedings of the 2010 {ACM} Conference on Recommender Systems, RecSys 2010, Barcelona, Spain, September 26-30, 2010}}, \bibfield{editor}{\bibinfo{person}{Xavier Amatriain}, \bibinfo{person}{Marc Torrens}, \bibinfo{person}{Paul Resnick}, {and} \bibinfo{person}{Markus Zanker}} (Eds.). \bibinfo{publisher}{{ACM}}, \bibinfo{pages}{39--46}.
\newblock
\href{https://doi.org/10.1145/1864708.1864721}{doi:\nolinkurl{10.1145/1864708.1864721}}


\bibitem[{De Roure}(2014)]%
        {DeRoure2014}
\bibfield{author}{\bibinfo{person}{D. {De Roure}}.} \bibinfo{year}{2014}\natexlab{}.
\newblock \showarticletitle{{The future of scholarly communications}}.
\newblock \bibinfo{journal}{\emph{Insights}} \bibinfo{volume}{27}, \bibinfo{number}{3} (\bibinfo{date}{November} \bibinfo{year}{2014}), \bibinfo{pages}{233--238}.
\newblock


\bibitem[Du et~al\mbox{.}(2022)]%
        {DBLP:conf/sigir/0002ZCZG22}
\bibfield{author}{\bibinfo{person}{Yuntao Du}, \bibinfo{person}{Xinjun Zhu}, \bibinfo{person}{Lu Chen}, \bibinfo{person}{Baihua Zheng}, {and} \bibinfo{person}{Yunjun Gao}.} \bibinfo{year}{2022}\natexlab{}.
\newblock \showarticletitle{{HAKG:} Hierarchy-Aware Knowledge Gated Network for Recommendation}. In \bibinfo{booktitle}{\emph{{SIGIR} '22: The 45th International {ACM} {SIGIR} Conference on Research and Development in Information Retrieval, Madrid, Spain, July 11 - 15, 2022}}, \bibfield{editor}{\bibinfo{person}{Enrique Amig{\'{o}}}, \bibinfo{person}{Pablo Castells}, \bibinfo{person}{Julio Gonzalo}, \bibinfo{person}{Ben Carterette}, \bibinfo{person}{J.~Shane Culpepper}, {and} \bibinfo{person}{Gabriella Kazai}} (Eds.). \bibinfo{publisher}{{ACM}}, \bibinfo{pages}{1390--1400}.
\newblock
\href{https://doi.org/10.1145/3477495.3531987}{doi:\nolinkurl{10.1145/3477495.3531987}}


\bibitem[Fan et~al\mbox{.}(2022)]%
        {DBLP:conf/sigir/FanL0ZT022}
\bibfield{author}{\bibinfo{person}{Wenqi Fan}, \bibinfo{person}{Xiaorui Liu}, \bibinfo{person}{Wei Jin}, \bibinfo{person}{Xiangyu Zhao}, \bibinfo{person}{Jiliang Tang}, {and} \bibinfo{person}{Qing Li}.} \bibinfo{year}{2022}\natexlab{}.
\newblock \showarticletitle{Graph Trend Filtering Networks for Recommendation}. In \bibinfo{booktitle}{\emph{{SIGIR} '22: The 45th International {ACM} {SIGIR} Conference on Research and Development in Information Retrieval, Madrid, Spain, July 11 - 15, 2022}}, \bibfield{editor}{\bibinfo{person}{Enrique Amig{\'{o}}}, \bibinfo{person}{Pablo Castells}, \bibinfo{person}{Julio Gonzalo}, \bibinfo{person}{Ben Carterette}, \bibinfo{person}{J.~Shane Culpepper}, {and} \bibinfo{person}{Gabriella Kazai}} (Eds.). \bibinfo{publisher}{{ACM}}, \bibinfo{pages}{112--121}.
\newblock
\href{https://doi.org/10.1145/3477495.3531985}{doi:\nolinkurl{10.1145/3477495.3531985}}


\bibitem[{Ferrari Dacrema} et~al\mbox{.}(2021)]%
        {DBLP:journals/tois/DacremaBCJ21}
\bibfield{author}{\bibinfo{person}{Maurizio {Ferrari Dacrema}}, \bibinfo{person}{Simone Boglio}, \bibinfo{person}{Paolo Cremonesi}, {and} \bibinfo{person}{Dietmar Jannach}.} \bibinfo{year}{2021}\natexlab{}.
\newblock \showarticletitle{A Troubling Analysis of Reproducibility and Progress in Recommender Systems Research}.
\newblock \bibinfo{journal}{\emph{{ACM} Trans. Inf. Syst.}} \bibinfo{volume}{39}, \bibinfo{number}{2} (\bibinfo{year}{2021}), \bibinfo{pages}{20:1--20:49}.
\newblock
\href{https://doi.org/10.1145/3434185}{doi:\nolinkurl{10.1145/3434185}}


\bibitem[{Ferrari Dacrema} et~al\mbox{.}(2019)]%
        {DBLP:conf/recsys/DacremaCJ19}
\bibfield{author}{\bibinfo{person}{Maurizio {Ferrari Dacrema}}, \bibinfo{person}{Paolo Cremonesi}, {and} \bibinfo{person}{Dietmar Jannach}.} \bibinfo{year}{2019}\natexlab{}.
\newblock \showarticletitle{Are we really making much progress? {A} worrying analysis of recent neural recommendation approaches}. In \bibinfo{booktitle}{\emph{Proceedings of the 13th {ACM} Conference on Recommender Systems, RecSys 2019, Copenhagen, Denmark, September 16-20, 2019}}, \bibfield{editor}{\bibinfo{person}{Toine Bogers}, \bibinfo{person}{Alan Said}, \bibinfo{person}{Peter Brusilovsky}, {and} \bibinfo{person}{Domonkos Tikk}} (Eds.). \bibinfo{publisher}{{ACM}}, \bibinfo{pages}{101--109}.
\newblock
\href{https://doi.org/10.1145/3298689.3347058}{doi:\nolinkurl{10.1145/3298689.3347058}}


\bibitem[{Ferrari Dacrema} et~al\mbox{.}(2020)]%
        {DBLP:conf/cikm/DacremaPCJ20}
\bibfield{author}{\bibinfo{person}{Maurizio {Ferrari Dacrema}}, \bibinfo{person}{Federico Parroni}, \bibinfo{person}{Paolo Cremonesi}, {and} \bibinfo{person}{Dietmar Jannach}.} \bibinfo{year}{2020}\natexlab{}.
\newblock \showarticletitle{Critically Examining the Claimed Value of Convolutions over User-Item Embedding Maps for Recommender Systems}. In \bibinfo{booktitle}{\emph{{CIKM} '20: The 29th {ACM} International Conference on Information and Knowledge Management, Virtual Event, Ireland, October 19-23, 2020}}, \bibfield{editor}{\bibinfo{person}{Mathieu d'Aquin}, \bibinfo{person}{Stefan Dietze}, \bibinfo{person}{Claudia Hauff}, \bibinfo{person}{Edward Curry}, {and} \bibinfo{person}{Philippe Cudr{\'{e}}{-}Mauroux}} (Eds.). \bibinfo{publisher}{{ACM}}, \bibinfo{pages}{355--363}.
\newblock
\href{https://doi.org/10.1145/3340531.3411901}{doi:\nolinkurl{10.1145/3340531.3411901}}


\bibitem[Ferro(2017)]%
        {Ferro2016d}
\bibfield{author}{\bibinfo{person}{N. Ferro}.} \bibinfo{year}{2017}\natexlab{}.
\newblock \showarticletitle{{Reproducibility Challenges in Information Retrieval Evaluation}}.
\newblock \bibinfo{journal}{\emph{{ACM Journal of Data and Information Quality (JDIQ)}}} \bibinfo{volume}{8}, \bibinfo{number}{2} (\bibinfo{date}{February} \bibinfo{year}{2017}), \bibinfo{pages}{8:1--8:4}.
\newblock


\bibitem[Freire et~al\mbox{.}(2016)]%
        {zz-DagstuhlSeminar16041}
\bibfield{editor}{\bibinfo{person}{J. Freire}, \bibinfo{person}{N. Fuhr}, {and} \bibinfo{person}{A. Rauber}} (Eds.). \bibinfo{year}{2016}\natexlab{}.
\newblock \bibinfo{booktitle}{\emph{{Report from Dagstuhl Seminar 16041: Reproducibility of Data-Oriented Experiments in e-Science}}}. \bibinfo{publisher}{{Schloss Dagstuhl--Leibniz-Zentrum f\"{u}r Informatik, Germany}}.
\newblock


\bibitem[Gao et~al\mbox{.}(2022)]%
        {DBLP:conf/sigir/Gao0HCZFZ22}
\bibfield{author}{\bibinfo{person}{Yunjun Gao}, \bibinfo{person}{Yuntao Du}, \bibinfo{person}{Yujia Hu}, \bibinfo{person}{Lu Chen}, \bibinfo{person}{Xinjun Zhu}, \bibinfo{person}{Ziquan Fang}, {and} \bibinfo{person}{Baihua Zheng}.} \bibinfo{year}{2022}\natexlab{}.
\newblock \showarticletitle{Self-Guided Learning to Denoise for Robust Recommendation}. In \bibinfo{booktitle}{\emph{{SIGIR} '22: The 45th International {ACM} {SIGIR} Conference on Research and Development in Information Retrieval, Madrid, Spain, July 11 - 15, 2022}}, \bibfield{editor}{\bibinfo{person}{Enrique Amig{\'{o}}}, \bibinfo{person}{Pablo Castells}, \bibinfo{person}{Julio Gonzalo}, \bibinfo{person}{Ben Carterette}, \bibinfo{person}{J.~Shane Culpepper}, {and} \bibinfo{person}{Gabriella Kazai}} (Eds.). \bibinfo{publisher}{{ACM}}, \bibinfo{pages}{1412--1422}.
\newblock
\href{https://doi.org/10.1145/3477495.3532059}{doi:\nolinkurl{10.1145/3477495.3532059}}


\bibitem[Gibney(2020)]%
        {Gibney2020}
\bibfield{author}{\bibinfo{person}{E. Gibney}.} \bibinfo{year}{2020}\natexlab{}.
\newblock \showarticletitle{{This AI researcher is trying to ward off a reproducibility crisis}}.
\newblock \bibinfo{journal}{\emph{{Nature}}}  \bibinfo{volume}{577} (\bibinfo{date}{January} \bibinfo{year}{2020}), \bibinfo{pages}{14}.
\newblock


\bibitem[He et~al\mbox{.}(2023)]%
        {HeEtAl2023}
\bibfield{author}{\bibinfo{person}{Wei He}, \bibinfo{person}{Guohao Sun}, \bibinfo{person}{Jinhu Lu}, {and} \bibinfo{person}{Xiu~Susie Fang}.} \bibinfo{year}{2023}\natexlab{}.
\newblock \showarticletitle{Candidate-aware Graph Contrastive Learning for Recommendation}. In \bibinfo{booktitle}{\emph{Proceedings of the 46th International {ACM} {SIGIR} Conference on Research and Development in Information Retrieval, {SIGIR} 2023, Taipei, Taiwan, July 23-27, 2023}}, \bibfield{editor}{\bibinfo{person}{Hsin{-}Hsi Chen}, \bibinfo{person}{Wei{-}Jou~(Edward) Duh}, \bibinfo{person}{Hen{-}Hsen Huang}, \bibinfo{person}{Makoto~P. Kato}, \bibinfo{person}{Josiane Mothe}, {and} \bibinfo{person}{Barbara Poblete}} (Eds.). \bibinfo{publisher}{{ACM}}, \bibinfo{pages}{1670--1679}.
\newblock
\href{https://doi.org/10.1145/3539618.3591647}{doi:\nolinkurl{10.1145/3539618.3591647}}


\bibitem[He et~al\mbox{.}(2020)]%
        {DBLP:conf/sigir/0001DWLZ020}
\bibfield{author}{\bibinfo{person}{Xiangnan He}, \bibinfo{person}{Kuan Deng}, \bibinfo{person}{Xiang Wang}, \bibinfo{person}{Yan Li}, \bibinfo{person}{Yong{-}Dong Zhang}, {and} \bibinfo{person}{Meng Wang}.} \bibinfo{year}{2020}\natexlab{}.
\newblock \showarticletitle{LightGCN: Simplifying and Powering Graph Convolution Network for Recommendation}. In \bibinfo{booktitle}{\emph{Proceedings of the 43rd International {ACM} {SIGIR} conference on research and development in Information Retrieval, {SIGIR} 2020, Virtual Event, China, July 25-30, 2020}}, \bibfield{editor}{\bibinfo{person}{Jimmy~X. Huang}, \bibinfo{person}{Yi~Chang}, \bibinfo{person}{Xueqi Cheng}, \bibinfo{person}{Jaap Kamps}, \bibinfo{person}{Vanessa Murdock}, \bibinfo{person}{Ji{-}Rong Wen}, {and} \bibinfo{person}{Yiqun Liu}} (Eds.). \bibinfo{publisher}{{ACM}}, \bibinfo{pages}{639--648}.
\newblock
\href{https://doi.org/10.1145/3397271.3401063}{doi:\nolinkurl{10.1145/3397271.3401063}}


\bibitem[Hern{\'{a}}ndez{-}Lobato et~al\mbox{.}(2014)]%
        {DBLP:conf/nips/Hernandez-LobatoHG14}
\bibfield{author}{\bibinfo{person}{Jos{\'{e}}~Miguel Hern{\'{a}}ndez{-}Lobato}, \bibinfo{person}{Matthew~W. Hoffman}, {and} \bibinfo{person}{Zoubin Ghahramani}.} \bibinfo{year}{2014}\natexlab{}.
\newblock \showarticletitle{Predictive Entropy Search for Efficient Global Optimization of Black-box Functions}. In \bibinfo{booktitle}{\emph{Advances in Neural Information Processing Systems 27: Annual Conference on Neural Information Processing Systems 2014, December 8-13 2014, Montreal, Quebec, Canada}}, \bibfield{editor}{\bibinfo{person}{Zoubin Ghahramani}, \bibinfo{person}{Max Welling}, \bibinfo{person}{Corinna Cortes}, \bibinfo{person}{Neil~D. Lawrence}, {and} \bibinfo{person}{Kilian~Q. Weinberger}} (Eds.). \bibinfo{pages}{918--926}.
\newblock
\urldef\tempurl%
\url{https://proceedings.neurips.cc/paper/2014/hash/069d3bb002acd8d7dd095917f9efe4cb-Abstract.html}
\showURL{%
\tempurl}


\bibitem[Hidasi and Czapp(2023)]%
        {DBLP:conf/recsys/HidasiC23}
\bibfield{author}{\bibinfo{person}{Bal{\'{a}}zs Hidasi} {and} \bibinfo{person}{{\'{A}}d{\'{a}}m~Tibor Czapp}.} \bibinfo{year}{2023}\natexlab{}.
\newblock \showarticletitle{The Effect of Third Party Implementations on Reproducibility}. In \bibinfo{booktitle}{\emph{Proceedings of the 17th {ACM} Conference on Recommender Systems, RecSys 2023, Singapore, Singapore, September 18-22, 2023}}, \bibfield{editor}{\bibinfo{person}{Jie Zhang}, \bibinfo{person}{Li~Chen}, \bibinfo{person}{Shlomo Berkovsky}, \bibinfo{person}{Min Zhang}, \bibinfo{person}{Tommaso~Di Noia}, \bibinfo{person}{Justin Basilico}, \bibinfo{person}{Luiz Pizzato}, {and} \bibinfo{person}{Yang Song}} (Eds.). \bibinfo{publisher}{{ACM}}, \bibinfo{pages}{272--282}.
\newblock
\href{https://doi.org/10.1145/3604915.3609487}{doi:\nolinkurl{10.1145/3604915.3609487}}


\bibitem[Hu et~al\mbox{.}(2008)]%
        {DBLP:conf/icdm/HuKV08}
\bibfield{author}{\bibinfo{person}{Yifan Hu}, \bibinfo{person}{Yehuda Koren}, {and} \bibinfo{person}{Chris Volinsky}.} \bibinfo{year}{2008}\natexlab{}.
\newblock \showarticletitle{Collaborative Filtering for Implicit Feedback Datasets}. In \bibinfo{booktitle}{\emph{Proceedings of the 8th {IEEE} International Conference on Data Mining {(ICDM} 2008), December 15-19, 2008, Pisa, Italy}}. \bibinfo{publisher}{{IEEE} Computer Society}, \bibinfo{pages}{263--272}.
\newblock
\href{https://doi.org/10.1109/ICDM.2008.22}{doi:\nolinkurl{10.1109/ICDM.2008.22}}


\bibitem[{Joint Committee for Guides in Metrology (JCGM)}(2008)]%
        {VIM2008}
\bibfield{author}{\bibinfo{person}{{Joint Committee for Guides in Metrology (JCGM)}}.} \bibinfo{year}{2008}\natexlab{}.
\newblock \bibinfo{booktitle}{\emph{{International vocabulary of metrology -- Basic and general concepts and associated terms (VIM)}} (\bibinfo{edition}{3rd} ed.)}.
\newblock \bibinfo{publisher}{{JCGM 200:2012}}.
\newblock


\bibitem[Kharazmi et~al\mbox{.}(2016)]%
        {KharazmiEtAl2016}
\bibfield{author}{\bibinfo{person}{S. Kharazmi}, \bibinfo{person}{F. Scholer}, \bibinfo{person}{D. Vallet}, {and} \bibinfo{person}{M. Sanderson}.} \bibinfo{year}{2016}\natexlab{}.
\newblock \showarticletitle{{Examining Additivity and Weak Baselines}}.
\newblock \bibinfo{journal}{\emph{{ACM Transactions on Information Systems (TOIS)}}} \bibinfo{volume}{34}, \bibinfo{number}{4} (\bibinfo{date}{June} \bibinfo{year}{2016}), \bibinfo{pages}{23:1--23:18}.
\newblock


\bibitem[Knyazev(2001)]%
        {DBLP:journals/siamsc/Knyazev01}
\bibfield{author}{\bibinfo{person}{Andrew~V. Knyazev}.} \bibinfo{year}{2001}\natexlab{}.
\newblock \showarticletitle{Toward the Optimal Preconditioned Eigensolver: Locally Optimal Block Preconditioned Conjugate Gradient Method}.
\newblock \bibinfo{journal}{\emph{{SIAM} J. Sci. Comput.}} \bibinfo{volume}{23}, \bibinfo{number}{2} (\bibinfo{year}{2001}), \bibinfo{pages}{517--541}.
\newblock
\href{https://doi.org/10.1137/S1064827500366124}{doi:\nolinkurl{10.1137/S1064827500366124}}


\bibitem[Krichene and Rendle(2020)]%
        {DBLP:conf/kdd/KricheneR20}
\bibfield{author}{\bibinfo{person}{Walid Krichene} {and} \bibinfo{person}{Steffen Rendle}.} \bibinfo{year}{2020}\natexlab{}.
\newblock \showarticletitle{On Sampled Metrics for Item Recommendation}. In \bibinfo{booktitle}{\emph{{KDD} '20: The 26th {ACM} {SIGKDD} Conference on Knowledge Discovery and Data Mining, Virtual Event, CA, USA, August 23-27, 2020}}, \bibfield{editor}{\bibinfo{person}{Rajesh Gupta}, \bibinfo{person}{Yan Liu}, \bibinfo{person}{Jiliang Tang}, {and} \bibinfo{person}{B.~Aditya Prakash}} (Eds.). \bibinfo{publisher}{{ACM}}, \bibinfo{pages}{1748--1757}.
\newblock
\href{https://doi.org/10.1145/3394486.3403226}{doi:\nolinkurl{10.1145/3394486.3403226}}


\bibitem[Li et~al\mbox{.}(2023)]%
        {LiEtAl2023b}
\bibfield{author}{\bibinfo{person}{Chaoliu Li}, \bibinfo{person}{Lianghao Xia}, \bibinfo{person}{Xubin Ren}, \bibinfo{person}{Yaowen Ye}, \bibinfo{person}{Yong Xu}, {and} \bibinfo{person}{Chao Huang}.} \bibinfo{year}{2023}\natexlab{}.
\newblock \showarticletitle{Graph Transformer for Recommendation}. In \bibinfo{booktitle}{\emph{Proceedings of the 46th International {ACM} {SIGIR} Conference on Research and Development in Information Retrieval, {SIGIR} 2023, Taipei, Taiwan, July 23-27, 2023}}, \bibfield{editor}{\bibinfo{person}{Hsin{-}Hsi Chen}, \bibinfo{person}{Wei{-}Jou~(Edward) Duh}, \bibinfo{person}{Hen{-}Hsen Huang}, \bibinfo{person}{Makoto~P. Kato}, \bibinfo{person}{Josiane Mothe}, {and} \bibinfo{person}{Barbara Poblete}} (Eds.). \bibinfo{publisher}{{ACM}}, \bibinfo{pages}{1680--1689}.
\newblock
\href{https://doi.org/10.1145/3539618.3591723}{doi:\nolinkurl{10.1145/3539618.3591723}}


\bibitem[Liang et~al\mbox{.}(2018)]%
        {DBLP:conf/www/LiangKHJ18}
\bibfield{author}{\bibinfo{person}{Dawen Liang}, \bibinfo{person}{Rahul~G. Krishnan}, \bibinfo{person}{Matthew~D. Hoffman}, {and} \bibinfo{person}{Tony Jebara}.} \bibinfo{year}{2018}\natexlab{}.
\newblock \showarticletitle{Variational Autoencoders for Collaborative Filtering}. In \bibinfo{booktitle}{\emph{Proceedings of the 2018 World Wide Web Conference on World Wide Web, {WWW} 2018, Lyon, France, April 23-27, 2018}}, \bibfield{editor}{\bibinfo{person}{Pierre{-}Antoine Champin}, \bibinfo{person}{Fabien Gandon}, \bibinfo{person}{Mounia Lalmas}, {and} \bibinfo{person}{Panagiotis~G. Ipeirotis}} (Eds.). \bibinfo{publisher}{{ACM}}, \bibinfo{pages}{689--698}.
\newblock
\href{https://doi.org/10.1145/3178876.3186150}{doi:\nolinkurl{10.1145/3178876.3186150}}


\bibitem[Lin(2022)]%
        {Lin2022}
\bibfield{author}{\bibinfo{person}{J. Lin}.} \bibinfo{year}{2022}\natexlab{}.
\newblock \showarticletitle{{Building a Culture of Reproducibility in Academic Research}}.
\newblock \bibinfo{journal}{\emph{{arXiv.org, Information Retrieval (cs.IR)}}}  \bibinfo{volume}{arXiv:2212.13534} (\bibinfo{date}{December} \bibinfo{year}{2022}).
\newblock


\bibitem[Lin et~al\mbox{.}(2021)]%
        {DBLP:conf/sigir/LinCCMY21}
\bibfield{author}{\bibinfo{person}{Jimmy Lin}, \bibinfo{person}{Daniel Campos}, \bibinfo{person}{Nick Craswell}, \bibinfo{person}{Bhaskar Mitra}, {and} \bibinfo{person}{Emine Yilmaz}.} \bibinfo{year}{2021}\natexlab{}.
\newblock \showarticletitle{Significant Improvements over the State of the Art? {A} Case Study of the {MS} {MARCO} Document Ranking Leaderboard}. In \bibinfo{booktitle}{\emph{{SIGIR} '21: The 44th International {ACM} {SIGIR} Conference on Research and Development in Information Retrieval, Virtual Event, Canada, July 11-15, 2021}}, \bibfield{editor}{\bibinfo{person}{Fernando Diaz}, \bibinfo{person}{Chirag Shah}, \bibinfo{person}{Torsten Suel}, \bibinfo{person}{Pablo Castells}, \bibinfo{person}{Rosie Jones}, {and} \bibinfo{person}{Tetsuya Sakai}} (Eds.). \bibinfo{publisher}{{ACM}}, \bibinfo{pages}{2283--2287}.
\newblock
\href{https://doi.org/10.1145/3404835.3463034}{doi:\nolinkurl{10.1145/3404835.3463034}}


\bibitem[Lin and Zhang(2020)]%
        {DBLP:conf/ecir/LinZ20}
\bibfield{author}{\bibinfo{person}{Jimmy Lin} {and} \bibinfo{person}{Qian Zhang}.} \bibinfo{year}{2020}\natexlab{}.
\newblock \showarticletitle{Reproducibility is a Process, Not an Achievement: The Replicability of {IR} Reproducibility Experiments}. In \bibinfo{booktitle}{\emph{Advances in Information Retrieval - 42nd European Conference on {IR} Research, {ECIR} 2020, Lisbon, Portugal, April 14-17, 2020, Proceedings, Part {II}}} \emph{(\bibinfo{series}{Lecture Notes in Computer Science}, Vol.~\bibinfo{volume}{12036})}, \bibfield{editor}{\bibinfo{person}{Joemon~M. Jose}, \bibinfo{person}{Emine Yilmaz}, \bibinfo{person}{Jo{\~{a}}o Magalh{\~{a}}es}, \bibinfo{person}{Pablo Castells}, \bibinfo{person}{Nicola Ferro}, \bibinfo{person}{M{\'{a}}rio~J. Silva}, {and} \bibinfo{person}{Fl{\'{a}}vio Martins}} (Eds.). \bibinfo{publisher}{Springer}, \bibinfo{pages}{43--49}.
\newblock
\href{https://doi.org/10.1007/978-3-030-45442-5\_6}{doi:\nolinkurl{10.1007/978-3-030-45442-5\_6}}


\bibitem[Liu et~al\mbox{.}(2023)]%
        {LiuEtAl2023}
\bibfield{author}{\bibinfo{person}{Jiahao Liu}, \bibinfo{person}{Dongsheng Li}, \bibinfo{person}{Hansu Gu}, \bibinfo{person}{Tun Lu}, \bibinfo{person}{Peng Zhang}, \bibinfo{person}{Li Shang}, {and} \bibinfo{person}{Ning Gu}.} \bibinfo{year}{2023}\natexlab{}.
\newblock \showarticletitle{Triple Structural Information Modelling for Accurate, Explainable and Interactive Recommendation}. In \bibinfo{booktitle}{\emph{Proceedings of the 46th International {ACM} {SIGIR} Conference on Research and Development in Information Retrieval, {SIGIR} 2023, Taipei, Taiwan, July 23-27, 2023}}, \bibfield{editor}{\bibinfo{person}{Hsin{-}Hsi Chen}, \bibinfo{person}{Wei{-}Jou~(Edward) Duh}, \bibinfo{person}{Hen{-}Hsen Huang}, \bibinfo{person}{Makoto~P. Kato}, \bibinfo{person}{Josiane Mothe}, {and} \bibinfo{person}{Barbara Poblete}} (Eds.). \bibinfo{publisher}{{ACM}}, \bibinfo{pages}{1086--1095}.
\newblock
\href{https://doi.org/10.1145/3539618.3591779}{doi:\nolinkurl{10.1145/3539618.3591779}}


\bibitem[Liu et~al\mbox{.}(2022)]%
        {DBLP:conf/sigir/LiuWZS22}
\bibfield{author}{\bibinfo{person}{Xiaoming Liu}, \bibinfo{person}{Shaocong Wu}, \bibinfo{person}{Zhaohan Zhang}, {and} \bibinfo{person}{Chao Shen}.} \bibinfo{year}{2022}\natexlab{}.
\newblock \showarticletitle{Unify Local and Global Information for Top-N Recommendation}. In \bibinfo{booktitle}{\emph{{SIGIR} '22: The 45th International {ACM} {SIGIR} Conference on Research and Development in Information Retrieval, Madrid, Spain, July 11 - 15, 2022}}, \bibfield{editor}{\bibinfo{person}{Enrique Amig{\'{o}}}, \bibinfo{person}{Pablo Castells}, \bibinfo{person}{Julio Gonzalo}, \bibinfo{person}{Ben Carterette}, \bibinfo{person}{J.~Shane Culpepper}, {and} \bibinfo{person}{Gabriella Kazai}} (Eds.). \bibinfo{publisher}{{ACM}}, \bibinfo{pages}{1262--1272}.
\newblock
\href{https://doi.org/10.1145/3477495.3532070}{doi:\nolinkurl{10.1145/3477495.3532070}}


\bibitem[Lucic et~al\mbox{.}(2022)]%
        {LucicEtAl2022}
\bibfield{author}{\bibinfo{person}{A. Lucic}, \bibinfo{person}{M. Bleeker}, \bibinfo{person}{M. de Rijke}, \bibinfo{person}{K. Sinha}, \bibinfo{person}{S. Jullien}, {and} \bibinfo{person}{R. Stojnic}.} \bibinfo{year}{2022}\natexlab{}.
\newblock \showarticletitle{{Towards Reproducible Machine Learning Research in Information Retrieval}}, See \citeN{zz-SIGIR2022}, \bibinfo{pages}{3459--3461}.
\newblock


\bibitem[Lv et~al\mbox{.}(2021)]%
        {LvEtAl2021}
\bibfield{author}{\bibinfo{person}{Q. Lv}, \bibinfo{person}{M. Ding}, \bibinfo{person}{Q. Liu}, \bibinfo{person}{Y. Chen}, \bibinfo{person}{W. Feng}, \bibinfo{person}{S. He}, \bibinfo{person}{C. Zhou}, \bibinfo{person}{J. Jiang}, \bibinfo{person}{Y. Dong}, {and} \bibinfo{person}{J. Tang}.} \bibinfo{year}{2021}\natexlab{}.
\newblock \showarticletitle{{Are we really making much progress?: Revisiting, benchmarking and refining heterogeneous graph neural networks}}. In \bibinfo{booktitle}{\emph{{Proc. 27th ACM SIGKDD International Conference on Knowledge Discovery and Data Mining (KDD 2021)}}}, \bibfield{editor}{\bibinfo{person}{F.~Zhu}, \bibinfo{person}{B.~Chin~Ooi}, \bibinfo{person}{C.~Miao}, \bibinfo{person}{H.~Wang}, \bibinfo{person}{I.~Skrypnyk}, {and} \bibinfo{person}{W.~Hsu}} (Eds.). \bibinfo{publisher}{{ACM Press, New York, USA}}, \bibinfo{pages}{133--142}.
\newblock


\bibitem[MacAvaney et~al\mbox{.}(2021)]%
        {MacAvaneyEtAl2021}
\bibfield{author}{\bibinfo{person}{S. MacAvaney}, \bibinfo{person}{A. Yates}, \bibinfo{person}{S. Feldman}, \bibinfo{person}{D. Downey}, \bibinfo{person}{A. Cohan}, {and} \bibinfo{person}{N. Goharian}.} \bibinfo{year}{2021}\natexlab{}.
\newblock \showarticletitle{{Simplified Data Wrangling with ir\_datasets}}. In \bibinfo{booktitle}{\emph{{Proc. 44th Annual International ACM SIGIR Conference on Research and Development in Information Retrieval (SIGIR 2021)}}}, \bibfield{editor}{\bibinfo{person}{F.~Diaz}, \bibinfo{person}{C.~Shah}, \bibinfo{person}{T.~Suel}, \bibinfo{person}{P.~Castells}, \bibinfo{person}{R.~Jones}, \bibinfo{person}{T.~Sakai}, \bibinfo{person}{A.~Bellog{\'\i}n}, {and} \bibinfo{person}{M.~Yoshioka}} (Eds.). \bibinfo{publisher}{{ACM Press, New York, USA}}, \bibinfo{pages}{2429--2436}.
\newblock


\bibitem[Maistro et~al\mbox{.}(2023)]%
        {MaistroEtAl2023}
\bibfield{author}{\bibinfo{person}{M. Maistro}, \bibinfo{person}{T. Breuer}, \bibinfo{person}{P. Schaer}, {and} \bibinfo{person}{N. Ferro}.} \bibinfo{year}{2023}\natexlab{}.
\newblock \showarticletitle{{An in-depth Investigation on the Behavior of Measures to Quantify Reproducibility}}.
\newblock \bibinfo{journal}{\emph{{Information Processing \& Management}}} (\bibinfo{year}{2023}).
\newblock


\bibitem[{National Academies of Sciences, Engineering, and Medicine}(2019)]%
        {NAP2019}
\bibfield{author}{\bibinfo{person}{{National Academies of Sciences, Engineering, and Medicine}}.} \bibinfo{year}{2019}\natexlab{}.
\newblock \bibinfo{booktitle}{\emph{{Reproducibility and Replicability in Science}}}.
\newblock \bibinfo{publisher}{{The National Academies Press, Washington, USA}}.
\newblock


\bibitem[Ning and Karypis(2011)]%
        {DBLP:conf/icdm/NingK11}
\bibfield{author}{\bibinfo{person}{Xia Ning} {and} \bibinfo{person}{George Karypis}.} \bibinfo{year}{2011}\natexlab{}.
\newblock \showarticletitle{{SLIM:} Sparse Linear Methods for Top-N Recommender Systems}. In \bibinfo{booktitle}{\emph{11th {IEEE} International Conference on Data Mining, {ICDM} 2011, Vancouver, BC, Canada, December 11-14, 2011}}, \bibfield{editor}{\bibinfo{person}{Diane~J. Cook}, \bibinfo{person}{Jian Pei}, \bibinfo{person}{Wei Wang}, \bibinfo{person}{Osmar~R. Za{\"{\i}}ane}, {and} \bibinfo{person}{Xindong Wu}} (Eds.). \bibinfo{publisher}{{IEEE} Computer Society}, \bibinfo{pages}{497--506}.
\newblock
\href{https://doi.org/10.1109/ICDM.2011.134}{doi:\nolinkurl{10.1109/ICDM.2011.134}}


\bibitem[{NISO}(2021)]%
        {NISO-RP-31-2021}
\bibfield{author}{\bibinfo{person}{{NISO}}.} \bibinfo{year}{2021}\natexlab{}.
\newblock \bibinfo{booktitle}{\emph{{NISO RP-31-2021 -- Reproducibility Badging and Definitions}}}.
\newblock \bibinfo{publisher}{{National Information Standards Organization (NISO)}}.
\newblock


\bibitem[{Open Science Collaboration}(2015)]%
        {OSC2015}
\bibfield{author}{\bibinfo{person}{{Open Science Collaboration}}.} \bibinfo{year}{2015}\natexlab{}.
\newblock \showarticletitle{{Estimating the reproducibility of psychological science}}.
\newblock \bibinfo{journal}{\emph{{Science}}} \bibinfo{volume}{349}, \bibinfo{number}{6251} (\bibinfo{date}{August} \bibinfo{year}{2015}), \bibinfo{pages}{943--952}.
\newblock


\bibitem[Paudel et~al\mbox{.}(2017)]%
        {DBLP:journals/tiis/PaudelCNB17}
\bibfield{author}{\bibinfo{person}{Bibek Paudel}, \bibinfo{person}{Fabian Christoffel}, \bibinfo{person}{Chris Newell}, {and} \bibinfo{person}{Abraham Bernstein}.} \bibinfo{year}{2017}\natexlab{}.
\newblock \showarticletitle{Updatable, Accurate, Diverse, and Scalable Recommendations for Interactive Applications}.
\newblock \bibinfo{journal}{\emph{{ACM} Trans. Interact. Intell. Syst.}} \bibinfo{volume}{7}, \bibinfo{number}{1} (\bibinfo{year}{2017}), \bibinfo{pages}{1:1--1:34}.
\newblock
\href{https://doi.org/10.1145/2955101}{doi:\nolinkurl{10.1145/2955101}}


\bibitem[Peng et~al\mbox{.}(2022)]%
        {DBLP:conf/sigir/PengSM22}
\bibfield{author}{\bibinfo{person}{Shaowen Peng}, \bibinfo{person}{Kazunari Sugiyama}, {and} \bibinfo{person}{Tsunenori Mine}.} \bibinfo{year}{2022}\natexlab{}.
\newblock \showarticletitle{Less is More: Reweighting Important Spectral Graph Features for Recommendation}. In \bibinfo{booktitle}{\emph{{SIGIR} '22: The 45th International {ACM} {SIGIR} Conference on Research and Development in Information Retrieval, Madrid, Spain, July 11 - 15, 2022}}, \bibfield{editor}{\bibinfo{person}{Enrique Amig{\'{o}}}, \bibinfo{person}{Pablo Castells}, \bibinfo{person}{Julio Gonzalo}, \bibinfo{person}{Ben Carterette}, \bibinfo{person}{J.~Shane Culpepper}, {and} \bibinfo{person}{Gabriella Kazai}} (Eds.). \bibinfo{publisher}{{ACM}}, \bibinfo{pages}{1273--1282}.
\newblock
\href{https://doi.org/10.1145/3477495.3532014}{doi:\nolinkurl{10.1145/3477495.3532014}}


\bibitem[Pineau et~al\mbox{.}(2021)]%
        {PineauEtAl2021}
\bibfield{author}{\bibinfo{person}{J. Pineau}, \bibinfo{person}{P. Vincent-Lamarre}, \bibinfo{person}{k. Sinha}, \bibinfo{person}{V. Lariviere}, \bibinfo{person}{A. Beygelzimer}, \bibinfo{person}{F. d'Alche Buc}, \bibinfo{person}{E. Fox}, {and} \bibinfo{person}{H. Larochelle}.} \bibinfo{year}{2021}\natexlab{}.
\newblock \showarticletitle{{Improving Reproducibility in Machine Learning Research. A Report from the NeurIPS 2019 Reproducibility Program}}.
\newblock \bibinfo{journal}{\emph{{Journal of Machine Learning Research (JMLR)}}} \bibinfo{volume}{22}, \bibinfo{number}{164} (\bibinfo{date}{May} \bibinfo{year}{2021}), \bibinfo{pages}{1--20}.
\newblock


\bibitem[Plesser(2018)]%
        {Plesser2018}
\bibfield{author}{\bibinfo{person}{H.~E. Plesser}.} \bibinfo{year}{2018}\natexlab{}.
\newblock \showarticletitle{{Reproducibility vs. Replicability: A Brief History of a Confused Terminology}}.
\newblock \bibinfo{journal}{\emph{{Frontiers in Neuroinformatics}}}  \bibinfo{volume}{11} (\bibinfo{date}{January} \bibinfo{year}{2018}), \bibinfo{pages}{76:1--76:4}.
\newblock


\bibitem[Ren et~al\mbox{.}(2023a)]%
        {RenEtAl2023}
\bibfield{author}{\bibinfo{person}{Xubin Ren}, \bibinfo{person}{Lianghao Xia}, \bibinfo{person}{Jiashu Zhao}, \bibinfo{person}{Dawei Yin}, {and} \bibinfo{person}{Chao Huang}.} \bibinfo{year}{2023}\natexlab{a}.
\newblock \showarticletitle{Disentangled Contrastive Collaborative Filtering}. In \bibinfo{booktitle}{\emph{Proceedings of the 46th International {ACM} {SIGIR} Conference on Research and Development in Information Retrieval, {SIGIR} 2023, Taipei, Taiwan, July 23-27, 2023}}, \bibfield{editor}{\bibinfo{person}{Hsin{-}Hsi Chen}, \bibinfo{person}{Wei{-}Jou~(Edward) Duh}, \bibinfo{person}{Hen{-}Hsen Huang}, \bibinfo{person}{Makoto~P. Kato}, \bibinfo{person}{Josiane Mothe}, {and} \bibinfo{person}{Barbara Poblete}} (Eds.). \bibinfo{publisher}{{ACM}}, \bibinfo{pages}{1137--1146}.
\newblock
\href{https://doi.org/10.1145/3539618.3591665}{doi:\nolinkurl{10.1145/3539618.3591665}}


\bibitem[Ren et~al\mbox{.}(2023b)]%
        {RenEtAl2023b}
\bibfield{author}{\bibinfo{person}{Yuyang Ren}, \bibinfo{person}{Haonan Zhang}, \bibinfo{person}{Luoyi Fu}, \bibinfo{person}{Xinbing Wang}, {and} \bibinfo{person}{Chenghu Zhou}.} \bibinfo{year}{2023}\natexlab{b}.
\newblock \showarticletitle{Distillation-Enhanced Graph Masked Autoencoders for Bundle Recommendation}. In \bibinfo{booktitle}{\emph{Proceedings of the 46th International {ACM} {SIGIR} Conference on Research and Development in Information Retrieval, {SIGIR} 2023, Taipei, Taiwan, July 23-27, 2023}}, \bibfield{editor}{\bibinfo{person}{Hsin{-}Hsi Chen}, \bibinfo{person}{Wei{-}Jou~(Edward) Duh}, \bibinfo{person}{Hen{-}Hsen Huang}, \bibinfo{person}{Makoto~P. Kato}, \bibinfo{person}{Josiane Mothe}, {and} \bibinfo{person}{Barbara Poblete}} (Eds.). \bibinfo{publisher}{{ACM}}, \bibinfo{pages}{1660--1669}.
\newblock
\href{https://doi.org/10.1145/3539618.3591666}{doi:\nolinkurl{10.1145/3539618.3591666}}


\bibitem[Rendle et~al\mbox{.}(2009)]%
        {DBLP:conf/uai/RendleFGS09}
\bibfield{author}{\bibinfo{person}{Steffen Rendle}, \bibinfo{person}{Christoph Freudenthaler}, \bibinfo{person}{Zeno Gantner}, {and} \bibinfo{person}{Lars Schmidt{-}Thieme}.} \bibinfo{year}{2009}\natexlab{}.
\newblock \showarticletitle{{BPR:} Bayesian Personalized Ranking from Implicit Feedback}. In \bibinfo{booktitle}{\emph{{UAI} 2009, Proceedings of the Twenty-Fifth Conference on Uncertainty in Artificial Intelligence, Montreal, QC, Canada, June 18-21, 2009}}, \bibfield{editor}{\bibinfo{person}{Jeff~A. Bilmes} {and} \bibinfo{person}{Andrew~Y. Ng}} (Eds.). \bibinfo{publisher}{{AUAI} Press}, \bibinfo{pages}{452--461}.
\newblock
\urldef\tempurl%
\url{https://www.auai.org/uai2009/papers/UAI2009\_0139\_48141db02b9f0b02bc7158819ebfa2c7.pdf}
\showURL{%
\tempurl}


\bibitem[Rendle et~al\mbox{.}(2020)]%
        {DBLP:conf/recsys/RendleKZA20}
\bibfield{author}{\bibinfo{person}{Steffen Rendle}, \bibinfo{person}{Walid Krichene}, \bibinfo{person}{Li Zhang}, {and} \bibinfo{person}{John~R. Anderson}.} \bibinfo{year}{2020}\natexlab{}.
\newblock \showarticletitle{Neural Collaborative Filtering vs. Matrix Factorization Revisited}. In \bibinfo{booktitle}{\emph{RecSys 2020: Fourteenth {ACM} Conference on Recommender Systems, Virtual Event, Brazil, September 22-26, 2020}}, \bibfield{editor}{\bibinfo{person}{Rodrygo L.~T. Santos}, \bibinfo{person}{Leandro~Balby Marinho}, \bibinfo{person}{Elizabeth~M. Daly}, \bibinfo{person}{Li~Chen}, \bibinfo{person}{Kim Falk}, \bibinfo{person}{Noam Koenigstein}, {and} \bibinfo{person}{Edleno~Silva de~Moura}} (Eds.). \bibinfo{publisher}{{ACM}}, \bibinfo{pages}{240--248}.
\newblock
\href{https://doi.org/10.1145/3383313.3412488}{doi:\nolinkurl{10.1145/3383313.3412488}}


\bibitem[Resnick et~al\mbox{.}(1994)]%
        {DBLP:conf/cscw/ResnickISBR94}
\bibfield{author}{\bibinfo{person}{Paul Resnick}, \bibinfo{person}{Neophytos Iacovou}, \bibinfo{person}{Mitesh Suchak}, \bibinfo{person}{Peter Bergstrom}, {and} \bibinfo{person}{John Riedl}.} \bibinfo{year}{1994}\natexlab{}.
\newblock \showarticletitle{GroupLens: An Open Architecture for Collaborative Filtering of Netnews}. In \bibinfo{booktitle}{\emph{{CSCW} '94, Proceedings of the Conference on Computer Supported Cooperative Work, Chapel Hill, NC, USA, October 22-26, 1994}}, \bibfield{editor}{\bibinfo{person}{John~B. Smith}, \bibinfo{person}{F.~Donelson Smith}, {and} \bibinfo{person}{Thomas~W. Malone}} (Eds.). \bibinfo{publisher}{{ACM}}, \bibinfo{pages}{175--186}.
\newblock
\href{https://doi.org/10.1145/192844.192905}{doi:\nolinkurl{10.1145/192844.192905}}


\bibitem[Rossi et~al\mbox{.}(2021)]%
        {RossiEtAl2021}
\bibfield{author}{\bibinfo{person}{A. Rossi}, \bibinfo{person}{D. Barbosa}, \bibinfo{person}{D. Firmani}, \bibinfo{person}{A. Matinata}, {and} \bibinfo{person}{P. Merialdo}.} \bibinfo{year}{2021}\natexlab{}.
\newblock \showarticletitle{{Knowledge Graph Embedding for Link Prediction: A Comparative Analysis}}.
\newblock \bibinfo{journal}{\emph{{ACM Transactions on Knowledge Discovery from Data (TKDD)}}} \bibinfo{volume}{15}, \bibinfo{number}{2} (\bibinfo{date}{April} \bibinfo{year}{2021}), \bibinfo{pages}{14:1--14:49}.
\newblock


\bibitem[Sarwar et~al\mbox{.}(2001)]%
        {DBLP:conf/www/SarwarKKR01}
\bibfield{author}{\bibinfo{person}{Badrul~Munir Sarwar}, \bibinfo{person}{George Karypis}, \bibinfo{person}{Joseph~A. Konstan}, {and} \bibinfo{person}{John Riedl}.} \bibinfo{year}{2001}\natexlab{}.
\newblock \showarticletitle{Item-based collaborative filtering recommendation algorithms}. In \bibinfo{booktitle}{\emph{Proceedings of the Tenth International World Wide Web Conference, {WWW} 10, Hong Kong, China, May 1-5, 2001}}, \bibfield{editor}{\bibinfo{person}{Vincent~Y. Shen}, \bibinfo{person}{Nobuo Saito}, \bibinfo{person}{Michael~R. Lyu}, {and} \bibinfo{person}{Mary~Ellen Zurko}} (Eds.). \bibinfo{publisher}{{ACM}}, \bibinfo{pages}{285--295}.
\newblock
\href{https://doi.org/10.1145/371920.372071}{doi:\nolinkurl{10.1145/371920.372071}}


\bibitem[Shehzad et~al\mbox{.}(2025)]%
        {DBLP:conf/sigir/ShehzadDJ25}
\bibfield{author}{\bibinfo{person}{Faisal Shehzad}, \bibinfo{person}{Maurizio {Ferrari Dacrema}}, {and} \bibinfo{person}{Dietmar Jannach}.} \bibinfo{year}{2025}\natexlab{}.
\newblock \showarticletitle{A Worrying Reproducibility Study of Intent-Aware Recommendation Models}. In \bibinfo{booktitle}{\emph{Proceedings of the 48th International {ACM} {SIGIR} Conference on Research and Development in Information Retrieval, {SIGIR} 2025, Padua, Italy, July 13-18, 2025}}, \bibfield{editor}{\bibinfo{person}{Nicola Ferro}, \bibinfo{person}{Maria Maistro}, \bibinfo{person}{Gabriella Pasi}, \bibinfo{person}{Omar Alonso}, \bibinfo{person}{Andrew Trotman}, {and} \bibinfo{person}{Suzan Verberne}} (Eds.). \bibinfo{publisher}{{ACM}}, \bibinfo{pages}{3155--3164}.
\newblock
\href{https://doi.org/10.1145/3726302.3730307}{doi:\nolinkurl{10.1145/3726302.3730307}}


\bibitem[Shehzad and Jannach(2023)]%
        {DBLP:conf/recsys/ShehzadJ23}
\bibfield{author}{\bibinfo{person}{Faisal Shehzad} {and} \bibinfo{person}{Dietmar Jannach}.} \bibinfo{year}{2023}\natexlab{}.
\newblock \showarticletitle{Everyone's a Winner! On Hyperparameter Tuning of Recommendation Models}. In \bibinfo{booktitle}{\emph{Proceedings of the 17th {ACM} Conference on Recommender Systems, RecSys 2023, Singapore, Singapore, September 18-22, 2023}}, \bibfield{editor}{\bibinfo{person}{Jie Zhang}, \bibinfo{person}{Li~Chen}, \bibinfo{person}{Shlomo Berkovsky}, \bibinfo{person}{Min Zhang}, \bibinfo{person}{Tommaso~Di Noia}, \bibinfo{person}{Justin Basilico}, \bibinfo{person}{Luiz Pizzato}, {and} \bibinfo{person}{Yang Song}} (Eds.). \bibinfo{publisher}{{ACM}}, \bibinfo{pages}{652--657}.
\newblock
\href{https://doi.org/10.1145/3604915.3609488}{doi:\nolinkurl{10.1145/3604915.3609488}}


\bibitem[Shen et~al\mbox{.}(2021)]%
        {DBLP:conf/cikm/ShenWZSZLL21}
\bibfield{author}{\bibinfo{person}{Yifei Shen}, \bibinfo{person}{Yongji Wu}, \bibinfo{person}{Yao Zhang}, \bibinfo{person}{Caihua Shan}, \bibinfo{person}{Jun Zhang}, \bibinfo{person}{Khaled~B. Letaief}, {and} \bibinfo{person}{Dongsheng Li}.} \bibinfo{year}{2021}\natexlab{}.
\newblock \showarticletitle{How Powerful is Graph Convolution for Recommendation?}. In \bibinfo{booktitle}{\emph{{CIKM} '21: The 30th {ACM} International Conference on Information and Knowledge Management, Virtual Event, Queensland, Australia, November 1 - 5, 2021}}, \bibfield{editor}{\bibinfo{person}{Gianluca Demartini}, \bibinfo{person}{Guido Zuccon}, \bibinfo{person}{J.~Shane Culpepper}, \bibinfo{person}{Zi~Huang}, {and} \bibinfo{person}{Hanghang Tong}} (Eds.). \bibinfo{publisher}{{ACM}}, \bibinfo{pages}{1619--1629}.
\newblock
\href{https://doi.org/10.1145/3459637.3482264}{doi:\nolinkurl{10.1145/3459637.3482264}}


\bibitem[Steck(2019)]%
        {DBLP:conf/www/Steck19}
\bibfield{author}{\bibinfo{person}{Harald Steck}.} \bibinfo{year}{2019}\natexlab{}.
\newblock \showarticletitle{Embarrassingly Shallow Autoencoders for Sparse Data}. In \bibinfo{booktitle}{\emph{The World Wide Web Conference, {WWW} 2019, San Francisco, CA, USA, May 13-17, 2019}}, \bibfield{editor}{\bibinfo{person}{Ling Liu}, \bibinfo{person}{Ryen~W. White}, \bibinfo{person}{Amin Mantrach}, \bibinfo{person}{Fabrizio Silvestri}, \bibinfo{person}{Julian~J. McAuley}, \bibinfo{person}{Ricardo Baeza{-}Yates}, {and} \bibinfo{person}{Leila Zia}} (Eds.). \bibinfo{publisher}{{ACM}}, \bibinfo{pages}{3251--3257}.
\newblock
\href{https://doi.org/10.1145/3308558.3313710}{doi:\nolinkurl{10.1145/3308558.3313710}}


\bibitem[Steck et~al\mbox{.}(2021)]%
        {DBLP:journals/aim/SteckBELRB21}
\bibfield{author}{\bibinfo{person}{Harald Steck}, \bibinfo{person}{Linas Baltrunas}, \bibinfo{person}{Ehtsham Elahi}, \bibinfo{person}{Dawen Liang}, \bibinfo{person}{Yves Raimond}, {and} \bibinfo{person}{Justin Basilico}.} \bibinfo{year}{2021}\natexlab{}.
\newblock \showarticletitle{Deep Learning for Recommender Systems: {A} Netflix Case Study}.
\newblock \bibinfo{journal}{\emph{{AI} Mag.}} \bibinfo{volume}{42}, \bibinfo{number}{3} (\bibinfo{year}{2021}), \bibinfo{pages}{7--18}.
\newblock
\href{https://doi.org/10.1609/aimag.v42i3.18140}{doi:\nolinkurl{10.1609/aimag.v42i3.18140}}


\bibitem[Steck and Liang(2021)]%
        {DBLP:conf/recsys/SteckL21}
\bibfield{author}{\bibinfo{person}{Harald Steck} {and} \bibinfo{person}{Dawen Liang}.} \bibinfo{year}{2021}\natexlab{}.
\newblock \showarticletitle{Negative Interactions for Improved Collaborative Filtering: Don't go Deeper, go Higher}. In \bibinfo{booktitle}{\emph{RecSys '21: Fifteenth {ACM} Conference on Recommender Systems, Amsterdam, The Netherlands, 27 September 2021 - 1 October 2021}}, \bibfield{editor}{\bibinfo{person}{Humberto Jes{\'{u}}s~Corona Pamp{\'{\i}}n}, \bibinfo{person}{Martha~A. Larson}, \bibinfo{person}{Martijn~C. Willemsen}, \bibinfo{person}{Joseph~A. Konstan}, \bibinfo{person}{Julian~J. McAuley}, \bibinfo{person}{Jean Garcia{-}Gathright}, \bibinfo{person}{Bouke Huurnink}, {and} \bibinfo{person}{Even Oldridge}} (Eds.). \bibinfo{publisher}{{ACM}}, \bibinfo{pages}{34--43}.
\newblock
\href{https://doi.org/10.1145/3460231.3474273}{doi:\nolinkurl{10.1145/3460231.3474273}}


\bibitem[Stodden et~al\mbox{.}(2018)]%
        {stodden2018empirical}
\bibfield{author}{\bibinfo{person}{Victoria Stodden}, \bibinfo{person}{Jennifer Seiler}, {and} \bibinfo{person}{Zhaokun Ma}.} \bibinfo{year}{2018}\natexlab{}.
\newblock \showarticletitle{An empirical analysis of journal policy effectiveness for computational reproducibility}.
\newblock \bibinfo{journal}{\emph{Proceedings of the National Academy of Sciences}} \bibinfo{volume}{115}, \bibinfo{number}{11} (\bibinfo{year}{2018}), \bibinfo{pages}{2584--2589}.
\newblock


\bibitem[Strang(2014)]%
        {strang2014differential}
\bibfield{author}{\bibinfo{person}{Gilbert Strang}.} \bibinfo{year}{2014}\natexlab{}.
\newblock \bibinfo{booktitle}{\emph{Differential equations and linear algebra}}.
\newblock \bibinfo{publisher}{Wellesley-Cambridge Press Wellesley}.
\newblock


\bibitem[Tian et~al\mbox{.}(2022)]%
        {DBLP:conf/sigir/TianXLYZ22}
\bibfield{author}{\bibinfo{person}{Changxin Tian}, \bibinfo{person}{Yuexiang Xie}, \bibinfo{person}{Yaliang Li}, \bibinfo{person}{Nan Yang}, {and} \bibinfo{person}{Wayne~Xin Zhao}.} \bibinfo{year}{2022}\natexlab{}.
\newblock \showarticletitle{Learning to Denoise Unreliable Interactions for Graph Collaborative Filtering}. In \bibinfo{booktitle}{\emph{{SIGIR} '22: The 45th International {ACM} {SIGIR} Conference on Research and Development in Information Retrieval, Madrid, Spain, July 11 - 15, 2022}}, \bibfield{editor}{\bibinfo{person}{Enrique Amig{\'{o}}}, \bibinfo{person}{Pablo Castells}, \bibinfo{person}{Julio Gonzalo}, \bibinfo{person}{Ben Carterette}, \bibinfo{person}{J.~Shane Culpepper}, {and} \bibinfo{person}{Gabriella Kazai}} (Eds.). \bibinfo{publisher}{{ACM}}, \bibinfo{pages}{122--132}.
\newblock
\href{https://doi.org/10.1145/3477495.3531889}{doi:\nolinkurl{10.1145/3477495.3531889}}


\bibitem[Wang et~al\mbox{.}(2023)]%
        {WangEtAl2023}
\bibfield{author}{\bibinfo{person}{Jihu Wang}, \bibinfo{person}{Yuliang Shi}, \bibinfo{person}{Han Yu}, \bibinfo{person}{Xinjun Wang}, \bibinfo{person}{Zhongmin Yan}, {and} \bibinfo{person}{Fanyu Kong}.} \bibinfo{year}{2023}\natexlab{}.
\newblock \showarticletitle{Mixed-Curvature Manifolds Interaction Learning for Knowledge Graph-aware Recommendation}. In \bibinfo{booktitle}{\emph{Proceedings of the 46th International {ACM} {SIGIR} Conference on Research and Development in Information Retrieval, {SIGIR} 2023, Taipei, Taiwan, July 23-27, 2023}}, \bibfield{editor}{\bibinfo{person}{Hsin{-}Hsi Chen}, \bibinfo{person}{Wei{-}Jou~(Edward) Duh}, \bibinfo{person}{Hen{-}Hsen Huang}, \bibinfo{person}{Makoto~P. Kato}, \bibinfo{person}{Josiane Mothe}, {and} \bibinfo{person}{Barbara Poblete}} (Eds.). \bibinfo{publisher}{{ACM}}, \bibinfo{pages}{372--382}.
\newblock
\href{https://doi.org/10.1145/3539618.3591730}{doi:\nolinkurl{10.1145/3539618.3591730}}


\bibitem[Wang et~al\mbox{.}(2019a)]%
        {DBLP:conf/kdd/Wang00LC19}
\bibfield{author}{\bibinfo{person}{Xiang Wang}, \bibinfo{person}{Xiangnan He}, \bibinfo{person}{Yixin Cao}, \bibinfo{person}{Meng Liu}, {and} \bibinfo{person}{Tat{-}Seng Chua}.} \bibinfo{year}{2019}\natexlab{a}.
\newblock \showarticletitle{{KGAT:} Knowledge Graph Attention Network for Recommendation}. In \bibinfo{booktitle}{\emph{Proceedings of the 25th {ACM} {SIGKDD} International Conference on Knowledge Discovery {\&} Data Mining, {KDD} 2019, Anchorage, AK, USA, August 4-8, 2019}}, \bibfield{editor}{\bibinfo{person}{Ankur Teredesai}, \bibinfo{person}{Vipin Kumar}, \bibinfo{person}{Ying Li}, \bibinfo{person}{R{\'{o}}mer Rosales}, \bibinfo{person}{Evimaria Terzi}, {and} \bibinfo{person}{George Karypis}} (Eds.). \bibinfo{publisher}{{ACM}}, \bibinfo{pages}{950--958}.
\newblock
\href{https://doi.org/10.1145/3292500.3330989}{doi:\nolinkurl{10.1145/3292500.3330989}}


\bibitem[Wang et~al\mbox{.}(2019b)]%
        {DBLP:conf/sigir/Wang0WFC19}
\bibfield{author}{\bibinfo{person}{Xiang Wang}, \bibinfo{person}{Xiangnan He}, \bibinfo{person}{Meng Wang}, \bibinfo{person}{Fuli Feng}, {and} \bibinfo{person}{Tat{-}Seng Chua}.} \bibinfo{year}{2019}\natexlab{b}.
\newblock \showarticletitle{Neural Graph Collaborative Filtering}. In \bibinfo{booktitle}{\emph{Proceedings of the 42nd International {ACM} {SIGIR} Conference on Research and Development in Information Retrieval, {SIGIR} 2019, Paris, France, July 21-25, 2019}}, \bibfield{editor}{\bibinfo{person}{Benjamin Piwowarski}, \bibinfo{person}{Max Chevalier}, \bibinfo{person}{{\'{E}}ric Gaussier}, \bibinfo{person}{Yoelle Maarek}, \bibinfo{person}{Jian{-}Yun Nie}, {and} \bibinfo{person}{Falk Scholer}} (Eds.). \bibinfo{publisher}{{ACM}}, \bibinfo{pages}{165--174}.
\newblock
\href{https://doi.org/10.1145/3331184.3331267}{doi:\nolinkurl{10.1145/3331184.3331267}}


\bibitem[Wei et~al\mbox{.}(2023)]%
        {WeiEtAl2023}
\bibfield{author}{\bibinfo{person}{Tianjun Wei}, \bibinfo{person}{Jianghong Ma}, {and} \bibinfo{person}{Tommy W.~S. Chow}.} \bibinfo{year}{2023}\natexlab{}.
\newblock \showarticletitle{Collaborative Residual Metric Learning}. In \bibinfo{booktitle}{\emph{Proceedings of the 46th International {ACM} {SIGIR} Conference on Research and Development in Information Retrieval, {SIGIR} 2023, Taipei, Taiwan, July 23-27, 2023}}, \bibfield{editor}{\bibinfo{person}{Hsin{-}Hsi Chen}, \bibinfo{person}{Wei{-}Jou~(Edward) Duh}, \bibinfo{person}{Hen{-}Hsen Huang}, \bibinfo{person}{Makoto~P. Kato}, \bibinfo{person}{Josiane Mothe}, {and} \bibinfo{person}{Barbara Poblete}} (Eds.). \bibinfo{publisher}{{ACM}}, \bibinfo{pages}{1107--1116}.
\newblock
\href{https://doi.org/10.1145/3539618.3591649}{doi:\nolinkurl{10.1145/3539618.3591649}}


\bibitem[Wu et~al\mbox{.}(2023)]%
        {WuEtAl2023}
\bibfield{author}{\bibinfo{person}{S. Wu}, \bibinfo{person}{F. Sun}, \bibinfo{person}{W. Zhang}, \bibinfo{person}{X. Xie}, {and} \bibinfo{person}{B. Cui}.} \bibinfo{year}{2023}\natexlab{}.
\newblock \showarticletitle{{Graph Neural Networks in Recommender Systems: A Survey}}.
\newblock \bibinfo{journal}{\emph{{ACM Computing Surveys (CSUR)}}} \bibinfo{volume}{55}, \bibinfo{number}{5} (\bibinfo{date}{May} \bibinfo{year}{2023}), \bibinfo{pages}{97:1--97:37}.
\newblock


\bibitem[Wu et~al\mbox{.}(2022)]%
        {DBLP:conf/sigir/WuCSTC22}
\bibfield{author}{\bibinfo{person}{Yunfan Wu}, \bibinfo{person}{Qi Cao}, \bibinfo{person}{Huawei Shen}, \bibinfo{person}{Shuchang Tao}, {and} \bibinfo{person}{Xueqi Cheng}.} \bibinfo{year}{2022}\natexlab{}.
\newblock \showarticletitle{{INMO:} {A} Model-Agnostic and Scalable Module for Inductive Collaborative Filtering}. In \bibinfo{booktitle}{\emph{{SIGIR} '22: The 45th International {ACM} {SIGIR} Conference on Research and Development in Information Retrieval, Madrid, Spain, July 11 - 15, 2022}}, \bibfield{editor}{\bibinfo{person}{Enrique Amig{\'{o}}}, \bibinfo{person}{Pablo Castells}, \bibinfo{person}{Julio Gonzalo}, \bibinfo{person}{Ben Carterette}, \bibinfo{person}{J.~Shane Culpepper}, {and} \bibinfo{person}{Gabriella Kazai}} (Eds.). \bibinfo{publisher}{{ACM}}, \bibinfo{pages}{91--101}.
\newblock
\href{https://doi.org/10.1145/3477495.3532000}{doi:\nolinkurl{10.1145/3477495.3532000}}


\bibitem[Xia et~al\mbox{.}(2022)]%
        {DBLP:conf/sigir/XiaHXZYH22}
\bibfield{author}{\bibinfo{person}{Lianghao Xia}, \bibinfo{person}{Chao Huang}, \bibinfo{person}{Yong Xu}, \bibinfo{person}{Jiashu Zhao}, \bibinfo{person}{Dawei Yin}, {and} \bibinfo{person}{Jimmy~X. Huang}.} \bibinfo{year}{2022}\natexlab{}.
\newblock \showarticletitle{Hypergraph Contrastive Collaborative Filtering}. In \bibinfo{booktitle}{\emph{{SIGIR} '22: The 45th International {ACM} {SIGIR} Conference on Research and Development in Information Retrieval, Madrid, Spain, July 11 - 15, 2022}}, \bibfield{editor}{\bibinfo{person}{Enrique Amig{\'{o}}}, \bibinfo{person}{Pablo Castells}, \bibinfo{person}{Julio Gonzalo}, \bibinfo{person}{Ben Carterette}, \bibinfo{person}{J.~Shane Culpepper}, {and} \bibinfo{person}{Gabriella Kazai}} (Eds.). \bibinfo{publisher}{{ACM}}, \bibinfo{pages}{70--79}.
\newblock
\href{https://doi.org/10.1145/3477495.3532058}{doi:\nolinkurl{10.1145/3477495.3532058}}


\bibitem[Yang et~al\mbox{.}(2019)]%
        {YangEtAl2019}
\bibfield{author}{\bibinfo{person}{W. Yang}, \bibinfo{person}{K. Lu}, \bibinfo{person}{P. Yang}, {and} \bibinfo{person}{J. Lin}.} \bibinfo{year}{2019}\natexlab{}.
\newblock \showarticletitle{{Critically Examining the "Neural Hype": Weak Baselines and the Additivity of Effectiveness Gains from Neural Ranking Models}}. In \bibinfo{booktitle}{\emph{{Proc. 42nd Annual International ACM SIGIR Conference on Research and Development in Information Retrieval (SIGIR 2019)}}}, \bibfield{editor}{\bibinfo{person}{B.~Piwowarski}, \bibinfo{person}{M.~Chevalier}, \bibinfo{person}{E.~Gaussier}, \bibinfo{person}{Y.~Maarek}, \bibinfo{person}{J.-Y. Nie}, {and} \bibinfo{person}{F.~Scholer}} (Eds.). \bibinfo{publisher}{{ACM Press, New York, USA}}, \bibinfo{pages}{1129--1132}.
\newblock


\bibitem[Yang et~al\mbox{.}(2022)]%
        {DBLP:conf/sigir/YangHXL22}
\bibfield{author}{\bibinfo{person}{Yuhao Yang}, \bibinfo{person}{Chao Huang}, \bibinfo{person}{Lianghao Xia}, {and} \bibinfo{person}{Chenliang Li}.} \bibinfo{year}{2022}\natexlab{}.
\newblock \showarticletitle{Knowledge Graph Contrastive Learning for Recommendation}. In \bibinfo{booktitle}{\emph{{SIGIR} '22: The 45th International {ACM} {SIGIR} Conference on Research and Development in Information Retrieval, Madrid, Spain, July 11 - 15, 2022}}, \bibfield{editor}{\bibinfo{person}{Enrique Amig{\'{o}}}, \bibinfo{person}{Pablo Castells}, \bibinfo{person}{Julio Gonzalo}, \bibinfo{person}{Ben Carterette}, \bibinfo{person}{J.~Shane Culpepper}, {and} \bibinfo{person}{Gabriella Kazai}} (Eds.). \bibinfo{publisher}{{ACM}}, \bibinfo{pages}{1434--1443}.
\newblock
\href{https://doi.org/10.1145/3477495.3532009}{doi:\nolinkurl{10.1145/3477495.3532009}}


\bibitem[Yang et~al\mbox{.}(2023)]%
        {YangEtAl2023}
\bibfield{author}{\bibinfo{person}{Yonghui Yang}, \bibinfo{person}{Zhengwei Wu}, \bibinfo{person}{Le Wu}, \bibinfo{person}{Kun Zhang}, \bibinfo{person}{Richang Hong}, \bibinfo{person}{Zhiqiang Zhang}, \bibinfo{person}{Jun Zhou}, {and} \bibinfo{person}{Meng Wang}.} \bibinfo{year}{2023}\natexlab{}.
\newblock \showarticletitle{Generative-Contrastive Graph Learning for Recommendation}. In \bibinfo{booktitle}{\emph{Proceedings of the 46th International {ACM} {SIGIR} Conference on Research and Development in Information Retrieval, {SIGIR} 2023, Taipei, Taiwan, July 23-27, 2023}}, \bibfield{editor}{\bibinfo{person}{Hsin{-}Hsi Chen}, \bibinfo{person}{Wei{-}Jou~(Edward) Duh}, \bibinfo{person}{Hen{-}Hsen Huang}, \bibinfo{person}{Makoto~P. Kato}, \bibinfo{person}{Josiane Mothe}, {and} \bibinfo{person}{Barbara Poblete}} (Eds.). \bibinfo{publisher}{{ACM}}, \bibinfo{pages}{1117--1126}.
\newblock
\href{https://doi.org/10.1145/3539618.3591691}{doi:\nolinkurl{10.1145/3539618.3591691}}


\bibitem[Yu et~al\mbox{.}(2022)]%
        {DBLP:conf/sigir/YuY00CN22}
\bibfield{author}{\bibinfo{person}{Junliang Yu}, \bibinfo{person}{Hongzhi Yin}, \bibinfo{person}{Xin Xia}, \bibinfo{person}{Tong Chen}, \bibinfo{person}{Lizhen Cui}, {and} \bibinfo{person}{Quoc Viet~Hung Nguyen}.} \bibinfo{year}{2022}\natexlab{}.
\newblock \showarticletitle{Are Graph Augmentations Necessary?: Simple Graph Contrastive Learning for Recommendation}. In \bibinfo{booktitle}{\emph{{SIGIR} '22: The 45th International {ACM} {SIGIR} Conference on Research and Development in Information Retrieval, Madrid, Spain, July 11 - 15, 2022}}, \bibfield{editor}{\bibinfo{person}{Enrique Amig{\'{o}}}, \bibinfo{person}{Pablo Castells}, \bibinfo{person}{Julio Gonzalo}, \bibinfo{person}{Ben Carterette}, \bibinfo{person}{J.~Shane Culpepper}, {and} \bibinfo{person}{Gabriella Kazai}} (Eds.). \bibinfo{publisher}{{ACM}}, \bibinfo{pages}{1294--1303}.
\newblock
\href{https://doi.org/10.1145/3477495.3531937}{doi:\nolinkurl{10.1145/3477495.3531937}}


\bibitem[Zhu et~al\mbox{.}(2023b)]%
        {ZhuEtAl2023}
\bibfield{author}{\bibinfo{person}{Guanghui Zhu}, \bibinfo{person}{Wang Lu}, \bibinfo{person}{Chunfeng Yuan}, {and} \bibinfo{person}{Yihua Huang}.} \bibinfo{year}{2023}\natexlab{b}.
\newblock \showarticletitle{AdaMCL: Adaptive Fusion Multi-View Contrastive Learning for Collaborative Filtering}. In \bibinfo{booktitle}{\emph{Proceedings of the 46th International {ACM} {SIGIR} Conference on Research and Development in Information Retrieval, {SIGIR} 2023, Taipei, Taiwan, July 23-27, 2023}}, \bibfield{editor}{\bibinfo{person}{Hsin{-}Hsi Chen}, \bibinfo{person}{Wei{-}Jou~(Edward) Duh}, \bibinfo{person}{Hen{-}Hsen Huang}, \bibinfo{person}{Makoto~P. Kato}, \bibinfo{person}{Josiane Mothe}, {and} \bibinfo{person}{Barbara Poblete}} (Eds.). \bibinfo{publisher}{{ACM}}, \bibinfo{pages}{1076--1085}.
\newblock
\href{https://doi.org/10.1145/3539618.3591632}{doi:\nolinkurl{10.1145/3539618.3591632}}


\bibitem[Zhu et~al\mbox{.}(2023a)]%
        {ZhuEtAl2023b}
\bibfield{author}{\bibinfo{person}{Xinjun Zhu}, \bibinfo{person}{Yuntao Du}, \bibinfo{person}{Yuren Mao}, \bibinfo{person}{Lu Chen}, \bibinfo{person}{Yujia Hu}, {and} \bibinfo{person}{Yunjun Gao}.} \bibinfo{year}{2023}\natexlab{a}.
\newblock \showarticletitle{Knowledge-refined Denoising Network for Robust Recommendation}. In \bibinfo{booktitle}{\emph{Proceedings of the 46th International {ACM} {SIGIR} Conference on Research and Development in Information Retrieval, {SIGIR} 2023, Taipei, Taiwan, July 23-27, 2023}}, \bibfield{editor}{\bibinfo{person}{Hsin{-}Hsi Chen}, \bibinfo{person}{Wei{-}Jou~(Edward) Duh}, \bibinfo{person}{Hen{-}Hsen Huang}, \bibinfo{person}{Makoto~P. Kato}, \bibinfo{person}{Josiane Mothe}, {and} \bibinfo{person}{Barbara Poblete}} (Eds.). \bibinfo{publisher}{{ACM}}, \bibinfo{pages}{362--371}.
\newblock
\href{https://doi.org/10.1145/3539618.3591707}{doi:\nolinkurl{10.1145/3539618.3591707}}


\bibitem[Zou et~al\mbox{.}(2022)]%
        {DBLP:conf/sigir/Zou0MWQ0C22}
\bibfield{author}{\bibinfo{person}{Ding Zou}, \bibinfo{person}{Wei Wei}, \bibinfo{person}{Xian{-}Ling Mao}, \bibinfo{person}{Ziyang Wang}, \bibinfo{person}{Minghui Qiu}, \bibinfo{person}{Feida Zhu}, {and} \bibinfo{person}{Xin Cao}.} \bibinfo{year}{2022}\natexlab{}.
\newblock \showarticletitle{Multi-level Cross-view Contrastive Learning for Knowledge-aware Recommender System}. In \bibinfo{booktitle}{\emph{{SIGIR} '22: The 45th International {ACM} {SIGIR} Conference on Research and Development in Information Retrieval, Madrid, Spain, July 11 - 15, 2022}}, \bibfield{editor}{\bibinfo{person}{Enrique Amig{\'{o}}}, \bibinfo{person}{Pablo Castells}, \bibinfo{person}{Julio Gonzalo}, \bibinfo{person}{Ben Carterette}, \bibinfo{person}{J.~Shane Culpepper}, {and} \bibinfo{person}{Gabriella Kazai}} (Eds.). \bibinfo{publisher}{{ACM}}, \bibinfo{pages}{1358--1368}.
\newblock
\href{https://doi.org/10.1145/3477495.3532025}{doi:\nolinkurl{10.1145/3477495.3532025}}


\end{thebibliography}
\endgroup
\makeatother

\clearpage
\appendix

\providecommand{\titlefont}{\LARGE \bfseries}
\noindent{\titlefont Appendix\par}

\begingroup
  \setcounter{tocdepth}{2}
  \tableofcontents
\endgroup
\clearpage



\newcommand{\tableTime}[4]{
\begin{table}[h]
    \caption{Computation time for the algorithms in  the selected results for the #2 method on the #4 dataset.}
    \label{tab:#2-#3-train-time}
    \footnotesize
    \centering
    \input{sections/tables/#1_#3_time_latex_results.txt}
\end{table}
}

\newcommand{\tableArticleResultApp}[5]{
\begin{table}[h]
    \caption{Experimental results for the #2 method for the #4 dataset.}
    \label{tab:#2-#3-result}
    \footnotesize
    \centering
    
    \ifnum#5=0
        \input{sections/tables/#1_#3_article_metrics_latex_results.txt}
    \else
        \resizebox{\linewidth}{!}{%
        \input{sections/tables/#1_#3_article_metrics_latex_results.txt}
    }    
    \fi
\end{table}
}

\newcommand{\tableAllResult}[4]{
\begin{table}[h]
    \caption{Experimental results for the #2 method for the #4 dataset on all available metrics.}
    \label{tab:#2-#3-all-result}
    \footnotesize
    \centering
    \resizebox{\linewidth}{!}{%
    \input{sections/tables/#1_#3_all_metrics_latex_results.txt}
    }
\end{table}
}

\newcommand{\tableBeyondAccuracyResult}[4]{
\begin{table}[h]
    \caption{Experimental results for the #2 method for the #4 dataset on beyond accuracy metrics.}
    \label{tab:#2-#3-beyond-accuracy}
    \footnotesize
    \centering
    \input{sections/tables/#1_#3_beyond_accuracy_metrics_latex_results.txt}
\end{table}
}

\makeatletter
\newcommand{\switchCaption}[1]{%
  \ifnum\pdf@strcmp{#1}{KNN}=0
    our collaborative KNN baselines%
  \else\ifnum\pdf@strcmp{#1}{ML_graph}=0
    our non-neural machine learning and graph based baselines%
  \else\ifnum\pdf@strcmp{#1}{CBF}=0
    our content based KNN baselines%
  \else\ifnum\pdf@strcmp{#1}{CFCBF}=0
    our hybrid KNN baselines%
  \else\ifnum\pdf@strcmp{#1}{neural}=0
    the neural algorithm%
  \else
    ??%
  \fi\fi\fi\fi\fi}
\makeatother

\newcommand{\tableHyperParams}[3]{
\begin{table}[h]
    \caption{Hyperparameter values for \switchCaption{#3} on all datasets.}
    \label{tab:#2-#3-hyperparams}
    \footnotesize
    \centering
    \input{sections/tables/#1_latex_hyperparameters_#3.txt}
\end{table}
}

\section{Baselines}
Here we list all the 21 collaborative and 5 content-based and hybrid baseline algorithms used in each experiment, most of them are the same used by \citet{DBLP:journals/tois/DacremaBCJ21}: 


\paragraph{Non-personalized}
\begin{itemize}
    \item \textbf{Random}: non-personalized method recommending random items the user has not yet interacted with.
    \item \textbf{TopPop}: non-personalized method recommending to all users the most popular items the user has not yet interacted with.
    \item \textbf{Global Effects}: leverages global, item and user biases to recommend items.
\end{itemize}

\paragraph{Nearest-Neighbor Collaborative and Content-Based}
\begin{itemize}
    \item \textbf{UserKNN}: user-based nearest-neighbor algorithm~\cite{DBLP:conf/cscw/ResnickISBR94}, with cosine similarity and shrinkage~\cite{bell2007improved}.
    \item \textbf{ItemKNN}: item-based nearest-neighbor algorithm~\cite{DBLP:conf/www/SarwarKKR01}, with cosine similarity and shrinkage~\cite{bell2007improved}.
    \item \textbf{UserKNN CBF}: UserKNN computed on the user features. 
    \item \textbf{UserKNN CFCBF}: UserKNN computed on the concatenation of the user profile and the user features. A hyperparameter controls the weight of the content-based part.
    \item \textbf{ItemKNN CBF <attribute>}: ItemKNN computed on the item features. 
    \item \textbf{ItemKNN CFCBF <attribute>}: ItemKNN computed on the concatenation of the item interactions and the item features. A hyperparameter controls the weight of the content-based part.
\end{itemize}

\paragraph{Graph-based}
\begin{itemize}
    \item \textbf{\palpha}: graph based algorithms modeling random walk on the bipartite graph of users-items interactions.
    \item \textbf{\pbeta}: graph-based method that uses a two-steps random walk from users to items and vice-versa, where transition probabilities are computed from the normalized ratings~\cite{DBLP:journals/tiis/PaudelCNB17}.
    \item \textbf{GF-CF}: a graph-based method that is based on a low-pass filter and has a closed form solution~\cite{DBLP:conf/cikm/ShenWZSZLL21}.
\end{itemize}

\paragraph{Item-Based Machine Learning}
\begin{itemize}
    \item \textbf{\EASER}: An ``embarrassingly shallow'' linear model with strong connections with autoencoders and a closed form solution~\cite{DBLP:conf/www/Steck19}.\footnote{\EASER has a high memory requirements and often exceeds the 64GB RAM available on our server.}
    \item \textbf{SLIM}: item-based model that uses linear regression to compute the item similarity~\cite{DBLP:conf/icdm/NingK11}.
    \item \textbf{SLIM-BPR}: item-based model similar to SLIM that computes the item similarity optimizing the \emph{Bayesian Personalized Ranking} (BPR) loss~\cite{DBLP:conf/uai/RendleFGS09}.
    \item \textbf{NegHOSLIM}: linear full-rank model similar to SLIM that includes higher-order interactions as input-features~\cite{DBLP:conf/recsys/SteckL21}.
    \item \textbf{NegHOSLIM (EN)}: linear full-rank model similar to NegHOSLIM that optimizes the ElasticNet loss.\footnote{Due to the large memory requirement of the original NegHOSLIM we trained this version by using an ElasticNet loss which reduces memory requirement but sacrifices some effectiveness.}
\end{itemize}

\paragraph{Matrix Factorization}
\begin{itemize}
    \item \textbf{MF-BPR}: matrix factorization method based on the \emph{Bayesian Personalized Ranking} (BPR) loss~\cite{DBLP:conf/uai/RendleFGS09}.
    \item \textbf{MF-WARP}: matrix factorization method based on the \emph{Weighted Approximate-Rank Pairwise} loss (WARP).
    \item \textbf{SVDpp}: matrix factorization method for rating prediction accounting for user biases \cite{DBLP:conf/recsys/LercheJ14}.\footnote{Note that to adapt SVDpp for the task of top-k recommendation we sample during training a certain quota of interactions that did not occur and attribute them a rating of zero. The specific quota is a hyperparameter.}
    \item \textbf{PureSVD}: Matrix factorization method based on the truncated SVD decomposition of the user-item interaction matrix \cite{DBLP:conf/recsys/CremonesiKT10}.\footnote{We use a standard SVD decomposition method provided in the \small{\texttt{scikit-learn}}  \footnotesize package for Python.}
    \item \textbf{NMF}: matrix factorization method that decomposes ratings matrix into two non-negative matrices \cite{DBLP:journals/ieicet/CichockiP09}.\footnote{We use a standard NMF decomposition method provided in the \small{\texttt{scikit-learn}} \footnotesize  package for Python.}
    \item \textbf{\iALS}: matrix factorization method for ranking tasks based on alternating least-squares \cite{DBLP:conf/icdm/HuKV08}.

\end{itemize}

\paragraph{Other Machine Learning}
\begin{itemize}
    \item \textbf{MultVAE}: variational autoencoder that assumes a multinomial likelihood for user-item interactions \cite{DBLP:conf/www/LiangKHJ18}.
    \item \textbf{LightFM CF}: factorization machine method that uses only collaborative data.\footnote{We use the LightFM library, \url{https://github.com/lyst/lightfm}}
    \item \textbf{LightFM ItemHybrid <attribute>}: factorization machine method that uses a combination of collaborative and item features.
    \item \textbf{LightFM UserHybrid <attribute>}: factorization machine method that uses a combination of collaborative and user features.
\end{itemize}

Note that occasionally the results for \textbf{GF-CF}, \textbf{\EASER}, \textbf{SLIM-BPR} and \textbf{NegHOSLIM} may be missing due to their memory requirements exceeding the 64GB available on our server.

\section{Less is More: Reweighting Important Spectral Graph Features for Recommendation}

\citet{DBLP:conf/sigir/PengSM22} analyzes the spectral properties of Graph Convolutional Networks and observe that the frequencies (\idest eigenvalues) that contribute the most to the recommendation accuracy are both the highest and lowest ones, with the intermediate ones being less important. This effect is attributed to the different semantics of the two, with higher frequencies representing differences between users while the lower ones representing the commonalities. The article proposes \emph{Graph Denoising Encoder} (GDE) which acts as a band-pass filter selecting high and low frequencies while removing intermediate ones. The proposed method is claimed to be substantially faster compared to LightGCN. The original implementation is available on GitHub.\footnote{\url{https://github.com/tanatosuu/GDE}}

\subsection{Datasets} 
The evaluation is performed on the datasets described and processed as follows, their statistics are reported in Table \ref{tab:GDE-dataset_stats}.
All existing interactions are made implicit and assigned a value of 1.

\begin{description}
\item[MovieLens:] Is a movie recommendation dataset, the explicit ratings (1-5) are all transformed in implicit ratings of value 1.
\item[CiteULike-a:] Is a dataset collected from CiteULike, which is an online service providing users with a digital catalogue to save and share academic papers. If the user has saved the article in their library it will be associated to a rating of 1.
\item[Pinterest:] refers to the well known social network which allows users to save or pin an image to their board. If a user has pinned an image on the board it will be associated to a rating of 1. 
\item[Gowalla:] A dataset collected from a social network where users check-in locations they visited.
\end{description}

\begin{table}[h]
    \small
    \begin{tabular}{lcccc}
    \toprule
    Dataset			& Interactions	& Items		& Users 	& Sparsity	\\
    \midrule
    CiteULike       &        210504 &  16980 &   5551 &  $ 0.9978 $ \\
    Gowalla         &       1027370 &  40981 &  29858 &  $ 0.9992 $ \\
    MovieLens1M     &       1000209 &   3952 &   6040 &  $ 0.9581 $ \\
    MovieLens100k   &        100000 &   1682 &    943 &  $ 0.9370 $ \\
    Pinterest       &       1000154 &   9836 &  37501 &  $ 0.9973 $ \\
	\bottomrule
   	\end{tabular}
    \caption{Dataset statistics for GDE.}
    \label{tab:GDE-dataset_stats}
\end{table}

\subsection{Results} 
The hyperparameter values used in our experiments are reported in Table \ref{tab:GDE-hyperparameter_values} and the results for all the datasets and baseline algorithms are reported in Table 
\ref{tab:GDE-citeulike-result} (CiteULike), 
\ref{tab:GDE-movielens1m-result} (MovieLens1M),  
\ref{tab:GDE-movielens100k-result} (MovieLens100k),  
\ref{tab:GDE-pinterest-result} (Pinterest), and
\ref{tab:GDE-gowalla-result} (Gowalla).

\begin{table}[h]
    \begin{minipage}{\textwidth}
    \centering
    \footnotesize
    \begin{tabular}{lc|cccccc}
    \toprule
    Hyperparameter	& Described in	& \multicolumn{6}{c}{Value}	\\
                	&               & All datasets	& CiteULike & ML-1M & ML-100K & Pinterest & Gowalla\\    
    \midrule
    Embedding size	 	& Paper	        & 64    & -     & -     & -     & -          & -  	\\
    Regularization rate & Paper	        & 0.01  & -     & -     & -     & -          & -  	\\
    Learning rate       & Source code   & 0.03  & 0.02  & 7.5   & 2.0   & 0.85/0.12  & 0.03 \\
    Dropout rate        & Source code   & 0.1   & 0.3   & 0.5   & 0.2   & 0.2        & 0.1  \\
    Epochs              & Source code   & 400   & 200   & 90    & 50    & >200       & 160  \\
    Batch size          & Paper         & 256   & -     & -     & -     & -          & -  	\\
    $\beta$             & Source code   & -     & 5.0\footnote{The paper reports that the optimal value should be 4.5.}   & 4.0   & 4.0\footnote{The paper reports that the optimal value should be 4.5.}   & 4.0/5.0    & 5.0  \\
    Loss type           & Source code   & adaptive      & adaptive  & adaptive    & bpr  & adaptive  & adaptive  \\
    Smooth ratio        & Source code   & 0.1   & 0.3   & 0.05  & 0.2   & 0.2        & 0.1  \\
    Rough ratio         & Source code   & 0.0   & 0.0   & 0.005 & 0.002 & 0.0        & 0.0  \\
    Feature type\footnote{If the value is "smoothed" only the smooth features (low frequencies) are used, the value is "both" rough features (high frequencies) are used as well.}        & Source code   & smoothed   & smoothed   & both & both & smoothed        & smoothed  \\
	\bottomrule
   	\end{tabular}
   	\end{minipage}
    \caption{Hyperparameter values for GDE.}
    \label{tab:GDE-hyperparameter_values}
\end{table}

\tableArticleResultApp{GDE}{GDE}{citeulike}{CiteULike}{0}
\tableArticleResultApp{GDE}{GDE}{movielens1m}{MovieLens1M}{0}
\tableArticleResultApp{GDE}{GDE}{movielens100k}{MovieLens100k}{0}
\tableArticleResultApp{GDE}{GDE}{pinterest}{Pinterest}{0}
\tableArticleResultApp{GDE}{GDE}{gowalla}{Gowalla}{0}

\clearpage
\section{Are Graph Augmentations Necessary? Simple Graph Contrastive Learning for Recommendation}
\citet{DBLP:conf/sigir/YuY00CN22} propose \emph{Simple Graph Contrastive Learning} (SimGCL). The paper claims that in constrastive learning based recommendations the main contribution to the recommendation quality is not the graph augmentation (\eg random edge dropout) but rather the constrastive learning loss function (\idest InfoNCE). The effect of the InfoNCE loss is to increase the separation between positive and negative samples for each user. SimGCL uses random perturbations of the embeddings instead of graph augmentations. In practice, SimCL is a LightGCM \cite{DBLP:conf/sigir/0001DWLZ020} with random embedding perturbations, a regularizing loss and the aggregated user and item embeddings that start from layer 1, therefore excluding layer zero (\idest $E^{(0)}$). 
The original implementation is available on GitHub.\footnote{\url{https://github.com/Coder-Yu/QRec} we use the pytorch implementation available from the authors here \url{https://github.com/Coder-Yu/SELFRec}}

\subsection{Datasets} 
The evaluation is performed on the datasets described and processed as follows, their statistics are reported in Table \ref{tab:SimGCL-dataset_stats}.

\begin{description}
\item[DoubanBook:] Is a dataset of relations for the Douban Book service, with ratings in the range 1-5. Ratings greater or equal to 4 are transformed in implicit interactions with value 1, the other ratings are removed.
\item[Yelp2018:] Is a business reviews dataset. The split is the same used in LightGCN \cite{DBLP:conf/sigir/0001DWLZ020}, see section \ref{sec:lightGCN}.
\item[Amazon-Book:] Is a dataset of book purchases on Amazon. The split is the same used in LightGCN \cite{DBLP:conf/sigir/0001DWLZ020}, see section \ref{sec:lightGCN}.
\end{description}

\begin{table}[h]
    \small
    \begin{tabular}{lcccc}
    \toprule
    Dataset			& Interactions	& Items		& Users 	& Sparsity	\\
    \midrule
    Amazon-Book &       2984108 &  91599 &  52643 &  $ 0.9994 $ \\
    DoubanBook &        598420 &  22348 &  13025 &  $ 0.9979 $ \\
    Yelp2018 &          1561406 &  38048 &  31668 &  $ 0.9987 $ \\
	\bottomrule
   	\end{tabular}
    \caption{Dataset statistics for SimGCL.}
    \label{tab:SimGCL-dataset_stats}
\end{table}

\subsection{Results} 
The hyperparameter values used in our experiments are reported in Table \ref{tab:SimGCL-hyperparameter_values} and the results for all the datasets and baseline algorithms are reported in Table 
\ref{tab:SimGCL-amazon-book_original-result} (Amazon-Book Original Split), 
\ref{tab:SimGCL-amazon-book_ours-result} (Amazon-Book Our Split), 
\ref{tab:SimGCL-doubanbook_original-result} (DoubanBook),  
\ref{tab:SimGCL-yelp2018_original-result} (Yelp2018 Original Split), and
\ref{tab:SimGCL-yelp2018_ours-result} (Yelp2018 Our Split).

\begin{table}[h]
    \begin{minipage}{\textwidth}
    \centering
    \footnotesize
    \begin{tabular}{lccccc}
    \toprule
    Hyperparameter	& Described in	& \multicolumn{4}{c}{Value}	\\
                	&               & All datasets & 	DoubanBook & 	Yelp2018 & 	Amazon-Book\\    
    \midrule
    $\lambda$ (contrastive loss weight)	 	& Paper\footnote{From a section discussing hyperparameter sensitivity.}	    & -  	    & 0.2  	& 0.5  	& 2  	\\
    $\tau$ (contrastive loss temperature)	 	            & Paper		& 0.2	    & -  	& -  	& -  	\\
    $\epsilon$ (noise magnitude)       & Paper\footnote{From a section discussing hyperparameter sensitivity.}		& 0.1 	    & -  	& -  	& -  	\\
    Batch size	 	        & Paper		& 2048	    & -  	& -  	& -  	\\
    Number of layers	 	& Paper\footnote{From a table comparing the result for different number of layers.}	    & 3	 	    & -  	& -  	& -  	\\
    Learning rate	 	    & Paper		& $10^{-3}$ & -  	& -  	& -  	\\
    Adaptive gradient 	    & Paper		& Adam      & -  	& -  	& -  	\\
    Embedding size	 	    & Paper		& 64	 	& -  	& -  	& -  	\\
    $L_2$ regularization	& Paper		& $10^{-4}$	& -  	& -  	& -  	\\
    Epochs      	 	    & Paper\footnote{From a section that discusses a plot showing when the models converge with Recall and BPR loss.}		& -	 	    & 25  	& 11  	& 10  	\\
	\bottomrule
   	\end{tabular}
   	\end{minipage}
    \caption{Hyperparameter values for SimGCL.}
    \label{tab:SimGCL-hyperparameter_values}
\end{table}

\tableArticleResultApp{SimGCL}{SimGCL}{amazon-book_original}{Amazon-Book Original Split}{0}
\tableArticleResultApp{SimGCL}{SimGCL}{amazon-book_ours}{Amazon Book Our Split}{0}
\tableArticleResultApp{SimGCL}{SimGCL}{doubanbook_original}{DoubanBook}{0}
\tableArticleResultApp{SimGCL}{SimGCL}{yelp2018_original}{Yelp2018 Original Split}{0}
\tableArticleResultApp{SimGCL}{SimGCL}{yelp2018_ours}{Yelp 2018 Our Split}{0}

\clearpage
\section{Learning to Denoise Unreliable Interactions for Graph Collaborative Filtering}
\citet{DBLP:conf/sigir/TianXLYZ22} presents \emph{Robust Graph Collaborative Filtering} (RGCF) based on the LighgGCN message passing architecture. RGCF consists of two steps, first a graph denoising module removes interactions that are estimated as being noisy and assigns a reliability weight to the other ones. This step is performed via the cosine similarity of the learned embeddings. Then, a diversity preserving module builds new interaction graphs (\idest adjacency matrix) based on the denoised one. A certain number of random user-item candidates are sampled, the prediction computed using the learned embeddings and those with high score (the paper calls it reliability) are added to the interaction graph. RGCF is trained with BPR with a second loss added to pull the representation of nodes learned with the augmented graphs close to each other, this is done with the contrastive loss InfoNCE. The original implementation is available on GitHub.\footnote{\url{https://github.com/ChangxinTian/RGCF}}

\subsection{Datasets} 
The evaluation is performed on the datasets described and processed as follows, their statistics are reported in Table \ref{tab:RGCF-dataset_stats}.

\begin{description}
\item[Amazon-Book:] Is a dataset of book purchases on Amazon. Only users and items with at least 15 interactions are retained, this corresponds to the 15-cores subgraph.
\item[MovieLens1M:] Is a movie recommendation dataset. Ratings $\geq 4$ are transformed into implicit interactions with value 1.
\item[Yelp:] Is a business reviews dataset. Only users and items with at least 15 interactions are retained, this corresponds to the 15-cores subgraph.
\end{description}

\begin{table}[h]
    \small
    \begin{tabular}{lcccc}
    \toprule
    Dataset			& Interactions	& Items		& Users 	& Sparsity	\\
    \midrule
    Amazon-Book  &       2517437 &  58051 &  58144 &  $ 0.9993 $ \\
    MovieLens1M &        836478 &   3883 &   6040 &  $ 0.9643 $ \\
    Yelp        &       1730025 &  31731 &  45160 &  $ 0.9988 $ \\
	\bottomrule
   	\end{tabular}
    \caption{Dataset statistics for RGCF.}
    \label{tab:RGCF-dataset_stats}
\end{table}

\subsection{Results} 
The hyperparameter values used in our experiments are reported in Table \ref{tab:RGCF-hyperparameter_values} and the results for all the datasets and baseline algorithms are reported in Table 
\ref{tab:RGCF-movielens1m-result} (MovieLens1M).

\begin{table}[h]
    \begin{minipage}{\textwidth}
    \centering
    \footnotesize
    \begin{tabular}{lcc}
    \toprule
    Hyperparameter	& Described in	& \multicolumn{1}{c}{Value}	\\
                	&               & All datasets	\\    
    \midrule
    epochs	 	        & Source code	    & 500  	\\
    K	 	            & Source code		& 2	 	\\
    batch size	 	    & Paper		        & 4096	 	\\
    embedding size	    & Paper		        & 64	 	\\
    prune threshold beta	 	& Source code		& 0.02	 	\\
    contrastive loss temperature tau	 	& Source code		& 0.2	 	\\
    contrastive loss weight	 	& Source code		& 1e-06	 	\\
    augmentation ratio	 	    & Source code		& 0.1	 	\\
    learning rate        	    & Source code		& 4e-5	 	\\
    l2 reg	 	                & Paper		        & 1e-05	 	\\
    optimizer	 	            & Paper		        & Adam	 	\\
	\bottomrule
   	\end{tabular}
   	\end{minipage}
    \caption{Hyperparameter values for RGCF.}
    \label{tab:RGCF-hyperparameter_values}
\end{table}

\tableArticleResultApp{RGCF}{RGCF}{movielens1m}{MovieLens1M}{0}



\clearpage
\section{INMO: A Model-Agnostic and Scalable Module for Inductive Collaborative Filtering}
\citet{DBLP:conf/sigir/WuCSTC22} presents \emph{Inductive Embedding Module for collaborative filtering} (INMO), that aims to improve the effectiveness of matrix factorization models to recommend to new users. The paper focuses on matrix factorization models that are \emph{transductive} (\idest memory-based, such as SVDpp, MF-BPR etc.) and proposes an \emph{inductive}  representation (\idest model-based) of the user and item embeddings as a function of the embeddings of a selected subset of template user and items. Due to this, the number of learnable parameters used in INMO can be lower compared to memory-based matrix factorization models. INMO includes an annealing process for normalization as a hyperparameter.
The original implementation is available on GitHub and the datasets are available in a Google Drive folder.\footnote{\url{https://github.com/WuYunfan/igcn_cf}}

\subsection{Datasets} 
The evaluation is performed on the datasets described and processed as follows, their statistics are reported in Table \ref{tab:INMO-dataset_stats}.

\begin{description}
\item[Amazon-Book:] Is a dataset of book purchases on Amazon. Ratings $\ge4$ are transformed into implicit interactions with value 1, then a 10-core subgraph selection is applied.
\item[Gowalla:] Is a dataset collected from a social network where users check-in locations they visited. No details are provided on the preprocessing.
\item[Yelp2018:] Is a business reviews dataset. Ratings $\ge4$ are transformed into implicit interactions with value 1, then a 10-core subgraph selection is applied.
\end{description}

\begin{table}[h]
    \small
    \begin{tabular}{lcccc}
    \toprule
    Dataset			& Interactions	& Items		& Users 	& Sparsity	\\
    \midrule
    Amazon-Book &       2780441 &  96421 & 109730 &  $ 0.9997 $ \\
    Gowalla     &        900713 &  40988 &  29858 &  $ 0.9993 $ \\
    Yelp2018    &       1680930 &  42706 &  75173 &  $ 0.9995 $ \\
	\bottomrule
   	\end{tabular}
    \caption{Dataset statistics for INMO.}
    \label{tab:INMO-dataset_stats}
\end{table}

\subsection{Results} 
The hyperparameter values used in our experiments are reported in Table \ref{tab:INMO-hyperparameter_values} and the results for all the datasets and baseline algorithms are reported in Table 
\ref{tab:INMO-gowalla-result} (Gowalla),
\ref{tab:INMO-amazon-book-result} (Amazon-Book), and
\ref{tab:INMO-yelp2018-result} (Yelp2018).

\begin{table}[h]
    \begin{minipage}{\textwidth}
    \centering
    \footnotesize
    \begin{tabular}{lccccc}
    \toprule
    Hyperparameter	& Described in	& \multicolumn{4}{c}{Value}	\\
                	&               & All datasets	& Amazon-Book  & Gowalla  &   Yelp2018 \\    
    \midrule
    embedding size	     & Source code	    & 64  	        & -  	& -  	& -  	\\
    batch size	 	     & Source code		& 2048	        & -  	& -  	& -  	\\
    K	 	             & Source code		& 3	 	        & -  	& -  	& -  	\\
    optimizer	 	     & Source code		& Adam 	        & -  	& -  	& -  	\\
    epochs	             & Source code		& 1000 (max)	& -  	& -  	& -  	\\
    learning rate  	 	 & Source code		& $10^{-3}$	 	& -  	& -  	& -  	\\
    template loss weight & Source code		& $10^{-2}$	 	& -  	& -  	& -  	\\
    $\lambda_2$	 	     & Source code		& 0.0	 	    & - 	& -  	& -  	\\    
    dropout rate	 	 & Source code		& -	 	        & 0.0  	& 0.3  	& 0.3  	\\
    feature ratio   	 & Source code		& -	 	        & 1.0  	& 1.0 	& 0.7  	\\
    normalization decay	 & Source code		& 0.99 	        & -  	& -  	& -  	\\
    template node ranking & Source code		& cardinality   & -  	& -  	& -  	\\
	\bottomrule
   	\end{tabular}
   	\end{minipage}
    \caption{Hyperparameter values for INMO.}
    \label{tab:INMO-hyperparameter_values}
\end{table}

\tableArticleResultApp{INMO}{INMO}{gowalla}{Gowalla}{0}
\tableArticleResultApp{INMO}{INMO}{amazon-book}{Amazon-Book}{0}
\tableArticleResultApp{INMO}{INMO}{yelp2018}{Yelp2018}{0}



\clearpage
\section{Hypergraph Contrastive Collaborative Filtering}
\citet{DBLP:conf/sigir/XiaHXZYH22} presents \emph{Hypergraph Contrastive Collaborative Filtering} (HCCF), based on the LightGCN paradigm adds several components: besides the message passing done on the user-item adjacency matrix as in LightGCN, but with the addition of a nonlinear aggregation function, HCCF incorporates one layer of message passing done on a hypergraph whose adjacency matrix is learnable and decomposed as the product of two lower dimensionality matrices. There is an additional step called Hierarchical Hypergraph Mapping which does the usual message passing but on the learned hypergraph adjacency matrix. The model is trained with contrastive learning using the InfoNCE loss, the goal is to push the embeddings learned via the message passing on the user-item adjacency matrix to be close to those obtained by learning the low dimensional approximation of the hypergraph adjacency.
The original implementation is available on GitHub.\footnote{\url{https://github.com/akaxlh/HCCF}}

\subsection{Datasets} 
The evaluation is performed on the datasets described and processed as follows, their statistics are reported in Table \ref{tab:HCCF-dataset_stats}.

\begin{description}
\item[Yelp:] Is a business reviews dataset. The preprocessing is a 10-cores subgraph selection.
\item[MovieLens10M:] Is a movie recommendation dataset. The preprocessing is a 10-cores subgraph selection.
\item[Amazon-Book:] Is a dataset of book purchases on Amazon. The preprocessing is a 20-cores subgraph selection.
\end{description}

\begin{table}[h]
    \small
    \begin{tabular}{lcccc}
    \toprule
    Dataset			& Interactions	& Items		& Users 	& Sparsity	\\
    \midrule
    Yelp	 	    &       1527326 &  24734 &  29601 &  $ 0.9979 $ \\
    MovieLens10M	&       9998816 &  10196 &  69878 &  $ 0.9860 $ \\
    Amazon-Book     &       3200224 &  77801 &  78578 &  $ 0.9995 $ \\
	\bottomrule
   	\end{tabular}
    \caption{Dataset statistics for HCCF.}
    \label{tab:HCCF-dataset_stats}
\end{table}

\subsection{Results} 
The hyperparameter values used in our experiments are reported in Table \ref{tab:HCCF-hyperparameter_values} and the results for all the datasets and baseline algorithms are reported in Table 
\ref{tab:HCCF-yelp2018-result} (Yelp2018),
\ref{tab:HCCF-amazon-book-result} (Amazon-Book), and
\ref{tab:HCCF-movielens10m-result} (MovieLens10M).

\begin{table}[h]
    \begin{minipage}{\textwidth}
    \centering
    \footnotesize
    \begin{tabular}{lccccc}
    \toprule
    Hyperparameter	& Described in	& \multicolumn{4}{c}{Value}	\\
                  &               & All datasets	    & MovieLens10M	   & Yelp	& Amazon-Book	\\    
    \midrule
    epochs	 	      & Paper		    & 100	 	& -  	& -  	& -  	\\
    sgd mode 	      & Paper		    & Adam	 	& -  	& -  	& -  	\\
    learning rate	  & Paper		    & $10^{-3}$	& -  	& -  	& -  	\\
    embedding size	  & Paper		    & 32	 	& -  	& -  	& -  	\\
    learning rate decay & Paper         & 0.96	& -  	& -  	& -  	\\
    GNN layers (K)	  & Paper		    & 2	 	    & -  	& -  	& -  	\\
    hyperedge size	  & Paper		    & 128	 	& -  	& -  	& -  	\\
    hypergraph mapping layers (C)	    & Paper		& 3	 	& -  	& -  	& -  	\\
    batch size	 	  & Source code		& 256	 	& -  	& - 	& 2048 \footnote{The paper states the optimal value is 256 but in the experiments we use 2048 for Amazon-Book due to the very large computational cost of this model. On Amazon-Book a batch size of 256 results in a training time of 45 minutes per epoch, hence a total of 3 days.} 	\\
    dropout	 	      & Source code		& 0.5 	    & 0.0  	& -  	& -  	\\
    contrastive loss weight ($\lambda_1$)	& Source code	        & $10^{-4}$	 	& $10^{-6}$  	& $10^{-4}$  	& $10^{-7}$  	\\
    $\lambda_2$	 	  & Source code		& $10^{-5}$ & -  	& $10^{-3}$  	& $10^{-2}$  	\\
    contrastive loss temperature ($\tau$)	 	    & Paper / Source code        & 1.0	 	& 0.1  	& 1.0  	& 0.1  	\\
    leaky relu slope  & Paper		        & 0.5	 	& -  	& -  	& -  	\\
    
	\bottomrule
   	\end{tabular}
   	\end{minipage}
    \caption{Hyperparameter values for HCCF.}
    \label{tab:HCCF-hyperparameter_values}
\end{table}

\tableArticleResultApp{HCCF}{HCCF}{yelp2018}{Yelp2018}{0}
\tableArticleResultApp{HCCF}{HCCF}{amazon-book}{Amazon-Book}{0}
\tableArticleResultApp{HCCF}{HCCF}{movielens10m}{MovieLens10M}{0}



\clearpage
\section{HAKG: Hierarchy-Aware Knowledge Gated Network for Recommendation}
\citet{DBLP:conf/sigir/0002ZCZG22} presents \emph{Hierarchy-Aware Knowledge Gated Network} (HAKG), which aims to combine graphs obtained with collaborative interactions as well as  knowledge-based. The goal of the paper is to exploit the hierarchical structure of knowledge graphs as well as the "higher order" relations in collaborative data. The paper claims that it is not sufficient to use a Euclidean space for this purpose, and therefore the embeddings are represented in hyperbolic space. The paper proposes a hierarchy-aware modeling strategy which includes an aggregation function for hyperbolic embeddings and a constraint on the angles generated by the embeddings involved, aiming at better preserving their hierarchical structure. The aggregation function is computed in Euclidean space, so the embeddings are converted from hyperbolic to Euclidean space, then aggregated, and then converted back to Hyperbolic space.
Knowledge-based and collaborative embeddings are separate (dual embeddings) and are fused with a "learnable gating fusion unit", which learns a weight matrix. 
The final prediction is computed with the cosine similarity of embeddings. 
The original implementation is available on GitHub.\footnote{\url{https://github.com/zealscott/HAKG}}

\subsection{Datasets} 
The evaluation is performed on the datasets described and processed as follows, their statistics are reported in Table \ref{tab:HAKG-dataset_stats}.

\begin{description}
\item[Alibaba-iFashion:] Is a datasets of outfits for garment recommendation. The data is preprocessed with 10-cores subgraph selection.
\item[Yelp2018:] Is a business reviews dataset. The data is preprocessed with 10-cores subgraph selection.
\item[Last-FM:] Is a dataset for song recommendation. The split is the same used in KGAT, including the knowledge base \cite{DBLP:conf/kdd/Wang00LC19}. The preprocessing filters the data retaining the interactions from Jan 2015 to June 2015, followed by 10-cores subgraph selection.
\end{description}

The two-hop neighbor entities of items in the knowledge base are used to construct the item knowledge graph for each dataset. All existing relations are considered as hierarchical. A 10-core subgraph selection is applied on the entities in the knowledge base as well.

\begin{table}[h]
    \small
    \begin{tabular}{lcccc}
    \toprule
    Dataset			& Interactions	& Items		& Users 	& Sparsity	\\
    \midrule
    Alibaba-iFashion &       1781093 &  30040 & 114737 &  $ 0.9995 $ \\
    Yelp2018 &       1183610 &  45538 &  45919 &  $ 0.9994 $ \\
    Last-FM &       1542856 &  48123 &  23566 &  $ 0.9986 $ \\
	\bottomrule
   	\end{tabular}
    \caption{Dataset statistics for HAKG.}
    \label{tab:HAKG-dataset_stats}
\end{table}

\subsection{Results} 
The hyperparameter values used in our experiments are reported in Table \ref{tab:HAKG-hyperparameter_values} and the results for all the datasets and baseline algorithms are reported in Table 
\ref{tab:HAKG-alibaba-ifashion_original-result} (Alibaba-iFashion),
\ref{tab:HAKG-yelp2018_original-result} (Yelp2018), and
\ref{tab:HAKG-last-fm_original-result} (Last-FM).

\begin{table}[h]
    \begin{minipage}{\textwidth}
    \centering
    \footnotesize
    \begin{tabular}{lccccc}
    \toprule
    Hyperparameter	& Described in	& \multicolumn{4}{c}{Value}	\\
                	&                      & All datasets	& Alibaba-iFashion   & Yelp2018      & Last-FM\\    
    \midrule
    embedding size	 	              & Paper      & 64  	  & -  	   & -  	& -  	\\
    optimizer	 	                  & Paper 	   & Adam	  & -  	   & -  	& -  	\\
    batch size	 	                  & Paper      & 4096	  & -  	   & -  	& -  	\\
    weight of angle loss w	  & Source code & $5\cdot 10^{-3}$	      & -  	& -  	& -  	\\
    learning rate	 	              & Source code & -	 	  & $10^{-4}$ 	    & $5\cdot 10^{-4}$  	& $10^{-4}$  	\\
    GNN layers	 	                  & Source code & -       & 3  	   & 2  	& 3  	\\
    negative samples $|M_u|$	      & Paper      & -	      & 200    & 400  	& 400  	\\
    margin $m$	 	                  & Paper      & -        & 0.6   & 	 0.8  	& 0.7  	\\
	\bottomrule
   	\end{tabular}
   	\end{minipage}
    \caption{Hyperparameter values for HAKG.}
    \label{tab:HAKG-hyperparameter_values}
\end{table}

\tableArticleResultApp{HAKG}{HAKG}{alibaba-ifashion_original}{Alibaba iFashion}{0}
\tableArticleResultApp{HAKG}{HAKG}{yelp2018_original}{Yelp2018}{0}
\tableArticleResultApp{HAKG}{HAKG}{last-fm_original}{Last-FM}{0}

\clearpage
\section{Graph Trend Filtering Networks for Recommendation}
\citet{DBLP:conf/sigir/FanL0ZT022} presents \emph{Graph Trend Filtering Networks for Recommendation} (GTN), which proposes a method to adaptively capture the reliability of interactions. This is done with a new \emph{smoothness} constraint on the embeddings, which in practice penalizes the occurrence of interactions between users and items with very different embeddings. The paper then proposes to use the Proximal Alternating Predictor-Corrector method and formulates an iterative solver requiring three steps. 
The original implementation is available on GitHub.\footnote{\url{https://github.com/wenqifan03/GTN-SIGIR2022}}

\subsection{Datasets} 
The evaluation is performed on the datasets described and processed as follows, their statistics are reported in Table \ref{tab:GTN-dataset_stats}.

\begin{description}
\item[Gowalla:] Is a dataset collected from a social network where users check-in locations they
visited. The split is the same used in LightGCN \cite{DBLP:conf/sigir/0001DWLZ020}, see section \ref{sec:lightGCN}.
\item[Yelp2018:] Is a business reviews dataset. The split is the same used in LightGCN \cite{DBLP:conf/sigir/0001DWLZ020}, see section \ref{sec:lightGCN}.
\item[Amazon-Book:] Is a dataset of book purchases on Amazon. The split is the same used in LightGCN \cite{DBLP:conf/sigir/0001DWLZ020}, see section \ref{sec:lightGCN}.
\item[Last-FM:] Is a dataset for song recommendation. The split is the same used in KGAT, including the knowledge base \cite{DBLP:conf/kdd/Wang00LC19}. The preprocessing filters the data retaining the interactions from Jan 2015 to June 2015, followed by 10-cores subgraph selection.
\end{description}

\begin{table}[h]
    \small
    \begin{tabular}{lcccc}
    \toprule
    Dataset			& Interactions	& Items		& Users 	& Sparsity	\\
    \midrule
    Gowalla     &       1027370 &  40981 &  29858 &  $ 0.9992 $ \\
    Yelp2018   &       1561406 &  38048 &  31668 &  $ 0.9987 $ \\
    Amazon-Book &       2984108 &  91599 &  52643 &  $ 0.9994 $ \\
    Last-FM	 	&       1542856 &  48123 &  23566 &  $ 0.9986 $ \\
	\bottomrule
   	\end{tabular}
    \caption{Dataset statistics for GTN.}
    \label{tab:GTN-dataset_stats}
\end{table}

\subsection{Results} 
The hyperparameter values used in our experiments are reported in Table \ref{tab:GTN-hyperparameter_values} and the results for all the datasets and baseline algorithms are reported in Table 
\ref{tab:GTN-yelp2018_original-result} (Yelp2018),
\ref{tab:GTN-amazon-book_original-result} (Amazon-Book),
\ref{tab:GTN-gowalla_original-result} (Gowalla), and
\ref{tab:GTN-last-fm_original-result} (Last-FM).

\begin{table}[h]
    \begin{minipage}{\textwidth}
    \centering
    \footnotesize
    \begin{tabular}{lcccccc}
    \toprule
    Hyperparameter	& Described in	& \multicolumn{5}{c}{Value}	\\
                	&               & All datasets	& Gowalla     &  Yelp2018    &  Amazon-Book &  Last-FM   \\    
    \midrule
    embedding size	 	              & Source code    & 256  	  & -  	   & -  	& -  	& -  	\\
    optimizer	 	                  & Paper 	       & Adam	  & -  	   & -  	& -  	& -  	\\
    batch size	 	                  & Source code    & 2048     & -  	   & -  	& -  	& -  	\\
    epochs	 	                      & Paper          & 1000 	  & -  	   & -  	& -  	& -  	\\
    learning rate	 	              & Source code    & $10^{-3}$  & -    & -  	& -  	& -  	\\
    GNN layers	 	                  & Paper          & -        & 3  	   & 3  	& 2  	& 3  	\\
    embedding smoothness weight\footnote{In the source code it is called \emph{lambda2}.}       & Paper          & 3        & -  	   & -  	& -  	& -  	\\  
    l2 regularization	 	          & Source code    & $10^{-4}$  & -    & -  	& -  	& -  	\\
    dropout rate LightGCN\footnote{In the source code it is called \emph{keep\_prob} and is 0.6, hence dropout is 0.4.}    & Source code    & 0.4 	  & -  	   & -  	& -  	& -  	\\
    dropout rate GTN\footnote{In the source code is called \emph{prop\_dropout}.}     & Source code    & 0.1 	  & -  	   & -  	& -  	& -  	\\

    ogb    & Paper 	       & True	  & -  	   & -  	& -  	& -  	\\
    incnorm\_para      & Paper 	       & True	  & -  	   & -  	& -  	& -  	\\
    \bottomrule
    \end{tabular}
    \end{minipage}
    \caption{Hyperparameter values for GTN.}
    \label{tab:GTN-hyperparameter_values}
\end{table}

\tableArticleResultApp{GTN}{GTN}{yelp2018_original}{Yelp2018}{0}
\tableArticleResultApp{GTN}{GTN}{amazon-book_original}{Amazon-Book}{0}
\tableArticleResultApp{GTN}{GTN}{gowalla_original}{Gowalla}{0}
\tableArticleResultApp{GTN}{GTN}{last-fm_original}{Last-FM}{0}

\clearpage
\section{Knowledge Graph Contrastive Learning for Recommendation}
\citet{DBLP:conf/sigir/YangHXL22} presents \emph{Knowledge Graph Contrastive Learning framework} (KGCL), aiming to reduce the impact of noisy knowledge bases, this is done with a knowledge graph augmentation schema that guides a contrastive learning process. KGCL uses a parameterized attention matrix on the concatenation of the user and item embeddings to calculate an estimation of relevance between the two. KGCL also uses TransE, which is a translation aware loss function aiming to ensure that the embedding of the head entity + the embedding of the relation is close to the embedding of the tail entity (\idest $e_h + e_r \approx e_t$). The training is done with contrastive learning and multiple views are created with a graph augmentation scheme which aims to identify items that are less sensitive to structure (edges) variations, the contrastive learning process is also guided by the knowledge based. 
The original implementation is available on GitHub.\footnote{\url{https://github.com/yuh-yang/KGCL-SIGIR22}}

\subsection{Datasets} 
The evaluation is performed on the datasets described and processed as follows, their statistics are reported in Table \ref{tab:KGCL-dataset_stats}.

\begin{description}
\item[Yelp2018:] Is a business reviews dataset. The preprocessing is a 10-cores subgraph selection. The split is the same used in HAKG \cite{DBLP:conf/sigir/0002ZCZG22}. The entities are collected in the same way as KGAT \cite{DBLP:conf/kdd/Wang00LC19}
\item[Amazon-Book:] Is a dataset of book purchases on Amazon. The preprocessing is a 10-cores subgraph selection. The entities are collected in the same way as KGAT \cite{DBLP:conf/kdd/Wang00LC19}
\item[MIND:] is a news recommendation dataset. The data and knowledge base are collected in the same way as \cite{DBLP:conf/sigir/TianYRWWWL21}, by randomly sampling one million users who had at least 5 news clicks during six weeks (\idest October 12 to November 22, 2019).
\end{description}

\begin{table}[h]
    \small
    \begin{tabular}{lcccc}
    \toprule
    Dataset			& Interactions	& Items		& Users 	& Sparsity	\\
    \midrule
    Amazon-Book     &        846434 &  24915 &  70679 &  $ 0.9995 $ \\
    Yelp2018        &       1183610 &  45538 &  45919 &  $ 0.9994 $ \\
    MIND            &       2545327 &  48957 & 300000 &  $ 0.9998 $ \\
	\bottomrule
   	\end{tabular}
    \caption{Dataset statistics for KGCL.}
    \label{tab:KGCL-dataset_stats}
\end{table}

\subsection{Results} 
The hyperparameter values used in our experiments are reported in Table \ref{tab:KGCL-hyperparameter_values} and the results for all the datasets and baseline algorithms are reported in Table 
\ref{tab:KGCL-amazon-book_original-result} (Amazon-Book),
\ref{tab:KGCL-yelp2018_original-result} (Yelp2018), and
\ref{tab:KGCL-MIND_original-result} (MIND).

\begin{table}[h]
    \begin{minipage}{\textwidth}
    \centering
    \footnotesize
    \begin{tabular}{lccccc}
    \toprule
    Hyperparameter	& Described in	& \multicolumn{4}{c}{Value}	\\
                	&               & All datasets	& Amazon-Book  & Yelp2018  & MIND\\    
    \midrule
    embedding size	 	& Paper	            & 64  	        & -  	& -  	& -  	\\
    learning date	 	& Paper		        & $10^{-3}$  	& -  	& -  	& $5 \cdot 10^{-4}$  	\\
    batch size	 	    & Paper		        & 2048  	    & -  	& -  	& -  	\\
    self supervised loss weight $\lambda_1$	& Paper	    & 0.1  	        & -  	& -  	& 0.06\footnote{For this dataset the source code uses a different value compared to the paper.}  	\\
    contrastive loss temperature $\tau$	 & Paper    & 0.2  	& -  	& -  	& -  	\\
    optimizer	 	    & Source code		& Adam  	    & -  	& -  	& -  	\\
    epochs	 	        & Source code	    & 1000 	        & -  	& -  	& -  	\\
    GNN layers K     	& Source code		& 3  	        & -  	& -  	& -  	\\
    GNN dropout rate 	& Source code		& 0.2  	        & 0.2  	& 0.2  	& 0.4  	\\
    entities per head	& Source code		& 10  	        & -  	& -  	& 6  	\\
    knowledge graph dropout rate	 	    & Source code		& 0.5  	        & -  	& -  	& 0.5  	\\
    user interaction dropout rate	 	    & Source code		& 0.001	        & 0.05  & 0.1  	& 0.4  	\\
    mix\_ratio\footnote{This hyperparameter appears to be used to add random samples as part of the user interaction dropout process, only when \emph{uicontrast} is "weighted-mix".}	 	    & Source code		& -  	        & 0.75\footnote{This hyperparameter is never used because it is not used when uicontrast is "weighted".}  & -  	& 0.6\footnote{Defined as 1-ui\_p\_drop.}  	\\
    uicontrast\footnote{This hyperparameter could impact how the graph augmentations are generated for the contrastive learning part, but the values are not described in the paper.}	 	    & Source code		& -             & "WEIGHTED"  	& "WEIGHTED"  	& "WEIGHTED-MIX"  	\\
    l2 regularization	& Source code		& $10^{-4}$\footnote{The TransR learning part had a hardcoded l2 regularization weight of $10^{-3}$, the ported version uses the one provided as hyperparameter.}     & -  	& -  	& $10^{-3}$  	\\
    learning rate milestones & Source code	& -  	        & [1500, 2500]\footnote{This hyperparameter has no impact because the epochs never reach 1500.}  	& [1500, 2500]\footnote{This hyperparameter has no impact because the epochs never reach 1500.}  	& [5, 10]  	\\
    \midrule
    min number of epochs\footnote{The patience and minimum number of epochs are different across the datasets, but the paper does not describe how were those values determined.}   & Source code        & -  	        & 15  	& 25  	& 1  	\\
    earlystopping patience & Source code        & -  	        & 5  	& 5  	& 3  	\\
	\bottomrule
   	\end{tabular}
   	\end{minipage}
    \caption{Hyperparameter values for KGCL.}
    \label{tab:KGCL-hyperparameter_values}
\end{table}

\tableArticleResultApp{KGCL}{KGCL}{amazon-book_original}{Amazon-Book}{0}
\tableArticleResultApp{KGCL}{KGCL}{yelp2018_original}{Yelp2018}{0}
\tableArticleResultApp{KGCL}{KGCL}{MIND_original}{MIND}{0}

\clearpage
\section{LightGCN: Simplifying and Powering Graph Convolution Network for Recommendation}
\label{sec:lightGCN}

\citet{DBLP:conf/sigir/0001DWLZ020} proposes LightGCN, a graph-based collaborative filtering method in which the user and item embeddings are propagated according to the graph adjacency matrix. LightGCN is presented as a "light" model based on message-passing, compared to previous more complex architectures.
The original implementation is available on Github.\footnote{\url{https://github.com/gusye1234/LightGCN-PyTorch}}

\subsection{Datasets} 
The evaluation is performed on the datasets described and processed as follows, their statistics are reported in Table \ref{tab:LightGCN_dataset}

\begin{description}
\item[Amazon Book:] Is a dataset of book purchases on Amazon. The preprocessing applies a 10-cores subgraph selection.
\item[Gowalla:] Is a dataset collected from a social network where users check-in locations they
visited. The preprocessing applies a 10-cores subgraph selection.
\item[Yelp2018:] Is a business reviews dataset. The preprocessing applies a 10-cores subgraph selection.
\end{description}

\begin{table}[h]
    \small
    \begin{tabular}{lcccc}
    \toprule
    Dataset			& Interactions	& Items		& Users 	& Sparsity	\\
    \midrule
    Amazon Book &       2984108 &  91599 &  52643 &  $ 0.9994 $ \\
    Gowalla     &       1027370 &  40981 &  29858 &  $ 0.9992 $ \\
    Yelp2018    &       1561406 &  38048 &  31668 &  $ 0.9987 $ \\
	\bottomrule
   	\end{tabular}
    \caption{Dataset statistics for LightGCN.}
    \label{tab:LightGCN_dataset}
\end{table}

\subsection{Results} 
The hyperparameter values used in our experiments are reported in Table \ref{tab:LightGCN-hyperparameter_values} and the results for all the datasets and baseline algorithms are reported in Table 
\ref{tab:LightGCN-gowalla_original-result} (Gowalla),
\ref{tab:LightGCN-amazon-book_original-result} (Amazon-Book Original Split),
\ref{tab:LightGCN-amazon-book_ours-result} (Amazon-Book Our Split),
\ref{tab:LightGCN-yelp2018_original-result} (Yelp 2018 Original Split), and
\ref{tab:LightGCN-yelp2018_ours-result} (Yelp 2018 Our Split).

\begin{table}[h]
    \begin{minipage}{\textwidth}
    \centering
    \footnotesize
    \begin{tabular}{lccccc}
    \toprule
    Hyperparameter	& Described in	& \multicolumn{4}{c}{Value}	\\
                	&               & All datasets     & Amazon Book   & Gowalla   & Yelp2018	\\    
    \midrule
    embedding size	 	& Paper	    & 64  	        & -  	& -  	& -  	 \\
    optimizer	 	    & Paper		& Adam	 	    & -  	& -  	& -  	 \\
    learning rate	 	& Paper		& $10^{-3}$	 	& -  	& -  	& -  	 \\
    batch size	 	    & Paper		& 1024	 	    & 2048 	& -  	& -  	 \\
    l2 reg	 	        & Paper		& $10^{-4}$	 	& -  	& -  	& -  	 \\
    dropout	 	        & Source code	& 0.0	 	& -  	& -  	& -  	 \\
    epochs	 	        & Paper		& 1000 (max) 	& -  	& -  	& -  	 \\
    GNN layers K	    & Paper		& 3		        & -  	& -  	& -  	 \\    
    $\alpha_k$	 	    & Paper		& $\frac{1}{1+K}$     & -  	& -  	& -  	 \\    
	\bottomrule
   	\end{tabular}
   	\end{minipage}
    \caption{Hyperparameter values for LightGCN.}
    \label{tab:LightGCN-hyperparameter_values}
\end{table}

\tableArticleResultApp{LightGCN}{LightGCN}{gowalla_original}{Gowalla}{0}
\tableArticleResultApp{LightGCN}{LightGCN}{amazon-book_original}{Amazon Book Original Split}{0}
\tableArticleResultApp{LightGCN}{LightGCN}{amazon-book_ours}{Amazon Book Our Split}{0}
\tableArticleResultApp{LightGCN}{LightGCN}{yelp2018_original}{Yelp 2018 Original Split}{0}
\tableArticleResultApp{LightGCN}{LightGCN}{yelp2018_ours}{Yelp 2018 Our Split}{0}



\clearpage
\section{Comparison of the Analyzed Methods of SIGIR 2022}
\label{sec:comparison}
This section reports the details of the experimental protocol of the comparative analysis between all SIGIR 2022 methods we analyze.

\subsection{Datasets}
The evaluation is performed on the datasets described and processed as follows, their statistics are reported in Table \ref{tab:ALL_dataset_stats}:
\begin{description}
\item[Amazon Book:] Is a dataset of book purchases on Amazon. The preprocessing applies a 10-cores subgraph selection. The entities for KGCL and HAKG are collected in the same way as KGAT \cite{DBLP:conf/kdd/Wang00LC19}.
\item[Yelp2018:] Is a business reviews dataset. The preprocessing applies a 10-cores subgraph selection. The entities for KGCL and HAKG are collected in the same way as KGAT \cite{DBLP:conf/kdd/Wang00LC19}.
\end{description}

\begin{table}[h]
    \centering
    \begin{tabular}{ccccc}
    \toprule
    Dataset		& Interactions	& Items	 & Users 	& Sparsity	\\
    \midrule
    Amazon Book & 846434 &  24915 &  70679 &  0.9995 \\
    Yelp2018    &       1183610 &  45538 &  45919 &  0.9994 \\
    \bottomrule
    \end{tabular}
    \caption{Datasets statistics for the comparison of the analyzed methods of SIGIR 2022.}
    \label{tab:ALL_dataset_stats}
\end{table}

\subsection{Hyperparameter Ranges}
In this section we report the hyperparameter ranges and distribution for all the GNN algorithms we analyze, see Table \ref{tab:hyperparameters_GNN_models} for the purely collaborative models and Table \ref{tab:hyperparameters_GNN_models-KB} for those including a Knowledge Base. Notice that some hyperparameters are searched for all models and are labeled as \emph{Common Hyperparameters} in Table \ref{tab:hyperparameters_GNN_models}. Overall, for each model there are between 9 and 16 hyperparameters.

\begin{table}[h]
    \begin{minipage}{\textwidth}
    \footnotesize
    \centering
    \begin{tabular}{cl|cccc}
    \toprule
    Algorithm	& Hyperparameter	&  Range	 & Type     & Distribution	\\ 
    \midrule
    \multirow{6}{*}{\begin{tabular}{c}Common  \\ Hyperparameters\end{tabular}}  	
    				&epochs	            & 1000   	& Categorical   & - 	\\ 
    				&batch size	        & 256, 512, 1024, 2048, 4096 & 	 Categorical   & - 	\\ 
    				&learning rate	    & $10^{-6}$ - $10^{-1}$ 	& Real   & log-uniform 	\\ 
                    &embedding size	    & 2 - 350   	& Integer   & uniform 	\\ 
                    &optimizer	        & Adam   	    & -   & - 	\\ 
    \midrule
    \multirow{6}{*}{\begin{tabular}{c}GDE\end{tabular}}  	
    				&beta	            & $10^{-1}$ - $10^{+2}$ 	& Real   & log-uniform 	\\ 
    				&feature type	    & smoothed, both 	        & Categorical   &  -	\\ 
    				&smooth ratio	    & $\frac{1}{\text{min(n\_users, n\_items)-1}}$ - $0.1$ 	& Real   & log-uniform 	\\ 
    				&rough ratio	    & $\frac{1}{\text{min(n\_users, n\_items)-1}}$ - $0.1$ 	& Real   & log-uniform 	\\ 
                    &loss type	         & adaptive, bpr  	& Categorical   & - 	\\
    				&dropout rate	    & $0.1$ - $0.9$ 	& Real   & uniform 	\\ 
    				&regularization rate	& $10^{-6}$ - $10^{-1}$ 	& Real   & log-uniform 	\\  
    \midrule
    \multirow{5}{*}{\begin{tabular}{c}GTN\end{tabular}}  	
                    &GNN layers K	                  & 1 - 6   	   & Integer   & uniform 	\\ 
                    &embedding smoothness weight	  & 1 - 15   	   & Integer   & uniform 	\\ 
    				&l2 reg	                & $10^{-6}$ - $10^{-1}$ 	& Real   & log-uniform 	\\ 
    				&dropout rate GTN	    & $0.1$ - $0.9$ 	& Real   & uniform 	\\ 
    				&dropout rate LightGCN  & $0.1$ - $0.9$ 	& Real   & uniform 	\\ 
    \midrule
    \multirow{8}{*}{\begin{tabular}{c}HCCF\end{tabular}}  
                    &GNN layers K	            & 1 - 6   	        & Integer       & uniform 	\\ 
                    &HYP layers C	            & 1 - 4   	        & Integer       & uniform 	\\ 	
                    &hyperedge size	            & 2, 350   	        & Integer       & uniform 	\\ 	
    				&dropout rate	            & $0.1$ - $0.9$ 	& Real          & uniform 	\\    
                    &l2 reg	                    & $10^{-6}$ - $10^{-1}$ 	& Real  & log-uniform 	\\   
                    &contrastive loss temperature $\tau$   & $10^{-2}$ - $10^{0}$ 	& Real  & log-uniform 	\\ 
                    &contrastive loss weight    & $10^{-7}$ - $10^{-1}$ 	& Real  & log-uniform 	\\   
                    &leaky relu slope	        & $0.01$   	    & -   & - 	\\  
    \midrule
    \multirow{7}{*}{\begin{tabular}{c}INMO\end{tabular}}  
                    &GNN layers K	            & 1 - 6   	        & Integer       & uniform 	\\    
                    &l2 reg	                    & $10^{-6}$ - $10^{-1}$ 	& Real  & log-uniform 	\\      
                    &template loss weight       & $10^{-4}$ - $10^{-1}$ 	& Real  & log-uniform 	\\   
                    &template node ranking metric    & degree, sort, page rank  	    & Categorical   & - 	\\
    				&dropout rate	            & $0.1$ - $0.9$ 	& Real          & uniform 	\\ 
    				&template ratio	            & $0.1$ - $1.0$ 	& Real          & uniform 	\\ 
                    &normalization decay	    & $0.99$   	    & -   & - 	\\  
    \midrule
    \multirow{6}{*}{\begin{tabular}{c}RGCF\end{tabular}}  
                    &GNN layers K	                & 1 - 6   	            & Integer       & uniform 	\\  
                    &prune threshold $\beta$        & $10^{-3}$ - $10^{0}$ 	& Real          & log-uniform 	\\  
                    &contrastive loss temperature $\tau$   & $10^{-2}$ - $10^{0}$ & Real    & log-uniform 	\\ 
                    &contrastive loss weight        & $10^{-7}$ - $10^{-1}$ 	  & Real    & log-uniform 	\\  
    				&augmentation ratio             & $0.01$ - $0.3$ 	          & Real    & log-uniform 	\\  
                    &l2 reg	                        & $10^{-6}$ - $10^{-1}$ 	  & Real    & log-uniform 	\\     
    \midrule
    \multirow{5}{*}{\begin{tabular}{c}SimGCL\end{tabular}}  
                    &GNN layers K	                & 1 - 6   	                  & Integer & uniform 	\\    
                    &noise magnitude $\epsilon$     & $10^{-2}$ - $10^{0}$        & Real    & log-uniform 	\\   
                    &contrastive loss temperature $\tau$   & $10^{-2}$ - $10^{0}$ & Real    & log-uniform 	\\ 
                    &contrastive loss weight        & $10^{-7}$ - $10^{-1}$ 	  & Real    & log-uniform 	\\  
                    &l2 reg	                        & $10^{-6}$ - $10^{-1}$ 	& Real  & log-uniform 	\\  
    \midrule
    \multirow{3}{*}{\begin{tabular}{c}LightGCN\end{tabular}}  
                    &GNN layers K	        & 1 - 6   	                & Integer   & uniform 	\\    
                    &l2 reg	                & $10^{-6}$ - $10^{-1}$     & Real      & log-uniform 	\\   
    				&dropout rate           & $0.1$ - $0.9$ 	        & Real      & uniform 	\\                    
	\bottomrule
   	\end{tabular}
   	\end{minipage}
    \caption{Hyperparameter ranges and distributions for the purely collaborative GNN models.}
    \label{tab:hyperparameters_GNN_models}
\end{table}

\begin{table}[h]
    \begin{minipage}{\textwidth}
    \footnotesize
    \centering
    \begin{tabular}{cl|cccc}
    \toprule
    Algorithm	& Hyperparameter	&  Range	 & Type     & Distribution	\\ 
    \midrule
    \multirow{9}{*}{\begin{tabular}{c}HAKG\end{tabular}}  	
                    &GNN layers K	            & 1 - 6   	        & Integer       & uniform 	\\ 
                    &angle loss weight	        & $10^{-6}$ - $10^{-1}$ 	& Real  & log-uniform 	\\     
                    &l2 reg	                    & $10^{-6}$ - $10^{-1}$ 	& Real  & log-uniform 	\\ 
                    &add KB inverse relation    & True, False  	    & Categorical   & - 	\\
    				&dropout rate node	        & $0.1$ - $0.9$ 	& Real          & uniform 	\\ 
    				&dropout rate mess	        & $0.1$ - $0.9$ 	& Real          & uniform 	\\ 
    				&dropout rate angle	        & $0.1$ - $0.9$ 	& Real          & uniform 	\\ 
                    &n negative samples M       & 100, 500  	    & Integer       & uniform 	\\  
    				&contrastive loss margin	& $0.1$ - $0.9$ 	& Real          & uniform 	\\ 
    \midrule
    \multirow{10}{*}{\begin{tabular}{c}KGCL\end{tabular}}  
                    &GNN layers K	                & 1 - 6   	        & Integer       & uniform 	\\    
                    &contrastive loss temperature $\tau$   & $10^{-2}$ - $10^{0}$ 	& Real  & log-uniform 	\\  
    				&GNN dropout rate	            & $0.1$ - $0.9$ 	& Real          & uniform 	\\ 
    				&knowledge graph dropout rate   & $0.1$ - $0.9$ 	& Real          & uniform 	\\ 
    				&user interaction dropout rate  & $0.1$ - $0.9$ 	& Real          & uniform 	\\ 
    				&mix ratio                      & $0.1$ - $0.9$ 	& Real          & uniform 	\\ 
                    &uicontrast                     & weighted, weighted-mix  	        & Categorical   & - 	\\
                    &entities per head              & 1, 20  	        & Integer       & uniform 	\\ 
                    &l2 reg	                        & $10^{-6}$ - $10^{-1}$ 	& Real  & log-uniform 	\\   
                    &self supervised loss weight    & $10^{-4}$ - $10^{-1}$ 	& Real  & log-uniform 	\\        
	\bottomrule
   	\end{tabular}
   	\end{minipage}
    \caption{Hyperparameter ranges and distributions for the GNN models that include a knowledge base.}
    \label{tab:hyperparameters_GNN_models-KB}
\end{table}

\begin{table}[h]
    \caption{Selected hyperparameter values for our Nearest-Neighbor Collaborative baselines.}
    \label{tab:ALL-hyperparameters_optimized-KNN}
    \footnotesize
    \centering
    \begin{tabular}{ll|cc}
\toprule
Algorithm & Hyperparameter &	 Yelp2018 &	 Amazon-Book \\
\midrule
\multirow{5}{*}{UserKNN CF} & topK &                                     397 &                              636 \\
       & shrink &                                       0 &                             1000 \\
       & similarity &                                  cosine &                           cosine \\
       & normalize &                                    True &                             True \\
       & feature weighting &                                  TF-IDF &                           TF-IDF \\
\midrule
\multirow{5}{*}{ItemKNN CF} & topK &                                     996 &                             1000 \\
       & shrink &                                     993 &                             1000 \\
       & similarity &                                  cosine &                           cosine \\
       & normalize &                                    True &                             True \\
       & feature weighting &                                  TF-IDF &                           TF-IDF \\
\bottomrule
\end{tabular}
\end{table}

\begin{table}[h]
    \caption{Selected hyperparameter values for our Graph-based baselines.}
    \label{tab:ALL-hyperparameters_optimized-graph}
    \footnotesize
    \centering    \begin{tabular}{ll|cc}
\toprule
Algorithm & Hyperparameter &	 Yelp2018 &	 Amazon-Book \\
\midrule
\multirow{3}{*}{P3alpha} & topK &                                     401 &                              201 \\
       & alpha &                                  0.6278 &                           0.4115 \\
       & normalize similarity &                                   False &                             True \\
\midrule
\multirow{4}{*}{\pbeta} & topK &                                     927 &                              680 \\
       & alpha &                                  0.3973 &                           0.4485 \\
       & beta &                                  0.2520 &                           0.2321 \\
       & normalize similarity &                                   False &                            False \\
\midrule
GF-CF & - &                                       - &                                - \\
\bottomrule
\end{tabular}
\end{table}

\begin{table}[h]
    \caption{Selected hyperparameter values for our Item-based Machine Learning baselines.}
    \label{tab:ALL-hyperparameters_optimized-item_based_ML}
    \footnotesize
    \centering    \begin{tabular}{ll|cc}
\toprule
Algorithm & Hyperparameter &	 Yelp2018 &	 Amazon-Book \\
\midrule
\multirow{1}{*}{\EASER} & l2 norm &                                       - &                         7.60E+01 \\
\midrule
\multirow{3}{*}{SLIM} & topK &                                    1000 &                             1000 \\
       & l1 ratio &                                1.15E-03 &                         2.08E-04 \\
       & alpha &                                  0.0010 &                           0.0010 \\
\midrule
\multirow{7}{*}{SLIM BPR} & topK &                                     880 &                              733 \\
       & epochs &                                     200 &                              420 \\
       & symmetric &                                    True &                             True \\
       & sgd mode &                                 adagrad &                          adagrad \\
       & lambda i &                                1.93E-04 &                         1.00E-05 \\
       & lambda j &                                1.00E-05 &                         1.00E-02 \\
       & learning rate &                                4.50E-02 &                         4.99E-03 \\
\midrule
NegHOSLIM & - &                                       - &                                - \\
\midrule
\multirow{5}{*}{NegHOSLIM (EN)} & feature pairs n &                                       6 &                                1 \\
       & topK &                                     777 &                             1000 \\
       & l1 ratio &                                3.38E-05 &                         1.81E-04 \\
       & alpha &                                  0.0273 &                           0.0010 \\
\bottomrule
\end{tabular}
\end{table}

\begin{table}[h]
    \caption{Selected hyperparameter values for our Matrix Factorization baselines.}
    \label{tab:ALL-hyperparameters_optimized-MF}
    \footnotesize
    \centering
    \begin{tabular}{ll|cc}
\toprule
Algorithm & Hyperparameter &	 Yelp2018 &	 Amazon-Book \\
\midrule
\multirow{7}{*}{MF-BPR} & sgd mode &                                 adagrad &                          adagrad \\
       & epochs &                                     540 &                             1500 \\
       & num factors &                                     200 &                              200 \\
       & batch size &                                    1024 &                             1024 \\
       & positive reg &                                2.78E-04 &                         2.55E-04 \\
       & negative reg &                                1.00E-02 &                         1.00E-02 \\
       & learning rate &                                4.31E-02 &                         1.00E-01 \\
\bottomrule
\end{tabular}
\end{table}

\begin{table}[h]
    \caption{Selected hyperparameter values for our Autoencoder baselines.}
    \label{tab:ALL-hyperparameters_optimized-AE}
    \footnotesize
    \centering
    \begin{tabular}{ll|cc}
\toprule
Algorithm & Hyperparameter &	 Yelp2018 &	 Amazon-Book \\
\midrule
\multirow{11}{*}{MultVAE} & epochs &                                     400 &                              500 \\
       & learning rate &                                9.22E-05 &                         3.68E-05 \\
       & sgd mode &                                 rmsprop &                          rmsprop \\
       & l2 reg &                                1.00E-06 &                         4.89E-05 \\
       & dropout &                                  0.6449 &                           0.2031 \\
       & anneal steps &                                  357709 &                           118728 \\
       & anneal cap &                                  0.1076 &                           0.3358 \\
       & batch size &                                    1024 &                              256 \\
       & encoding size &                                     503 &                              512 \\
       & layer size multiplier &                                       6 &                               10 \\
       & max n hidden layers &                                       1 &                                4 \\
\bottomrule
\end{tabular}
\end{table}

\begin{table}[h]
    \caption{Selected hyperparameter values for the purely collaborative GNN models.}
    \label{tab:ALL-hyperparameters_optimized-GNN-CF-A}
    \footnotesize
    \centering
    \begin{tabular}{ll|cc}
\toprule
Algorithm & Hyperparameter &	 Yelp2018 &	 Amazon-Book \\
\midrule
\multirow{12}{*}{GDE} & epochs &                                     845 &                                5 \\
       & batch size &                                    4096 &                             4096 \\
       & learning rate &                                1.00E-06 &                         1.02E-03 \\
       & embedding size &                                     344 &                              168 \\
       & sgd mode &                                    adam &                             adam \\
       & beta &                                  0.1000 &                           3.7201 \\
       & feature type &                                    both &                             both \\
\bottomrule
\end{tabular}
\end{table}

\begin{table}[h]
    \caption{Selected hyperparameter values for the purely collaborative GNN models.}
    \label{tab:ALL-hyperparameters_optimized-GNN-CF-B}
    \footnotesize
    \centering
    \begin{tabular}{ll|cc}
\toprule
Algorithm & Hyperparameter &	 Yelp2018 &	 Amazon-Book \\
\midrule
\multirow{10}{*}{SimGCL} & epochs &                                     100 &                               20 \\
       & batch size &                                     512 &                             2048 \\
       & learning rate &                                1.56E-04 &                         7.68E-04 \\
       & embedding size &                                     322 &                              344 \\
       & sgd mode &                                    adam &                             adam \\
       & GNN layers K &                                       6 &                                1 \\
       & noise magnitude $\epsilon$ &                                  0.0476 &                           0.0379 \\
\bottomrule
\end{tabular}
\end{table}

\begin{table}[h]
    \caption{Selected hyperparameter values for the GNN models that include a knowledge base.}
    \label{tab:ALL-hyperparameters_optimized-GNN-KB}
    \footnotesize
    \centering
    \begin{tabular}{ll|cc}
\toprule
Algorithm & Hyperparameter &	 Yelp2018 &	 Amazon-Book \\
\midrule
HAKG & - &                                       - &                                - \\
\midrule
\multirow{15}{*}{KGCL} & epochs &                                      25 &                              630 \\
       & batch size &                                    1024 &                              512 \\
       & learning rate &                                1.81E-03 &                         1.43E-04 \\
       & sgd mode &                                    adam &                             adam \\
       & GNN layers K &                                       4 &                                5 \\
       & contrastive loss temperature $\tau$ &                                  0.0342 &                           0.0123 \\
       & GNN dropout rate &                                4.91E-01 &                         3.50E-01 \\
       & knowledge graph dropout rate &                                4.79E-01 &                         3.39E-01 \\
       & user interaction dropout rate &                                8.38E-01 &                         3.84E-01 \\
       & embedding size &                                     211 &                              323 \\
       & mix ratio &                                  0.8547 &                           0.7095 \\
       & uicontrast &                            weighted-mix &                     weighted-mix \\
       & entities per head &                                      20 &                                7 \\
       & l2 reg &                                7.91E-05 &                         1.02E-04 \\
       & self supervised loss weight &                                  0.0122 &                           0.0002 \\
\bottomrule
\end{tabular}
\end{table}

\subsection{Results}
The values of the optimized hyperparameters for our baselines are reported in 
Table  \ref{tab:ALL-hyperparameters_optimized-KNN} (Nearest-Neighbor Collaborative), 
Table  \ref{tab:ALL-hyperparameters_optimized-graph} (Graph-based), 
Table  \ref{tab:ALL-hyperparameters_optimized-item_based_ML} (Item-based Machine Learning), 
Table  \ref{tab:ALL-hyperparameters_optimized-MF} (Matrix Factorization), and
Table  \ref{tab:ALL-hyperparameters_optimized-AE} (Autoencoder).
The values of the optimized hyperparameters for the GNN algorithms we analyze are reported in 
Table \ref{tab:ALL-hyperparameters_optimized-GNN-CF-A} and \ref{tab:ALL-hyperparameters_optimized-GNN-CF-B} (Collaborative), and 
Table  \ref{tab:ALL-hyperparameters_optimized-GNN-KB} (Knowledge Base).

The results for all the datasets and baseline algorithms are reported in \ref{tab:ALL-amazon-book-result} (Amazon-Book), and \ref{tab:ALL-yelp2018-result} (Yelp2018).

\begin{table}[h]
    \caption{Experimental results for all analyzed methods of SIGIR 2022 for the Amazon Book dataset.}
    \label{tab:ALL-amazon-book-result}
    \footnotesize
    \centering
    \begin{tabular}{l|cc|}
\toprule
{} & \multicolumn{2}{c}{Cutoff 20} \vline\\
{} &              Recall &             NDCG \\
\midrule
Random               &           0.0007 &           0.0003 \\
TopPop               &           0.0370 &           0.0168 \\
GlobalEffects        &           0.0003 &           0.0001 \\
\midrule
UserKNN CF           &           0.2301 &           0.1371 \\
ItemKNN CF           &           0.2436 &  \textbf{0.1496} \\
\palpha              &           0.2432 &  \textbf{0.1459} \\
\bottomrule
\end{tabular}
\end{table}

\begin{table}[h]
    \caption{Experimental results for all analyzed methods of SIGIR 2022 for the Yelp 2018 dataset.}
    \label{tab:ALL-yelp2018-result}
    \footnotesize
    \centering    \begin{tabular}{l|cc|}
\toprule
{} & \multicolumn{2}{c}{Cutoff 20} \vline\\
{} &              Recall &             NDCG \\
\midrule
Random               &           0.0004 &           0.0003 \\
TopPop               &           0.0213 &           0.0132 \\
GlobalEffects        &           0.0004 &           0.0002 \\
\midrule
UserKNN CF           &           0.0988 &           0.0668 \\
ItemKNN CF           &           0.1057 &           0.0718 \\
\palpha              &           0.1033 &           0.0682 \\
\bottomrule
\end{tabular}
\end{table}

\clearpage
\section{Baseline Hyperparameter Ranges}
\label{sec:hyperparameter_range}

In this section we report the hyperparameter ranges and distribution for all the baselines in our experiments, see Table \ref{tab:hyperparameters_our_baselines-KNN} (Nearest-Neighbor Collaborative and Content-Based),
\ref{tab:hyperparameters_our_baselines-graph_based} (Graph-based),
\ref{tab:hyperparameters_our_baselines-IB_ML} (Item-based Machine Learning), 
\ref{tab:hyperparameters_our_baselines-MF} (Matrix Factorization), 
\ref{tab:hyperparameters_our_baselines-FM} (Factorization Machines Collaborative and Hybrid), and
\ref{tab:hyperparameters_our_baselines-AE} (Autoencoder).

\makeatletter
\newcommand\footnoteref[1]{\protected@xdef\@thefnmark{\ref{#1}}\@footnotemark}
\makeatother

\begin{table}[h]
    \begin{minipage}{\textwidth}
    \footnotesize
    \centering
    \begin{tabular}{cl|cccc}
    \toprule
    Algorithm	& Hyperparameter	&  Range	 & Type     & Distribution	\\ 
    \midrule
    \multirow{5}{*}{\begin{tabular}{c}UserKNN, ItemKNN \\ UserKNN CBF\\ ItemKNN CBF\end{tabular}}  	
    				&topK	        & 5 - 1000 	& Integer   & uniform 	\\ 
    				&shrink	        & 0 - 1000 	& Integer   & uniform 	\\ 
    				&similarity	    & cosine 	& Categorical 	& 	\\ 
    				&normalize\footnote{\label{foot:knn_normalize}The \emph{normalize} hyperparameter in KNNs refers to the use of the denominator when computing the similarity.} 	    & True, False 	& Categorical 	& 	\\ 
    				&feature weighting	& none, TF-IDF, BM25 	& Categorical 	& 	\\ 
    \midrule
    \multirow{5}{*}{\begin{tabular}{c}UserKNN CFCBF \\ ItemKNN CFCBF\end{tabular}}  	
    				&topK	        & 5 - 1000 	& Integer   & uniform 	\\ 
    				&shrink	        & 0 - 1000 	& Integer   & uniform 	\\ 
    				&similarity	    & cosine 	& Categorical 	& 	\\ 
    				&normalize\footnoteref{foot:knn_normalize} 	    & True, False 	& Categorical 	& 	\\ 
    				&feature weighting	& none, TF-IDF, BM25 	& Categorical 	& 	\\ 
                    &ICM or UCM weight	&$10^{-2}$ - $10^{+2}$     & Real   & log-uniform 	\\
	\bottomrule
   	\end{tabular}
   	\end{minipage}
    \caption{Hyperparameter ranges and distributions for our Nearest-Neighbor Collaborative and Content-Based baselines.}
    \label{tab:hyperparameters_our_baselines-KNN}
\end{table}

\begin{table}[h]
    \begin{minipage}{\textwidth}
    \footnotesize
    \centering
    \begin{tabular}{cl|cccc}
    \toprule
    Algorithm	& Hyperparameter	&  Range	 & Type     & Distribution	\\ 
    \midrule
    \multirow{3}{*}{\palpha}  	
     				&topK	        & 5 - 1000 	& Integer   & uniform 	\\
     				&alpha	        & 0 - 2	& Real   & uniform 	\\ 
     				&normalize similarity\footnote{\label{foot:normalize_similarity}The \emph{normalize similarity} hyperparameter refers to applying L1 regularization on the rows of the similarity matrix.}	& True, False 	& Categorical 	& 	\\ 
    \midrule
    \multirow{4}{*}{\pbeta}  	
    				&topK	        & 5 - 1000 	& Integer   & uniform 	\\ 
    				&alpha	        & 0 - 2	& Real   & uniform 	\\ 
    				&beta	        & 0 - 2	& Real   & uniform 	\\ 
    				&normalize similarity\footnoteref{foot:normalize_similarity}	& True, False 	& Categorical 	& 	\\ 
    \midrule
    \multirow{3}{*}{GF-CF}  	
    				&topK	        & 5 - 5000 	 & Integer   & uniform 	\\ 
    				&alpha	        &$10^{-3}$ - $10^{+3}$     & Real   & log-uniform 	\\
    				&num factors	& 1 - 350  	 & Integer   & uniform 	\\ 
	\bottomrule
   	\end{tabular}
   	\end{minipage}
    \caption{Hyperparameter ranges and distributions for our Graph-based baselines.}
    \label{tab:hyperparameters_our_baselines-graph_based}
\end{table}

\begin{table}[h]
    \begin{minipage}{\textwidth}
    \footnotesize
    \centering
    \begin{tabular}{ll|ccl}
    \toprule
    Algorithm	& Hyperparameter	&  Range	 & Type     & Distribution	\\ 
    \midrule
    \multirow{1}{*}{\EASER}  	
    				&l2 norm	& $10^{0}$ - $10^{+7}$     & Real   & log-uniform 	\\
    \midrule
    \multirow{3}{*}{SLIM}  	
    				&topK	        & 5 - 1000 	& Integer   & uniform 	\\ 
    				&l1 ratio	    & $10^{-5}$ - $10^{0}$     & Real   & log-uniform 	\\ 
    				&alpha	        & $10^{-3}$ - $10^{0}$     & Real   & uniform 	\\ 
    \midrule
    \multirow{7}{*}{SLIM BPR }  	
     				&topK	        & 5 - 1000 	& Integer   & uniform 	\\ 
     				&epochs	        & 1 - 1500 	& Integer 	& early-stopping 	\\ 
     				&symmetric	    & True, False 	& Categorical 	& 	\\ 
     				&sgd mode	    & sgd, adam, adagrad    & Categorical 	& 	\\ 
     				&lambda i   	& $10^{-5}$ - $10^{-2}$     & Real   & log-uniform 	\\ 
     				&lambda j   	& $10^{-5}$ - $10^{-2}$     & Real   & log-uniform 	\\ 
     				&learning rate 	& $10^{-4}$ - $10^{-1}$     & Real   & log-uniform 	\\ 
    \midrule
    \multirow{4}{*}{NegHOSLIM}  
                    &epochs	        & 1 - 300\footnote{\label{foot:epochs_lower_slow}The number of epochs is lower due to the algorithm being slower, but converging in a lower number of epochs.} 	& Integer 	& early-stopping 	\\
    				&feature pairs n      & 1 - 1000 	& Integer   & uniform 	\\ 
                    &lambdaBB	          & $1$ - $10^{7}$     & Real   & log-uniform 	\\ 
                    &lambdaCC	          & $1$ - $10^{7}$     & Real   & log-uniform 	\\ 
                    &rho	              & $1$ - $10^{7}$     & Real   & log-uniform 	\\  
    \midrule\multirow{4}{*}{NegHOSLIM (EN)}  	
    				&feature pairs n      & 1 - 1000 	& Integer   & uniform 	\\ 
                    &topK	              & 5 - 1000 	& Integer   & uniform 	\\               
    				&l1 ratio	          & $10^{-5}$ - $10^{0}$     & Real   & log-uniform 	\\ 
    				&alpha	              & $10^{-3}$ - $10^{0}$     & Real   & uniform 	\\  
	\bottomrule
   	\end{tabular}
   	\end{minipage}
    \caption{Hyperparameter ranges and distributions for our Item-based Machine Learning baselines.}
    \label{tab:hyperparameters_our_baselines-IB_ML}
\end{table}

\begin{table}[h]
    \begin{minipage}{\textwidth}
    \footnotesize
    \centering
    \begin{tabular}{ll|ccl}
    \toprule
    Algorithm	& Hyperparameter	&  Range	 & Type     & Distribution	\\ 
    \midrule
    \multirow{7}{*}{MF-BPR}  	
    				&num factors	& 1 - 200\footnoteref{foot:num_factor_lower_slow}  	& Integer   & uniform 	\\ 
    				&epochs	        & 1 - 1500 	& Integer 	& early-stopping 	\\
    				&sgd mode	    & sgd, adam, adagrad    & Categorical 	& 	\\
    				&batch size	    & $2^{0}$ - $2^{10}$    & Integer 	& log-uniform	\\ 
    				&positive reg  	& $10^{-5}$ - $10^{-2}$     & Real   & log-uniform 	\\ 
    				&negative reg  	& $10^{-5}$ - $10^{-2}$     & Real   & log-uniform 	\\ 
    				&learning rate 	& $10^{-4}$ - $10^{-1}$     & Real   & log-uniform 	\\ 
    \midrule
    \multirow{7}{*}{MF-WARP}  	
    				&num factors	& 1 - 200\footnoteref{foot:num_factor_lower_slow}  	& Integer   & uniform 	\\ 
    				&epochs	        & 1 - 1500 	& Integer 	& early-stopping 	\\
    				&sgd mode	    & sgd, adam, adagrad    & Categorical 	& 	\\
    				&batch size	    & $2^{0}$ - $2^{10}$    & Integer 	& log-uniform	\\ 
    				&positive reg  	& $10^{-5}$ - $10^{-2}$     & Real   & log-uniform 	\\ 
    				&negative reg  	& $10^{-5}$ - $10^{-2}$     & Real   & log-uniform 	\\ 
    				&learning rate 	& $10^{-4}$ - $10^{-1}$     & Real   & log-uniform 	\\ 
                    &neg item attempts 	& 5, 10, 15, 20     & Categorical   & \\ 
     \midrule
     \multirow{9}{*}{SVDpp}  	
                    &num factors	& 1 - 200\footnote{\label{foot:num_factor_lower_slow}The number of factors is lower than PureSVD or NFM due to the algorithm being slower.} 	& Integer   & uniform 	\\ 
                    &epochs	        & 1 - 500\footnote{\label{foot:epochs_lower_slow}The number of epochs is lower than SLIM BPR or MF BPR due to the algorithm being slower.} 	& Integer 	& early-stopping 	\\
                    &use bias	    & True, False 	& Categorical 	& 	\\ 
    				&sgd mode	    & sgd, adam, adagrad    & Categorical 	& 	\\
    				&batch size	    & $2^{0}$ - $2^{10}$    & Integer 	& log-uniform	\\ 
    				&item reg   	& $10^{-5}$ - $10^{-2}$     & Real   & log-uniform 	\\ 
    				&user reg   	& $10^{-5}$ - $10^{-2}$     & Real   & log-uniform 	\\ 
    				&learning rate  & $10^{-4}$ - $10^{-1}$     & Real   & log-uniform 	\\ 
    				&negative quota\footnote{The \emph{negative quota} is the percentage of samples chosen among items unobserved by the user, having a target rating of 0.}	& 0.00 - 0.50 	& Real   & uniform 	\\ 
     \midrule
     \multirow{1}{*}{PureSVD}  	
     				&num factors	& 1 - 350 	& Integer   & uniform 	\\ 
     \midrule
     \multirow{4}{*}{NMF}  	
     				&num factors	& 1 - 350 	& Integer   & uniform 	\\ 
                        &init type	    & nndsvda, random                & Categorical 	& 	\\
     				&\shortstack{\phantom{0}\\solver	beta loss\\\phantom{0}}        & \shortstack{mult. update:frobenius, \\coord. descent:frobenius, \\coord. descent:kullback-leibler} 	 & \shortstack{\phantom{0}\\Categorical\\\phantom{0}} 	& 	\\
    \midrule
    \multirow{6}{*}{\iALS}  	
    				&num factors    & 1 - 200\footnote{\label{foot:num_factor_lower_slow}The number of factors is lower due to the algorithm being slower.} 	    & Integer       & uniform 	\\ 
    				&epochs	        & 1 - 500\footnote{\label{foot:epochs_lower_slow}The number of epochs is lower due to the algorithm being slower, but converging in a lower number of epochs.} 	& Integer 	& early-stopping 	\\
    				&confidence scaling	&linear, log    & Categorical 	& 	\\
    				&alpha	    & $10^{-3}$ - $5\cdot10^{+1}$ \footnote{\label{foot:hyperparameter_value_original_article}The maximum value of this hyperparameter had been suggested in the article proposing the algorithm.}      & Real   & log-uniform 	\\
    				&epsilon	& $10^{-3}$ - $10^{+1}$ \footnoteref{foot:hyperparameter_value_original_article}     & Real   & log-uniform 	\\
    				&reg	    & $10^{-5}$ - $10^{-2}$     & Real   & log-uniform 	\\ 
	\bottomrule
   	\end{tabular}
   	\end{minipage}
    \caption{Hyperparameter ranges and distributions for our Matrix Factorization baselines.}
    \label{tab:hyperparameters_our_baselines-MF}
\end{table}

\begin{table}[h]
    \begin{minipage}{\textwidth}
    \footnotesize
    \centering
    \begin{tabular}{ll|ccl}
    \toprule
    Algorithm	& Hyperparameter	&  Range	 & Type     & Distribution	\\ 
    \midrule        
    \multirow{7}{*}{LightFM CF}  	
     				&  epochs          	&   1 - 300       	& Integer          & early-stopping        	\\
                    &  n components     & 1 - 200          	&  Integer         & uniform        	\\ 
                    &  loss             &  Categorical      & bpr, warp, warp-kos     &  uniform      	\\ 
                    &  sgd mode         & Categorical       & adagrad, adadelta       & uniform       	\\ 
                    &  learning rate    & Real           	& $10^{-6} - 10^{-1}$     & log-uniform       	\\ 
                    &  item alpha       & Real          	& $10^{-5} - 10^{-2}$     &  log-uniform      	\\ 
                    &  user alpha       & Real           	& $10^{-5} - 10^{-2}$     &  log-uniform      	\\ 
    \midrule   
    \multirow{7}{*}{LightFM ItemHybrid}  	
     				&  epochs          	&   1 - 300       	& Integer          & early-stopping        	\\
                    &  n components     & 1 - 200          	&  Integer         & uniform        	\\ 
                    &  loss             &  Categorical      & bpr, warp, warp-kos     &  uniform      	\\ 
                    &  sgd mode         & Categorical       & adagrad, adadelta       & uniform       	\\ 
                    &  learning rate    & Real           	& $10^{-6} - 10^{-1}$     & log-uniform       	\\ 
                    &  item alpha       & Real          	& $10^{-5} - 10^{-2}$     &  log-uniform      	\\ 
                    &  user alpha       & Real           	& $10^{-5} - 10^{-2}$     &  log-uniform      	\\ 
	\bottomrule
   	\end{tabular}
   	\end{minipage}
    \caption{Hyperparameter ranges and distributions for our Factorization Machines Collaborative and Hybrid baselines.}
    \label{tab:hyperparameters_our_baselines-FM}
\end{table}

\begin{table}[h]
    \begin{minipage}{\textwidth}
    \footnotesize
    \centering
    \begin{tabular}{ll|ccl}
    \toprule
    Algorithm	& Hyperparameter	&  Range	 & Type     & Distribution	\\ 
    \midrule   
    \multirow{10}{*}{MultVAE}  	
                &epochs	        & 1 - 500\footnote{\label{foot:epochs_lower_slow}The number of epochs is lower due to the algorithm being slower, but converging in a lower number of epochs.} 	    & Integer 	& early-stopping 	\\
                &learning rate  & $10^{-6}$ - $10^{-2}$     & Real   & log-uniform 	\\ 
                &l2 reg         & $10^{-6}$ - $10^{-2}$     & Real   & log-uniform 	\\ 
                &dropout        & $0$ - $0.8$     & Real   & uniform 	\\ 
                &annealing steps  & $10^{5}$ - $6\cdot 10^{5}$     & Integer   & uniform \\ 
                &anneal cap     & $0$ - $0.6$     & Real   & uniform 	\\ 
                &batch size	    & 128, 256, 512, 1024                & Categorical 	& 	\\
                &encoding size  & 1 - 512     & Integer   & uniform 	\\ 
                &layer size multiplier\footnote{This hyperparameter is used to generate the decoder architecture. Starting from the encoding size the size of the next hidden layer is computer as the product of the previous one and the layer multiplier. The process terminates when either the desired number of hidden layers is reached or any further hidden layer added would exceed the size of the input data.}  & 2 - 10     & Integer   & uniform 	\\
                &max n hidden layers    & 2 - 4     & Integer   & uniform 	\\
	\bottomrule
   	\end{tabular}
   	\end{minipage}
    \caption{Hyperparameter ranges and distributions for our Autoencoder baselines.}
    \label{tab:hyperparameters_our_baselines-AE}
\end{table}

\end{document}